\pdfoutput=1
\documentclass[11pt,twoside,a4paper,cmspaper,final,collab]{cms-tdr}
\def\svnVersion{af18a9b}\def\svnDate{2023/06/16}\def\cmsCernNoTag{CERN-EP-2022-222}\def\cmsCernDate{\today}\def\cmsMessage{Published in the European Physical Journal C as \href{http://dx.doi.org/10.1140/epjc/s10052-023-11587-8}{\doi{10.1140/epjc/s10052-023-11587-8}.}}
\begin{document}\cmsNoteHeader{TOP-21-012}

\newlength\cmsFigWidth
\ifthenelse{\boolean{cms@external}}{\setlength\cmsFigWidth{0.45\textwidth}}{\setlength\cmsFigWidth{0.49\textwidth}}
\ifthenelse{\boolean{cms@external}}{\providecommand{\cmsLeft}{upper\xspace}}{\providecommand{\cmsLeft}{left\xspace}}
\ifthenelse{\boolean{cms@external}}{\providecommand{\cmsRight}{lower\xspace}}{\providecommand{\cmsRight}{right\xspace}}

\newcommand{\mjet}{\ensuremath{m_\text{jet}}\xspace}
\newcommand{\mtop}{\ensuremath{m_{\PQt}}\xspace}
\newcommand{\mw}{\ensuremath{m_{\PW}}\xspace}
\newcommand{\pti}{\ensuremath{p_{\text{T},i}}\xspace}
\newcommand{\ptjet}{\ensuremath{\pt^\text{jet}}\xspace}
\newcommand{\antikt}{anti-\kt}
\newcommand{\ptrel}{\ensuremath{p_\text{T}^{\text{rel}}}\xspace}
\newcommand{\tauratio}{\ensuremath{\tau_{32}}\xspace}
\newcommand{\fFSR}{\ensuremath{f_\text{FSR}}\xspace}
\newcommand{\muf}{\ensuremath{\mu_{\text{F}}}\xspace}
\newcommand{\mur}{\ensuremath{\mu_{\text{R}}}\xspace}
\newcommand{\hdamp}{\ensuremath{h_{\text{damp}}}\xspace}
\newcommand{\Wjets}{\ensuremath{\PW\text{+jets}}\xspace}
\newcommand{\pp}{\ensuremath{\Pp\Pp}\xspace}
\newcommand{\fb}{\unit{fb}}
\newcommand{\pb}{\unit{pb}}
\newcommand{\model}{\ensuremath{\,\text{(model)}}\xspace}
\newcommand{\exper}{\ensuremath{\,\text{(exp)}}\xspace}
\newcommand{\ptW}{\ensuremath{\pt^\PW}\xspace}
\newcommand{\ptratio}{\ensuremath{r_{\pt}}\xspace}
\newcommand{\fjec}{\ensuremath{f^\text{JEC}}\xspace}
\newcommand{\fxcone}{\ensuremath{f^\text{XCone}}\xspace}
\newcommand{\sigmajec}{\ensuremath{\sigma_\text{JEC}}\xspace}
\newcommand{\sigmaxcone}{\ensuremath{\sigma_\text{XC}}\xspace}
\newcommand{\cjec}{\ensuremath{c_\text{JEC}}\xspace}
\newcommand{\cxcone}{\ensuremath{c_\text{XC}}\xspace}
\newcommand{\asFSR}{\ensuremath{\alpS^{\text{FSR}}}\xspace}
\newcommand{\asFSRZ}{\ensuremath{\alpS^{\text{FSR}}(m_\PZ^2)}\xspace}
\newcommand{\asFSRmu}{\ensuremath{\alpS^{\text{FSR}}(\mu^2)}\xspace}
\newcommand{\dm}{\ensuremath{d_m}\xspace}
\newcommand{\Vm}{\ensuremath{V_m}\xspace}

\newlength\cmsTabSkip\setlength{\cmsTabSkip}{1ex}

\cmsNoteHeader{TOP-21-012}
\title{Measurement of the differential \texorpdfstring{\ttbar}{ttbar} production cross section as a function of the jet mass and 
extraction of the top quark mass in hadronic decays of boosted top quarks}

\titlerunning{Measurement of the jet mass and top quark mass in hadronic decays of boosted top quarks}

\date{\today}

\abstract{
{\tolerance=800
A measurement of the jet mass distribution in hadronic decays of Lorentz-boosted top quarks is presented. The measurement is performed in the lepton+jets channel of top quark pair production (\ttbar) events, where the lepton is an electron or muon. The products of the hadronic top quark decay are reconstructed using a single large-radius jet with transverse momentum greater than 400\GeV. The data were collected with the CMS detector at the LHC in proton-proton collisions and correspond to an integrated luminosity of 138\fbinv. The differential \ttbar production cross section as a function of the jet mass is unfolded to the particle level and is used to extract the top quark mass. The jet mass scale is calibrated using the hadronic \PW boson decay within the large-radius jet. The uncertainties in the modelling of the final state radiation are reduced by studying angular correlations in the jet substructure. These developments lead to a significant increase in precision, and a top quark mass of $173.06 \pm 0.84\GeV$.
\par}
}

\hypersetup{
pdfauthor={CMS Collaboration},
pdftitle={Measurement of the jet mass distribution and top quark mass in hadronic decays of Lorentz-boosted top quarks in proton-proton collisions at sqrt(s)=13 TeV},
pdfsubject={CMS},
pdfkeywords={CMS,  top quark mass, jet mass}}

\maketitle

\section{Introduction}
\label{sec:introduction}

The top quark is the most massive elementary particle discovered so far~\cite{Abe:1995hr, D0:1995jca}. 
Because of its high mass \mtop and its large Yukawa coupling 
it plays a crucial role in the electroweak sector of the standard model (SM) of particle physics.
Precise measurements of \mtop allow for stringent tests of the validity of the 
SM~\cite{ALEPH:2010aa, Haller:2018nnx, PDG:2020} and place constraints 
on the stability of the electroweak vacuum~\cite{Degrassi:2012ry, Bezrukov:2012sa, Bednyakov:2015sca}.

Direct measurements of \mtop using the top quark decay products have already achieved a 
precision of about 0.5\GeV~\cite{Aaboud:2016igd, Aaboud:2017mae, Aaboud:2018zbu,
Khachatryan:2015hba, Sirunyan:2017idq, Sirunyan:2018gqx, Sirunyan:2018mlv}.
In these measurements, observables with high sensitivity to the value of \mtop are constructed. 
Measured distributions in these observables are compared to detector level simulations 
to extract the value of \mtop that fits the data best.  
The predictions rely on a precise modelling of the parton shower and 
hadronisation process, which cannot be calculated from first principles, and are thus subject to 
corresponding systematic uncertainties. 
In addition, uncertainties of the size 0.5--1\GeV exist in the translation of \mtop extracted from event generators 
to a value of \mtop in a well-defined renormalisation scheme~\cite{Hoang:2017suc, Hoang:2020iah}, 
as used in precise analytic calculations in quantum field theory. 

A different approach is the determination of \mtop from cross section measurements corrected for 
detector effects. To facilitate a direct comparison to analytic calculations 
from first principles, these measurements have to be corrected to the parton level, 
which represents the \ttbar pair before its decay. The corrections applied need to include 
effects from the top quark decay and the hadronisation of its colour-charged decay products. 
The inclusive cross section of top quark pair (\ttbar) production can be measured precisely and has been 
used to extract a value of the top quark pole mass 
by a comparison to fixed-order calculations in perturbative quantum chromodynamics (QCD). 
Such measurements have been carried out by the  
\DZERO~\cite{Abazov:2011pta, Abazov:2016ekt}, 
ATLAS~\cite{Aad:2014kva, Aaboud:2017ujq, ATLAS:2019hau}, and   
CMS~\cite{Chatrchyan:2013haa, Khachatryan:2016mqs, Sirunyan:2018goh} Collaborations. 
These measurements of the total \ttbar cross section are sensitive to various sources of 
uncertainties, which can not be constrained in situ 
during the extraction of \mtop, resulting in a precision of about 2\GeV. 
Differential cross section measurements can also be used for 
measuring \mtop~\cite{ATLAS:2015pfy, ATLAS:2017dhr, ATLAS:2019guf, CMS:2022emx}.
A multi-differential cross section measurement has been performed by the CMS 
Collaboration, achieving an uncertainty of 0.8\GeV in the top quark pole mass~\cite{CMS:2019esx}. 
The shape of the measured distributions close to the \ttbar production threshold 
is sensitive to the value of \mtop, and a more precise result is achieved compared 
to the inclusive cross section measurements.

An alternative method which combines the advantages of the two approaches 
is the determination of \mtop from a measurement of the jet mass \mjet 
in events with Lorentz-boosted top quarks~\cite{Larkoski:2017jix, Asquith:2018igt, Kogler:2021kkw}. 
At high energies, the decay products of top quarks are Lorentz boosted and merge into a 
single large-radius jet. 
The peak position of the distribution in \mjet is sensitive to \mtop and allows for a precise 
measurement of \mtop~\cite{Hoang:2017kmk}. 
The unfolding of the data to the level of stable particles will allow 
for a comparison to analytic calculations in perturbative QCD, once these become available. 
This enables a measurement of the top quark pole mass from the shape of a 
distribution at the particle level.  
Presently, analytic calculations for \mjet are restricted to top quark transverse momenta 
$\pt>750\GeV$~\cite{Hoang:2017kmk}, a requirement which results in too few events in data for a 
differential cross section measurement using the current CERN LHC data sets. 
Previous measurements by the CMS Collaboration using proton-proton (\pp) collision data 
at $\sqrt{s}=8\TeV$~\cite{Sirunyan:2017yar} and 13\TeV~\cite{Sirunyan:2019rfa} 
with a top quark $\pt>400\GeV$, have reached an uncertainty of 2.5\GeV, 
where \mtop has been determined using event generators. The results are compatible with those obtained 
from \ttbar production at lower energy scales. 
In this article, we present a measurement of the differential cross section for \ttbar production 
as a function of the large-radius jet mass with significantly improved statistical and systematic uncertainties.  
The measurement is used to determine \mtop using event generators at next-to-leading order 
(NLO) precision in QCD. 
The approach is complementary to measurements close to threshold production 
with fully resolved final state objects. 
This provides a precise test of the validity of the approximations made 
in event generators and the corresponding systematic uncertainties. 

In the lepton+jets channel of \ttbar production, the final state 
is obtained from one top quark decaying to a \PQb quark and leptons, $\PQt \to \PQb\PW \to \PQb \Pell \PGn_{\!\ell}$, 
and the second decaying hadronically, $\PQt \to \PQb\PW \to \PQb \cPq \cPaq^\prime$. 
Here, the term lepton denotes an electron or muon. 
This final state combines the advantages of a clear signature from the leptonic \PW boson decay,
with a small background from events with jets from light-flavour quarks and gluons. 
The large \ttbar branching fraction for the lepton+jets channel also results in
large event samples. 
In addition, in case of \ttbar production with high top quark \pt, 
the hadronic decay allows the full reconstruction of the top quark decay within 
a single large-radius jet with $\pt>400\GeV$, provided that the decay products are produced within the detector acceptance. 
The lepton serves as a means to select \ttbar events, and the mass of the large-radius jet 
in the opposite hemisphere of the event is the measurable for this analysis. The lepton is not necessarily isolated, 
because the large Lorentz boost can result in particles from the fragmentation of the \PQb quark 
to be produced inside of the isolation cone around the lepton. The analysis strategy follows the one 
from the previous measurement at 13\TeV~\cite{Sirunyan:2019rfa}.

In this article, we analyse 13\TeV \pp collision data, 
recorded in the years 2016 to 2018 and corresponding to an integrated luminosity of 138\fbinv. 
Besides the improved statistical precision, the leading systematic uncertainties are reduced by 
using a dedicated calibration of the jet mass scale (JMS) and a detailed study of the effects 
from final state radiation (FSR) in large-radius jets. 

In the previous measurements of \mjet in boosted \ttbar events~\cite{Sirunyan:2017yar, Sirunyan:2019rfa},  
the uncertainties in the jet energy scale (JES) have been propagated to \mjet. 
For these the JES uncertainties are the leading experimental systematic uncertainties. 
While the JES, and therefore the jet momentum, can be 
determined precisely using the \pt balance or the MPF (missing transverse momentum projection fraction) 
methods~\cite{CMS:2011shu, Khachatryan:2016kdb}, these methods do not necessarily provide the most precise 
calibrations for \mjet. In this article, we calibrate the JMS by reconstructing the 
\PW boson mass from two subjets within the large-radius jet. A fit to data in the peak region of 
the jet mass results in a JMS with smaller uncertainties. 

The FSR is modelled by the parton showers in the event generators, which are matched to the 
simulation of the hard process. The value of the strong coupling used in the FSR shower, 
evaluated at the mass of the \PZ boson, \asFSRZ, is an important parameter that affects the amount of FSR. 
Changes in its value can cause large differences in the substructure of large-radius jets. 
Observables probing the angular distributions of the energy density within a jet, 
such as $N$-subjettiness~\cite{Thaler:2010tr,Thaler:2011gf} ratios, are very sensitive to the amount of FSR in the simulation. 
In this article, we measure distributions in $N$-subjettiness ratios calculated for large-radius jets, 
and use these to constrain the value of \asFSRZ used in the modelling of FSR. This leads to smaller uncertainties in 
\mjet from the FSR modelling compared to the usual variations of the scale $\mu$ in \asFSRmu~\cite{Skands:2014pea, Sirunyan:2019rfa, CMS:2018ypj}. 

{\tolerance=800
Tabulated results are provided in the HEPData record for this analysis~\cite{hepdata}.
\par}

\section{The CMS detector}
\label{sec:detector}

{\tolerance=800
The central feature of the CMS detector is a superconducting solenoid of 6\unit{m} internal diameter,
providing a magnetic field of 3.8\unit{T}. A silicon pixel and strip tracker, a lead tungstate crystal
electromagnetic calorimeter (ECAL), and a brass and scintillator hadron calorimeter (HCAL),
each composed of a central barrel and two endcap sections, reside within the solenoid volume.
Forward calorimeters extend the pseudorapidity ($\eta$) coverage provided by the barrel and endcap detectors.
Muons are detected in gas-ionisation chambers embedded in the steel flux-return yoke outside the solenoid.
A more detailed description of the CMS detector, together with a definition of the coordinate system,
can be found in Ref.~\cite{Chatrchyan:2008zzk}.
Between the 2016 and 2017 data taking runs, the CMS pixel detector was upgraded with additional layers 
in the barrel and endcap regions of the CMS detector. 
Details about the changes can be found in Ref.~\cite{CMSTrackerGroup:2020edz}.
\par}

Events of interest are selected using a two-tiered trigger system. The first
level, composed of custom hardware processors, uses information from the
calorimeters and muon detectors to select events at a rate of around
100\unit{kHz} within a fixed latency of about 4\mus~\cite{CMS:2020cmk}. The
second level, known as the high-level trigger, consists of a farm of
processors running a version of the full event reconstruction software optimised
for fast processing, and reduces the event rate to around 1\unit{kHz} before
data storage~\cite{Khachatryan:2016bia}.

\section{Data and simulated samples}
\label{sec:dataMC}

The measurement is performed in the lepton+jets final state of \ttbar production. 
The event selection is based on the presence of a single lepton which uses the 
data selected by single-lepton triggers~\cite{Khachatryan:2016bia, CMS:2020cmk}.
Muon candidates are required to have $\pt>50\GeV$ and $\abs{\eta}<2.4$, 
without any requirement on the isolation of the muon. 
In the electron channel, we use a combination of triggers. 
The first trigger requires electron candidates with $\abs{\eta}<2.5$ that are isolated and have a minimum 
\pt of 27, 35, or 32\GeV for the years 2016, 2017, and 2018, respectively.
A second trigger selects electron candidates with $\pt > 120\GeV$, without an isolation requirement. 
In addition a single-photon trigger is used for selecting electrons without 
a track requirement. This trigger selects photon candidates with a minimum \pt 
of 175\GeV in 2016, and 200\GeV in 2017 and 2018.
The photon trigger ensures a stable selection efficiency for electrons at high \pt 
because selection criteria applied to clusters in the ECAL are less strict 
than those used by the electron trigger.
In the offline analysis, muons and electrons are selected with $\abs{\eta}<2.4$ and $\pt>55\GeV$, 
ensuring that selected events are in the plateau region of the trigger efficiency. 
After this selection, the average efficiency of the muon trigger is 
91, 90, and 91\% for 2016, 2017, and 2018, respectively. 
The combination of the three electron and photon triggers provides high efficiency over the full 
range in \pt considered in this analysis, which is comparable to that obtained using the muon triggers.
For lepton $\pt<120\GeV$, the top quark decay is less collimated and the \PQb jet does 
not overlap with the lepton isolation cone. In this case, the event selection efficiency is 
greater than 90\% for triggers with an isolation requirement.
For $\pt>120\GeV$, the nonisolated electron trigger has an average efficiency of 95\%, 
increasing to nearly 100\% for $\pt>200\GeV$, where the high \pt efficiency is calculated 
in combination with the photon trigger. 
The total data set corresponds to an integrated luminosity of 138\fbinv, 
with 36.3\fbinv~\cite{lumi16}, 41.5\fbinv~\cite{lumi17}, and 59.7\fbinv~\cite{lumi18} 
recorded in the years 2016, 2017, and 2018, respectively.

For each of the three years of data taking, the processes relevant for this analysis are simulated 
individually using a Monte Carlo (MC) simulation technique and they are normalised to the integrated luminosity of each year. 
The \ttbar process is simulated at NLO using the
\POWHEG~v2~\cite{Nason:2004rx,Frixione:2007vw,Alioli:2010xd,Frixione:2007nw, Alioli:2009je,Re:2010bp} generator with 
a top quark mass of 172.5\GeV. We adjust the total cross section 
to 831.8\pb, obtained from a prediction at next-to-NLO~(NNLO) precision in QCD, 
including resummation of next-to-next-to-leading logarithmic 
soft gluon terms, using the computer program \TOPpp\,2.0~\cite{Czakon:2011xx}.
We simulate additional \ttbar samples with $\mtop=169.5$, 171.5, 173.5, and 175.5\GeV, 
which are used for studying the dependence of the measured cross section on the value of \mtop used in simulation, 
and for the extraction of \mtop. 
The background contribution from electroweak single \PQt production is generated at NLO 
using \POWHEG, and the \Wjets background is generated at leading order (LO) using \MGvATNLO v2.2.2~\cite{Alwall:2014hca, Frixione:2002ik}.
The cross section for single \PQt production in association with a \PW boson 
is adjusted to approximate NNLO calculations taken from Refs.~\cite{Kidonakis:2010ux,Kidonakis:2013zqa}.
The single top quark $s$- and $t$-channel cross sections are adjusted to predictions at NLO precision
obtained with \textsc{hathor}~v2.1~\cite{Aliev:2010zk}.
Events from Drell--Yan~(DY) production with additional jets are simulated at 
LO using \MGvATNLO and normalised to the NLO cross section~\cite{Li:2012wna}.
The production of two heavy gauge bosons with additional jets, and events in which jets are produced only
through QCD interactions are simulated at LO using the \PYTHIA event generator in 
version 8.212~\cite{Sjostrand:2014zea} for the simulation of 2016 data and version 8.230 for 2017 and 2018. 
The diboson and QCD multijet samples are referred to as ``other SM'' backgrounds in the following. 
The NNPDF3.0~\cite{Ball:2014uwa} parton distribution functions (PDFs) are used for
2016 simulations and the NNPDF3.1~\cite{NNPDF:2017mvq} PDFs are used for 2017 and 2018 simulations. 

{\tolerance=3650
In all processes, the hadronisation, parton showers, and multiple parton interactions are simulated with \PYTHIA.
In samples simulated with \MGvATNLO, the matrix element calculation is matched to the parton showers using the 
FxFx~\cite{Frederix:2012ps} and MLM~\cite{Alwall:2007fs} algorithms for NLO and LO, respectively.
In the simulation of 2016 data, \PYTHIA~8.212 is used with the underlying event~(UE) tune 
CUETP8M2T4~\cite{Sirunyan:2019dfx} for the simulation of \ttbar and single top quark production in the $t$ channel. 
In this tune, $\asFSRZ = 0.1365$ is used for the simulation of FSR. 
All other simulated samples in 2016 use the CUETP8M1~\cite{Khachatryan:2015pea,Skands:2014pea} tune. 
For the 2017 and 2018 data, \PYTHIA~8.230 is used with the CP5~\cite{Sirunyan:2019dfx} tune. 
Here, a value of $\asFSRZ = 0.118$ is used. 
The detector response is simulated with the \GEANTfour package~\cite{Agostinelli:2002hh, Allison:2006ve}. 
\par}

{\tolerance=800
Additional inelastic \pp collision events are simulated using \PYTHIA and superimposed on sim\-u\-lat\-ed
events to model the effect of additional \pp collisions within the same or adjacent bunch crossings (pileup). 
We use a total inelastic cross section of 69.2\unit{mb}~\cite{Sirunyan:2018nqx} to estimate the expected
number of \pp interactions per bunch crossing and correct the simulation to 
match the corresponding distribution to that observed in data.
\par}

\section{Event reconstruction}
\label{sec:reco}

The particle-flow (PF) algorithm~\cite{Sirunyan:2017ulk} aims to reconstruct and identify
each individual particle in an event, using an optimised combination of information from the
various elements of the CMS detector.
The candidate vertex with the largest sum of the square of the transverse momenta $\pt^2$ of the
physics objects is taken to be the leading primary \pp interaction vertex.
The physics objects are the jets, clustered using the \antikt jet finding 
algorithm~\cite{Cacciari:2008gp,Cacciari:2011ma} with a distance parameter of $R=0.4$ with tracks assigned
to candidate vertices as inputs, and the associated missing transverse momentum, 
taken as the negative vector sum of the \pt of those jets.
More details are given in Section~9.4.1 of Ref.~\cite{CMS-TDR-15-02}.

{\tolerance=800
Muons are reconstructed from tracks in the inner tracker and hits in the muon system using the PF algorithm. 
The muon momentum is obtained from the curvature of the corresponding track~\cite{Sirunyan:2018fpa}.
For electron reconstruction, clusters in the ECAL are connected to tracks in the inner tracker.
The electron energy is determined by a combination of the electron momentum at the primary interaction 
vertex as determined by the tracker,
the energy of the corresponding cluster in the ECAL, and the sum of all bremsstrahlung photons 
spatially compatible with originating from the elec\-tron track~\cite{CMS:2020uim}.  
The energy of photons is directly obtained from the ECAL measurement~\cite{CMS:2020uim}.  
Both muons and electrons have to pass tight quality criteria developed by the CMS Collaboration 
to ensure a proper reconstruction~\cite{Sirunyan:2018fpa, CMS:2020uim}.
The energy of charged hadrons is determined from a combination of their momentum measured 
in the tracker and the matching ECAL and HCAL energy deposits. 
Finally, the energy of neutral hadrons is obtained from the corresponding corrected 
ECAL and HCAL energy~\cite{Sirunyan:2017ulk}. 
\par}

Jets are reconstructed from PF candidates using the 
\antikt~\cite{Cacciari:2008gp} or the XCone (eXclusive Co\-ne)~\cite{Stewart:2015waa} algorithm as implemented in the
\FASTJET software package~\cite{Cacciari:2011ma}. Two sets of \antikt jets are obtained using 
distance parameters of $R=0.4$~(AK4 jets) and 0.8~(AK8 jets). In the jet clustering procedure, 
charged PF candidates are excluded if they are associated with pileup vertices. 
While AK4 jets are used mostly for the identification of \PQb jets in this analysis, 
AK8 jets are used to study the influence of FSR on the jet substructure as described in Section~\ref{sec:FSR}.
For the XCone jets, a specialized two-step clustering procedure~\cite{Thaler:2015xaa} is used.
Being an exclusive algorithm, XCone always returns a requested number of jets.  
This feature of the algorithm can be leveraged to efficiently reconstruct the boosted \ttbar final state.
At first, the XCone algorithm is run finding exactly two large-radius jets with a distance parameter of $R=1.2$. 
This step takes all PF candidates, after removing charged particles assigned to a pileup vertex, as an input
and aims to reconstruct the two top quark decays of the \ttbar process in separate jets.
As a second step, all PF candidates clustered into a large-radius XCone jet are input to the XCone algorithm again, 
which is now required to find three XCone subjets, $N_{\text{sub}}=3$, with a distance parameter $R_{\text{sub}}=0.4$. 
The second step aims to reconstruct the three-prong decay $\PQt \to \PQb \PW \to \PQb \cPq \cPaq^\prime$ 
while minimising the effects of uncorrelated soft radiation or additional energy deposits from pileup. 
The final XCone jets are then defined as the sum of the four-momenta of their respective subjets. 
In this way, all particles not clustered into the three subjets are removed from the large-radius XCone jets, 
similar to the trimming algorithm~\cite{Krohn:2009th}. 
The jet mass is calculated from the sum of the four-momenta of all particles clustered into the subjets.
Since no lepton selection has been applied at this stage, the XCone reconstruction 
will also reconstruct $\PQt \to \PQb \PW \to \PQb \Pell \PGn_{\!\ell}$ with three subjets.
We have verified that the difference from the more natural choice of two XCone subjets for the reconstruction of the 
leptonic decay does not significantly affect the identification of the leptonic XCone jet 
and the event reconstruction. 
The XCone jet with larger angular distance to the identified single lepton is selected as the 
measurement jet and is labelled ``XCone jet'' in the following.
The XCone jet closer to the lepton is referred to as ``second XCone jet''.
Here, the angular distance between two objects is defined as 
${\Delta}R = \sqrt{\smash[b]{({\Delta}\eta)^2+({\Delta}\phi)^2}}$, 
where $\phi$ is the azimuthal angle in radians.
The four-momenta of identified leptons are subtracted from AK4 jets and XCone subjets 
if they are within ${\Delta}R<0.4$ of the respective (sub)jet. 

Jet energy corrections (JECs)~\cite{Khachatryan:2016kdb} derived for AK4 
jets are applied to AK4 jets, as well as to XCone subjets in this analysis. 
These JECs include corrections for contributions from pileup, as derived for AK4 jets clustered 
after removing all charged particles assigned to a pileup vertex. 
Jet energies in simulated events are smeared to match the jet energy resolution (JER) observed in data.
The XCone subjets are corrected with the same procedure as in Ref.~\cite{Sirunyan:2019rfa}, 
where an additional XCone correction is derived because of residual differences to AK4 jets.
The correction is obtained from simulated samples of \ttbar in the all-jets channel 
and parametrised as a function of the XCone subjet \pt and $\abs{\eta}$. 
The XCone jet mass is calibrated as described in Section~\ref{sec:jms}. The JMS correction is applied 
to the four-momentum of the jet such that it changes only the mass but leaves the three-momentum unaltered. 

\section{Particle-level phase space}
\label{sec:particle}

The measurement of \mjet is carried out at the particle level in the fiducial region defined below. 
The particle level is defined by the set of all stable particles, \ie with a lifetime longer than $10^{-8}\unit{s}$ as provided by the event simulation. 
We develop an unfolding procedure to correct the data for detector and pileup effects. 
This procedure provides a measurement at the particle level.

\begin{figure}[tb]
  \centering
    \includegraphics[width=0.49\textwidth]{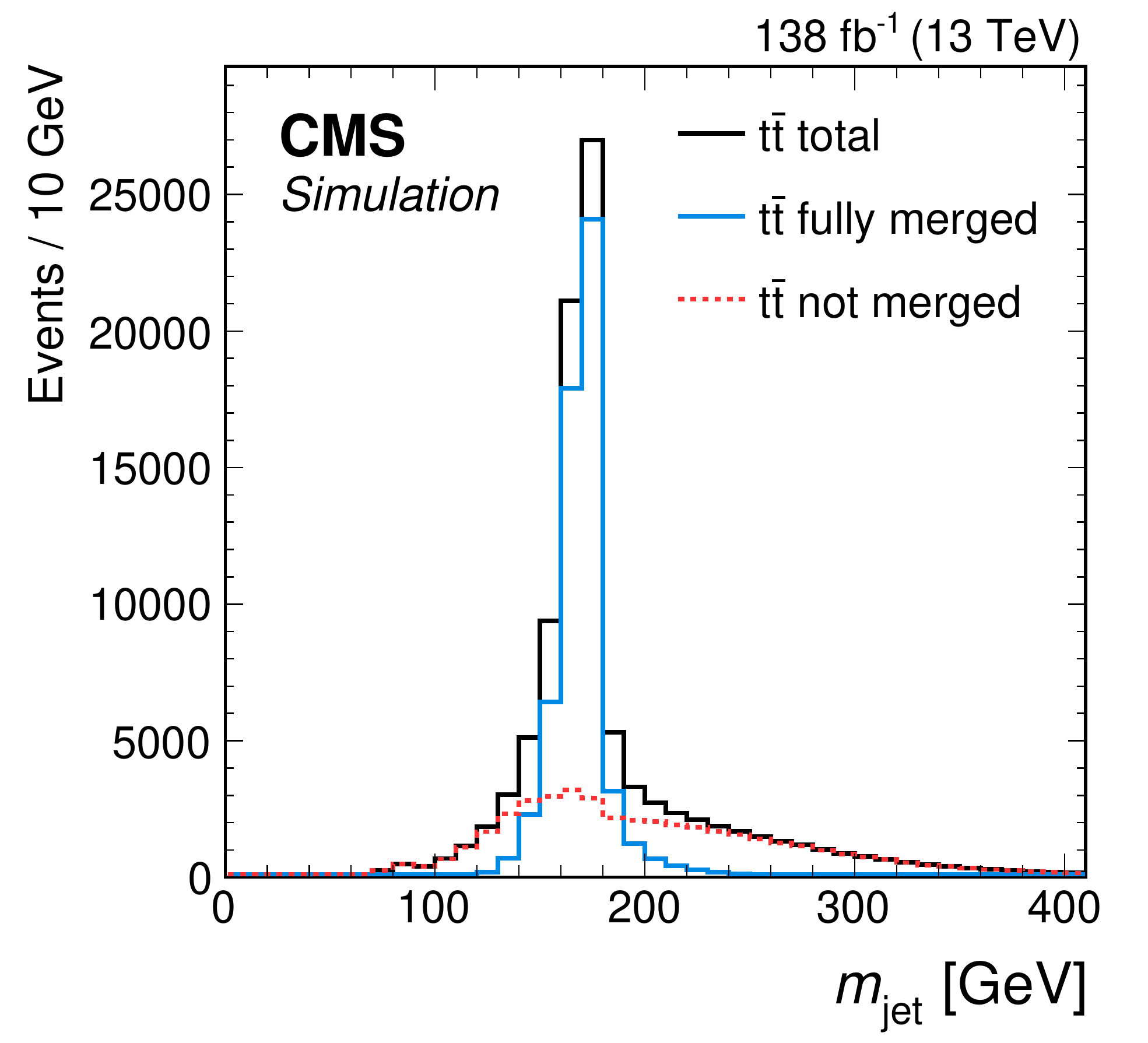}
    \caption{Distribution in \mjet at the particle level after the selection of the fiducial region in the lepton+jets 
    channel of \ttbar, simulated with \POWHEG.
    The contributions from fully merged events~(blue solid) and not merged events~(red dashed) are displayed, 
    as well as the sum of the two (black solid).
    \label{fig:particlelevel}}
\end{figure}
The fiducial region at the particle level is defined such that similar requirements 
can be used on data at the detector level, which helps to keep the corrections small 
in the unfolding step.  
In order to select the lepton+jets channel of the \ttbar process, 
exactly one prompt electron or muon with $\pt>60\GeV$ originating from the decay of a \PW boson must be present. 
Decays to \Pgt leptons contribute a small background. 
They are not selected and are treated as background in this analysis. 
The two-step XCone clustering procedure is performed similarly to the one at the reconstruction level, 
as explained in Section~\ref{sec:reco}, with all stable particles except for neutrinos as input. 
Decays of boosted top quarks must have an XCone jet with $\pt>400\GeV$. 
All three XCone subjets have to satisfy $\pt>30\GeV$ and $\abs{\eta}<2.5$. 
The requirement on $\abs{\eta}$ ensures that the XCone jet is reconstructed within the 
geometric acceptance of the detector. 
The second XCone jet has to have $\pt>10\GeV$ after the lepton four-momentum has been subtracted. 
This requirement rejects pathological cases, 
where the second XCone jet does not contain the \PQb subjet from the \PQt decay. 
We find that 6.7\% of all events are rejected by this requirement.
The XCone-jet mass \mjet has to be larger than the invariant mass of the sum of 
the second XCone jet and the selected lepton.
Since the neutrino from the leptonic decay is not reconstructed, 
this requirement is always fulfilled if all decay products of the hadronic decay are 
reconstructed within the XCone jet, referred to as ``fully merged events''.
This criterion helps to select fully merged decays without introducing a bias on the 
measurement XCone jet, which would be the case with additional requirements on its substructure. 
It removes about 32.6\% of the \ttbar events at the particle level, 
where a large fraction of the removed events consists of not fully merged events. 
Figure~\ref{fig:particlelevel} shows the distribution in \mjet at the particle level 
after the selection of the fiducial region. 
The distribution has a narrow peak, with the maximum close to \mtop. 
Contributions from the UE and FSR lead to a shift of the peak towards higher values. 
In the peak region, the contribution of fully merged top quark decays is about 87\%. 
Contributions from \ttbar events that are not fully merged 
dominate the regions to the left and right of the peak. 
Typically, in these events the top quark has only been partially  
reconstructed within the XCone jet, or the XCone jet originates 
from radiation not associated with the \ttbar system. 
With respect to the measurement at 8\TeV~\cite{Sirunyan:2017yar}, 
which used Cambridge--Aachen jets~\cite{CACluster1,CACluster2} with $R=1.2$ and no grooming, 
the width of the distribution in the peak region is reduced by a factor of two.  
This improvement is achieved by the two-step XCone clustering procedure which 
acts as a grooming algorithm~\cite{Kogler:2021kkw}, removing all particles in the XCone jet 
not clustered into the three subjets.

\section{Event selection}
\label{sec:selection}

At the detector level, the event selection aims to include a similar phase space as selected at the particle level.
Events must contain a single muon or single electron with $\pt>60\GeV$ and $\abs{\eta}<2.4$.
Leptons with $55<\pt<60\GeV$ and $\abs{\eta}<2.4$ are used to construct a sideband region when unfolding the data, 
as described in Section~\ref{sec:unfolding}. 
Electrons with $\pt<120\GeV$ must pass an isolation requirement~\cite{CMS:2020uim},
where the isolation is defined as the \pt sum of charged hadrons and neutral particles 
in a cone with radius $R=0.3$ around the electron. The isolation variable is corrected to 
mitigate the contribution from pileup.
Electron candidates with $\pt>120\GeV$ and muons are rejected if there is an AK4 jet 
within ${\Delta}R<0.4$ and $\ptrel<40\GeV$,
where \ptrel is the component of the lepton momentum 
orthogonal to the AK4-jet axis. 
The last criterion has high efficiency of selecting highly boosted 
$\PQt \to \PQb \PW (\to \Pell \PGn_{\!\ell})$ decays, where the lepton would not 
have passed an isolation requirement because of the angular proximity of the \PQb jet, 
while rejecting QCD multijet events~\cite{Sirunyan:2018ryr, Sirunyan:2018rfo}. 

In order to suppress non-\ttbar backgrounds, at least one AK4 jet is required to be \PQb tagged 
using the \textsc{DeepJet} algorithm~\cite{CMS-DP-2018-033, Bols:2020bkb}.
The candidate \PQb jets are required to have $\pt>30\GeV$ and $\abs{\eta}<2.4$, 
and must pass a selection on the \textsc{DeepJet} discriminator value corresponding to a 
misidentification rate of 0.1\% for light-flavour quark and gluon jets, and an efficiency of 68\%.

\begin{figure}[tb]
  \centering
    \includegraphics[width=0.49\textwidth]{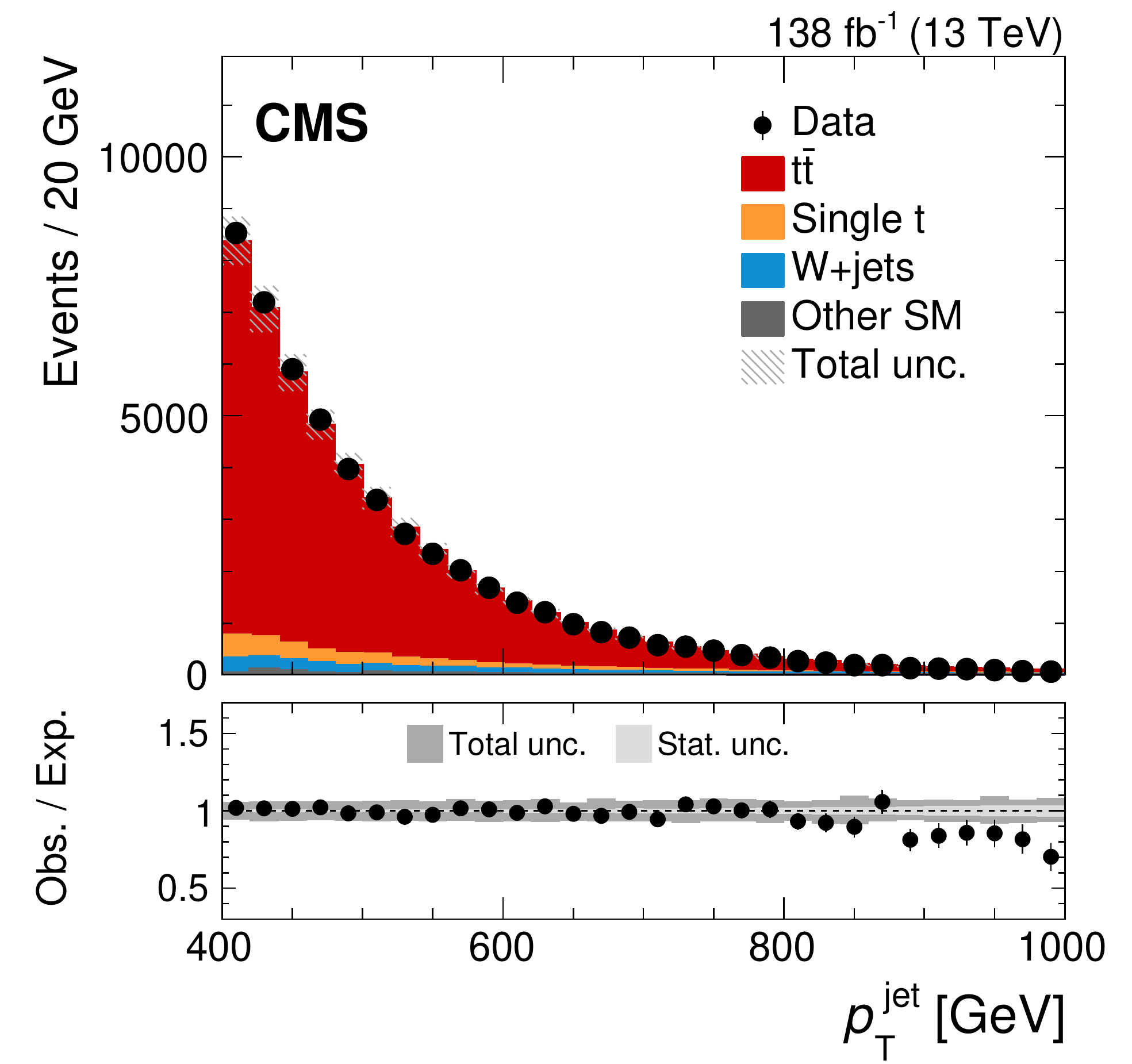}
    \includegraphics[width=0.49\textwidth]{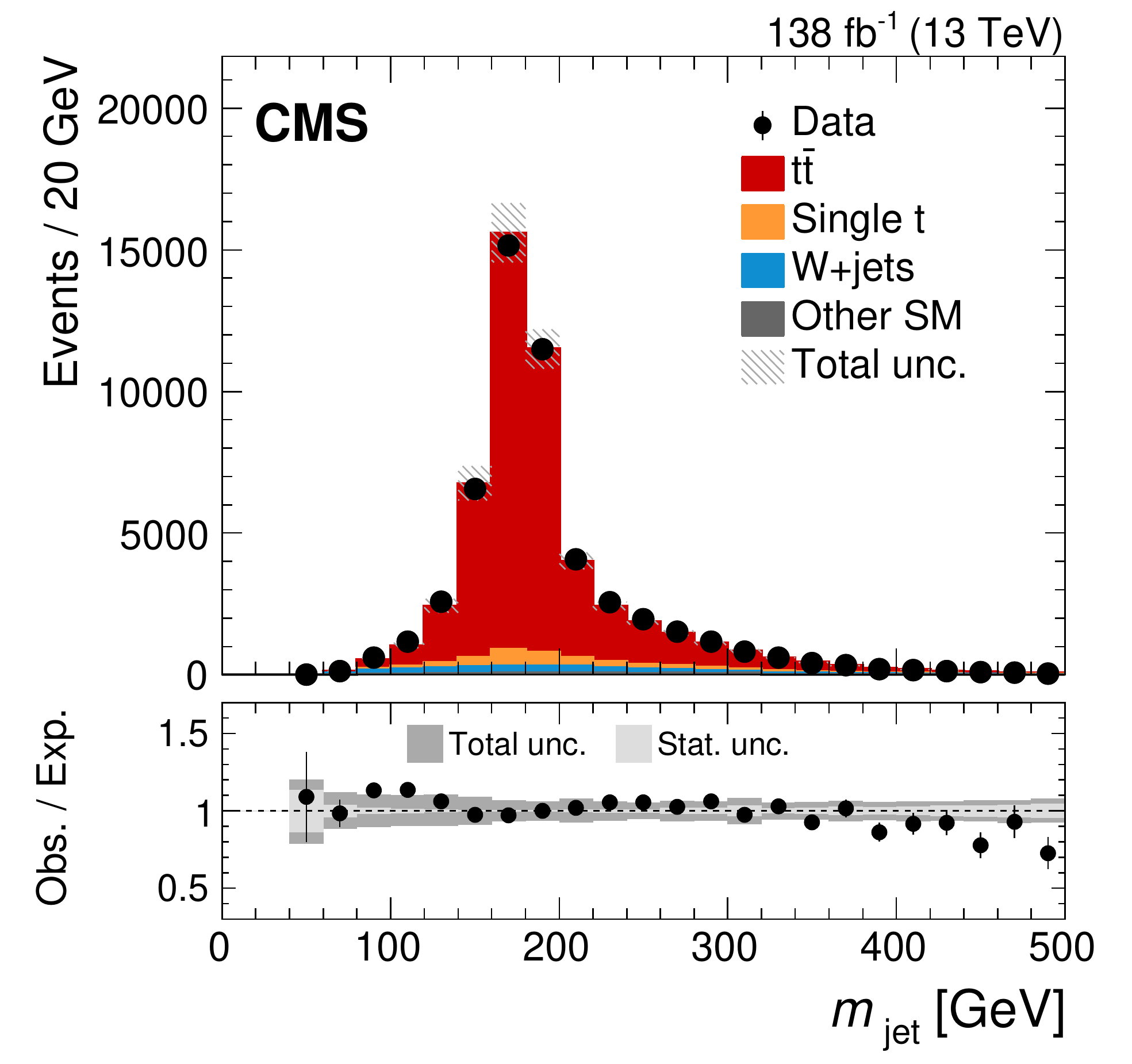}
    \caption{Distributions in the reconstructed XCone jet \pt~(\cmsLeft) and \mjet~(\cmsRight), 
      after the full event selection.
      The vertical bars on the markers show the statistical uncertainty.
      The hatched regions show the total uncertainty in the
      simulation, including the statistical and experimental systematic uncertainties. The lower panels
      show the ratio of the data to the simulation. The uncertainty bands include the experimental systematic uncertainties 
      and statistical uncertainties in the simulation.
      In the ratios, the statistical (light grey) and total (dark grey)
      uncertainties are shown separately.
    \label{fig:mjet}}
\end{figure}
\begin{figure*}[tb]
  \centering
  \includegraphics[width=0.49\textwidth]{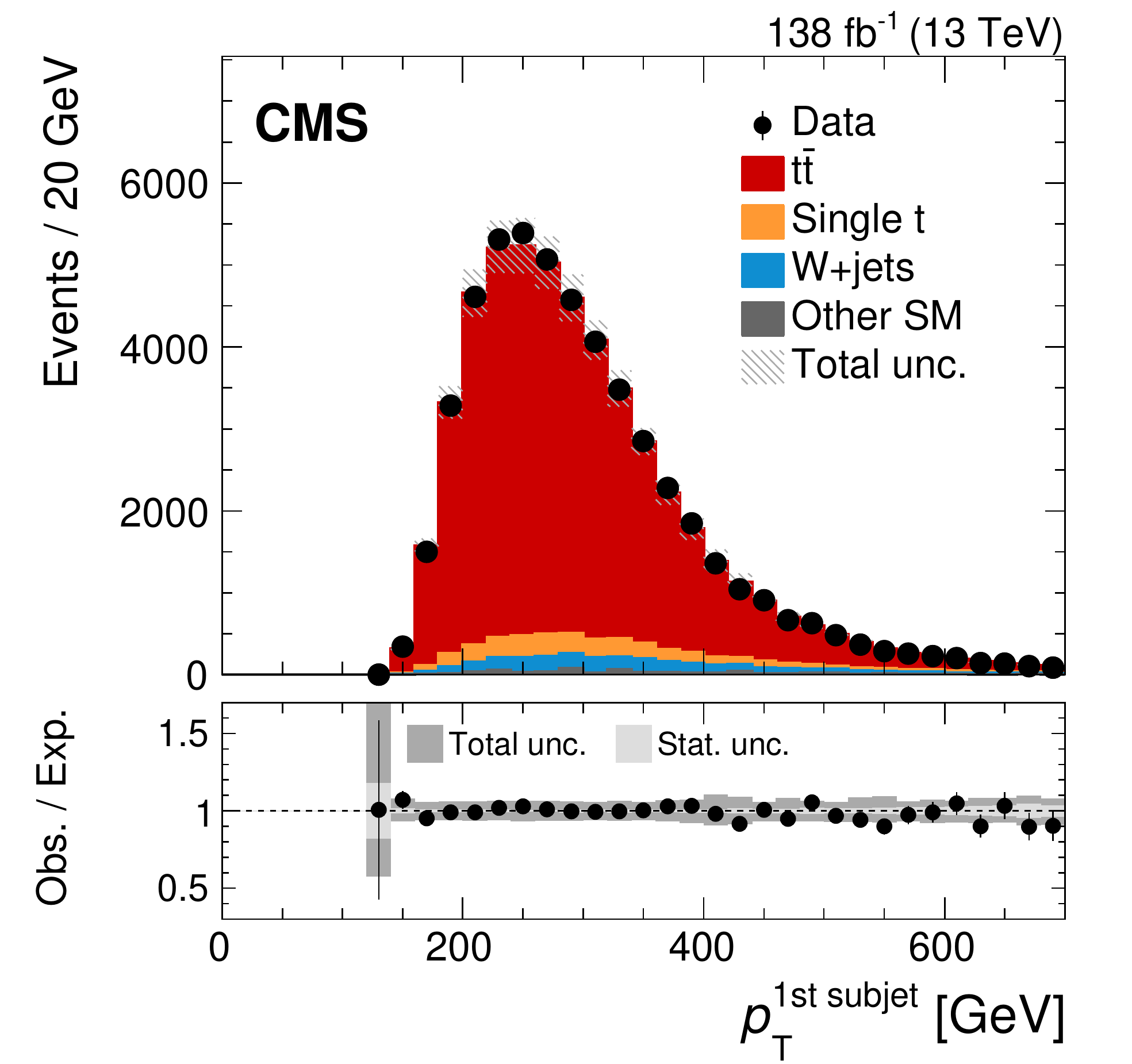}
  \includegraphics[width=0.49\textwidth]{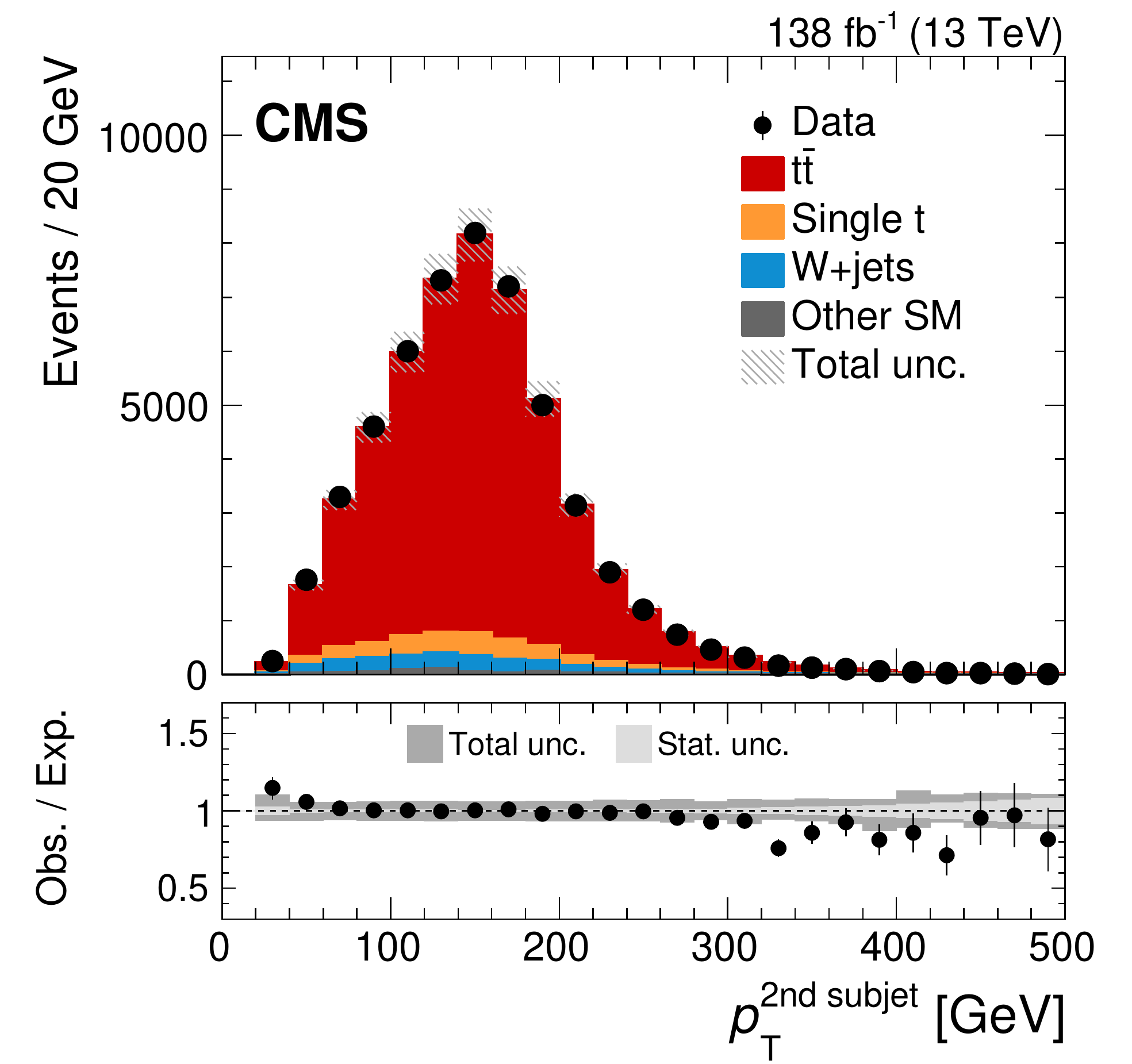} 
  \includegraphics[width=0.49\textwidth]{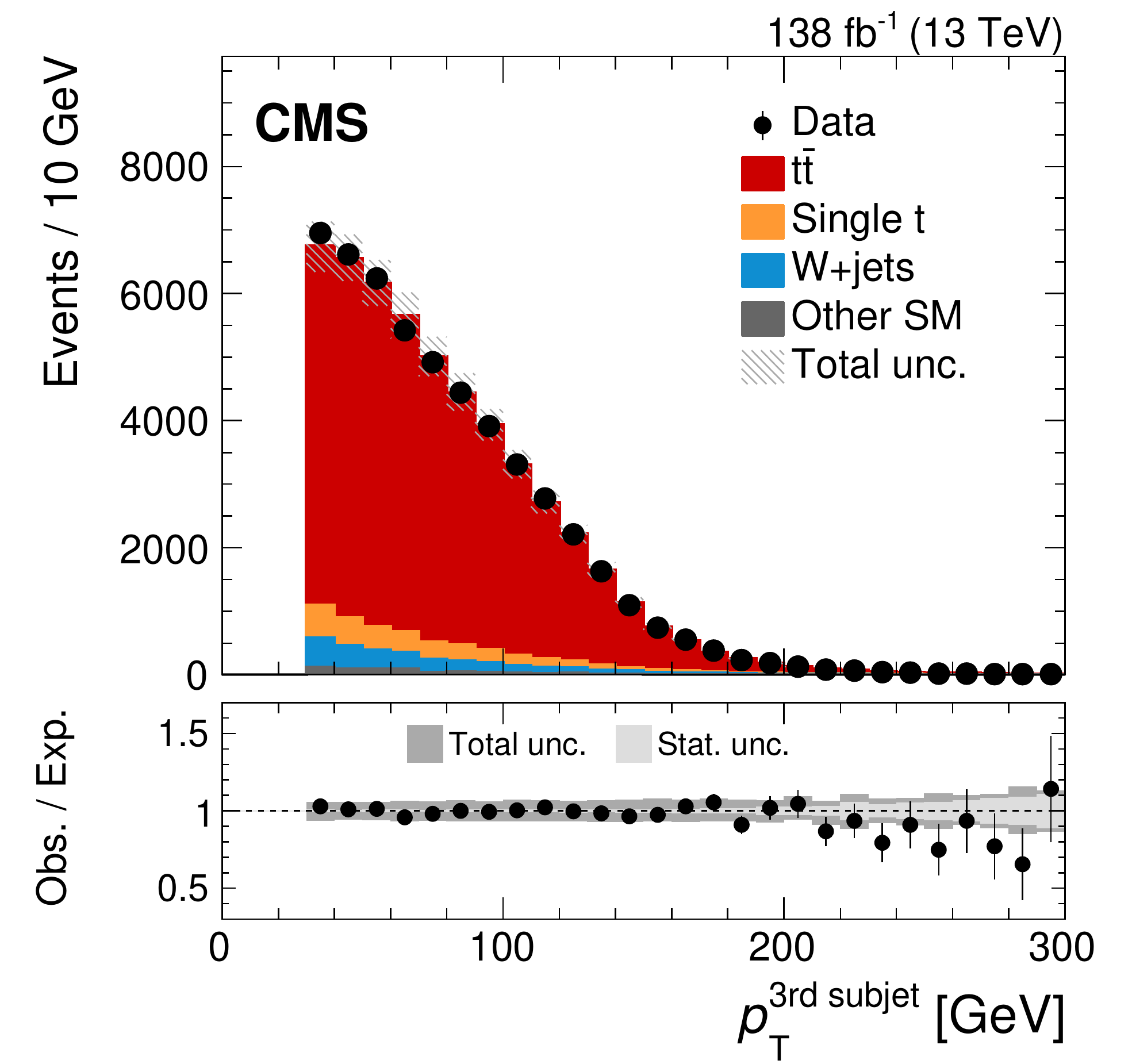}
  \caption{Distributions in reconstructed \pt of the \pt-leading XCone subjet~(upper left),
    second XCone subjet~(upper right) and third XCone subjet~(lower).
    The vertical bars on the markers show the statistical uncertainty.
    The hatched regions show the total uncertainty in the 
    simulation, including the statistical and experimental systematic uncertainties. The lower panels
    show the ratio of the data to the simulation. The uncertainty bands include the experimental systematic uncertainties 
    and statistical uncertainties in the simulation.
    In the ratios, the statistical (light grey) and total (dark grey) 
    uncertainties are shown separately. 
    \label{fig:ptsub}}
\end{figure*}

In addition, the magnitude of the negative vector sum of the transverse momenta 
of the PF candidates in an event~\cite{Sirunyan:2019kia}, \ptmiss,
has to be larger than 50\GeV. The energy scale corrections applied to AK4 jets are propagated to \ptmiss.
This requirement suppresses the contribution of multijet backgrounds 
from the production of light-flavour quarks and gluons.

The XCone jet is required to have $\pt>400\GeV$ and all three subjets must have $\pt>30\GeV$ and $\abs{\eta}<2.5$.
The second XCone jet has to have $\pt>10\GeV$ and the invariant mass of the 
system containing the second XCone jet and the lepton must not surpass \mjet.

{\tolerance=800
Figure~\ref{fig:mjet} shows the XCone jet \pt~(\cmsLeft) and \mjet~(\cmsRight) 
spectra at the detector level. 
Here, data from all three years and both lepton flavours are combined.
For the sake of comparing the shapes of these distributions, 
the \ttbar simulation has been scaled down such that the number of 
simulated events matches the number of events observed in the data. 
The distribution in \pt shows the characteristic falling behaviour above the 
400\GeV threshold, while the distribution in \mjet shows a narrow peak close to \mtop. 
We find reasonable agreement between data and simulation in the \pt and \mjet 
distributions when we use the JECs, and the XCone and JMS corrections 
described in Section~\ref{sec:jms}. 
For \pt above 900\GeV, we observe that the simulation predicts more events than 
observed in data, a feature which has been reported previously in differential \ttbar cross 
section measurements when comparing to NLO calculations~\cite{CMS:2021vhb, ATLAS:2022xfj, ATLAS:2022mlu}. 
Figure~\ref{fig:ptsub} shows the distributions in \pt of the XCone subjets. 
Because of the XCone-jet selection with $\pt>400\GeV$, the first subjet 
has a most probable \pt of about 250\GeV, and the second subjet has a value of about 150\GeV. 
The remaining subjet features a falling distribution, starting from the 
minimum value of 30\GeV. 
\par}

\section{Calibration of the jet mass scale}
\label{sec:jms}

The experimental precision in the measurement of \mjet in boosted top quark decays 
is limited by the calibration of the jet four-momentum. 
In our previous analysis~\cite{Sirunyan:2019rfa}, the uncertainty in the JES was  
propagated to \mjet and resulted in the dominant experimental systematic uncertainty.
In this article, we measure the JMS using the invariant mass of the two XCone subjets 
originating from the hadronic \PW boson decay. 
With this additional measurement, the uncertainty in the JES affects the jet three-momentum, 
while the uncertainty in the JMS affects \mjet. 
The JMS calibration is crucial for the improvement in the overall precision of this measurement.

\begin{figure*}[tb] 
  \centering
  \includegraphics[width=0.49\textwidth]{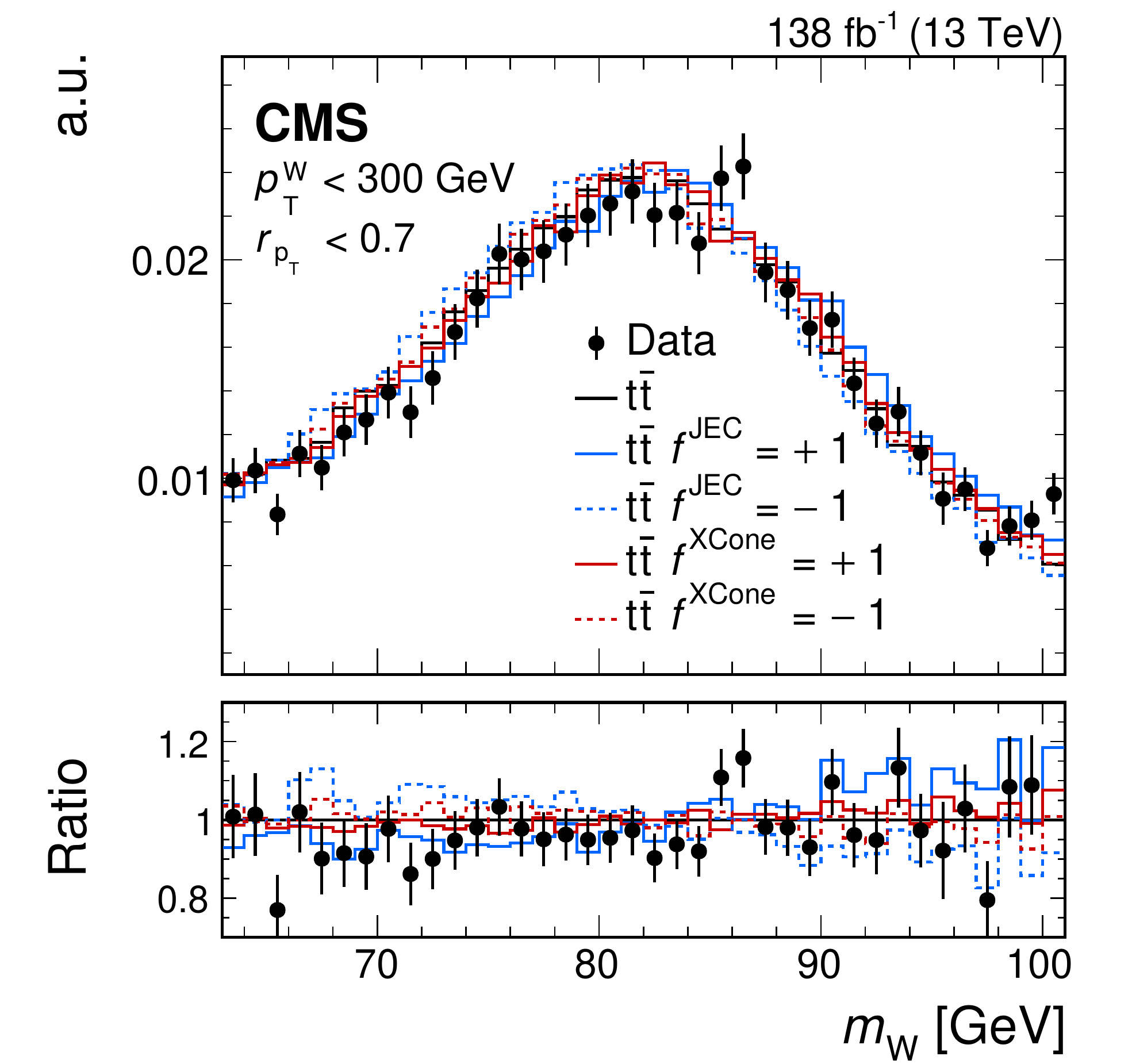}
  \includegraphics[width=0.49\textwidth]{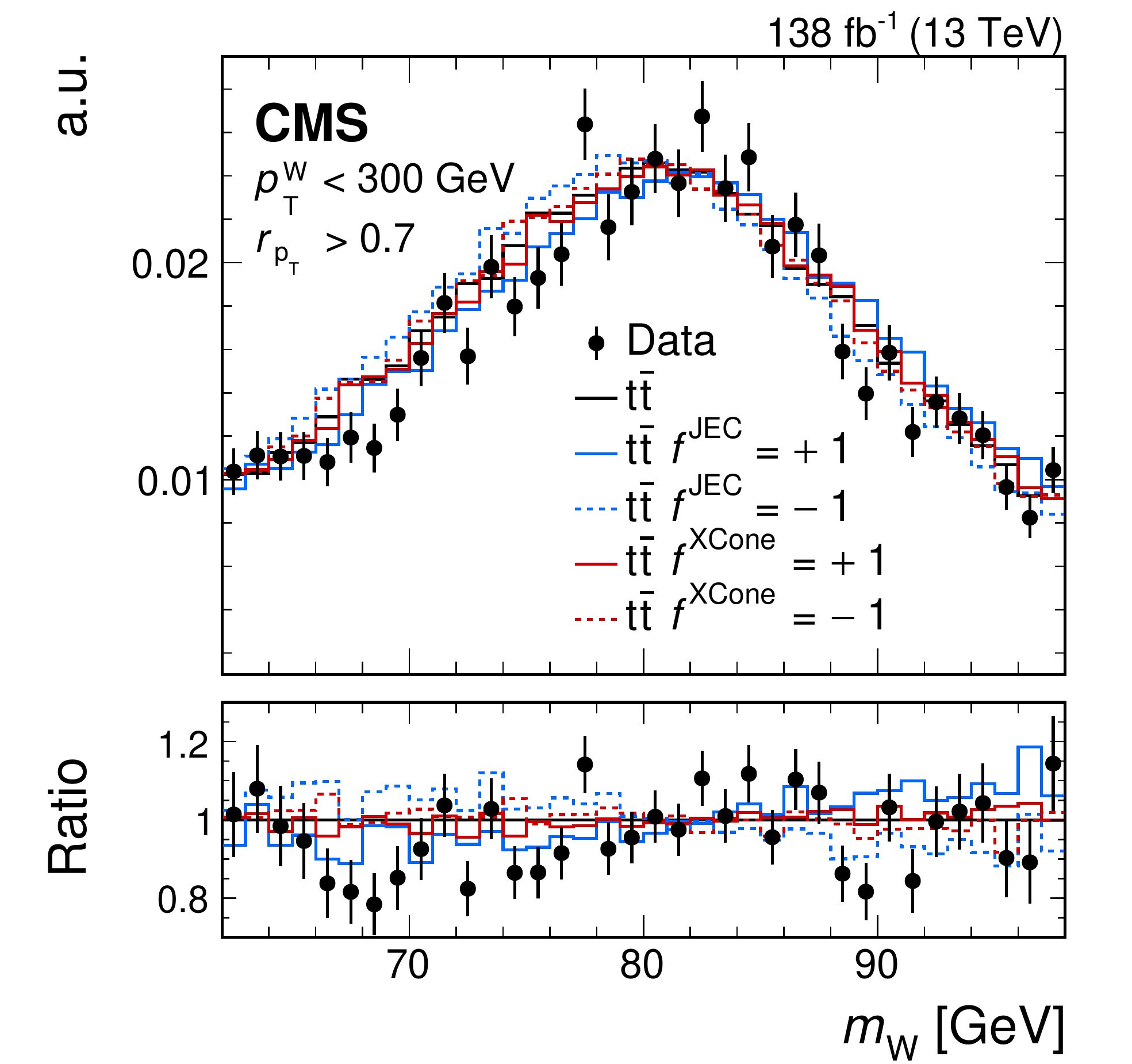}
  \includegraphics[width=0.49\textwidth]{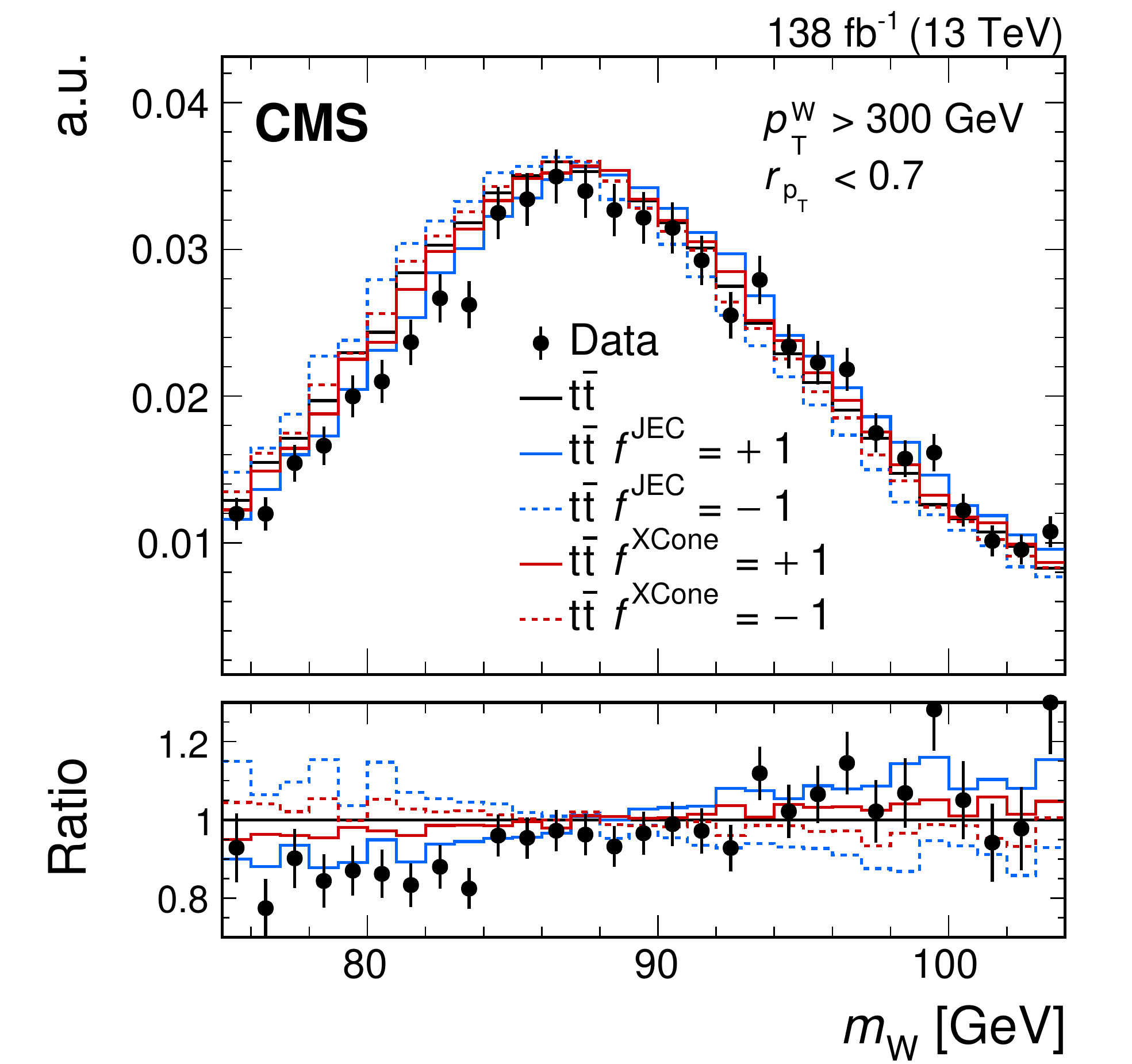}
  \includegraphics[width=0.49\textwidth]{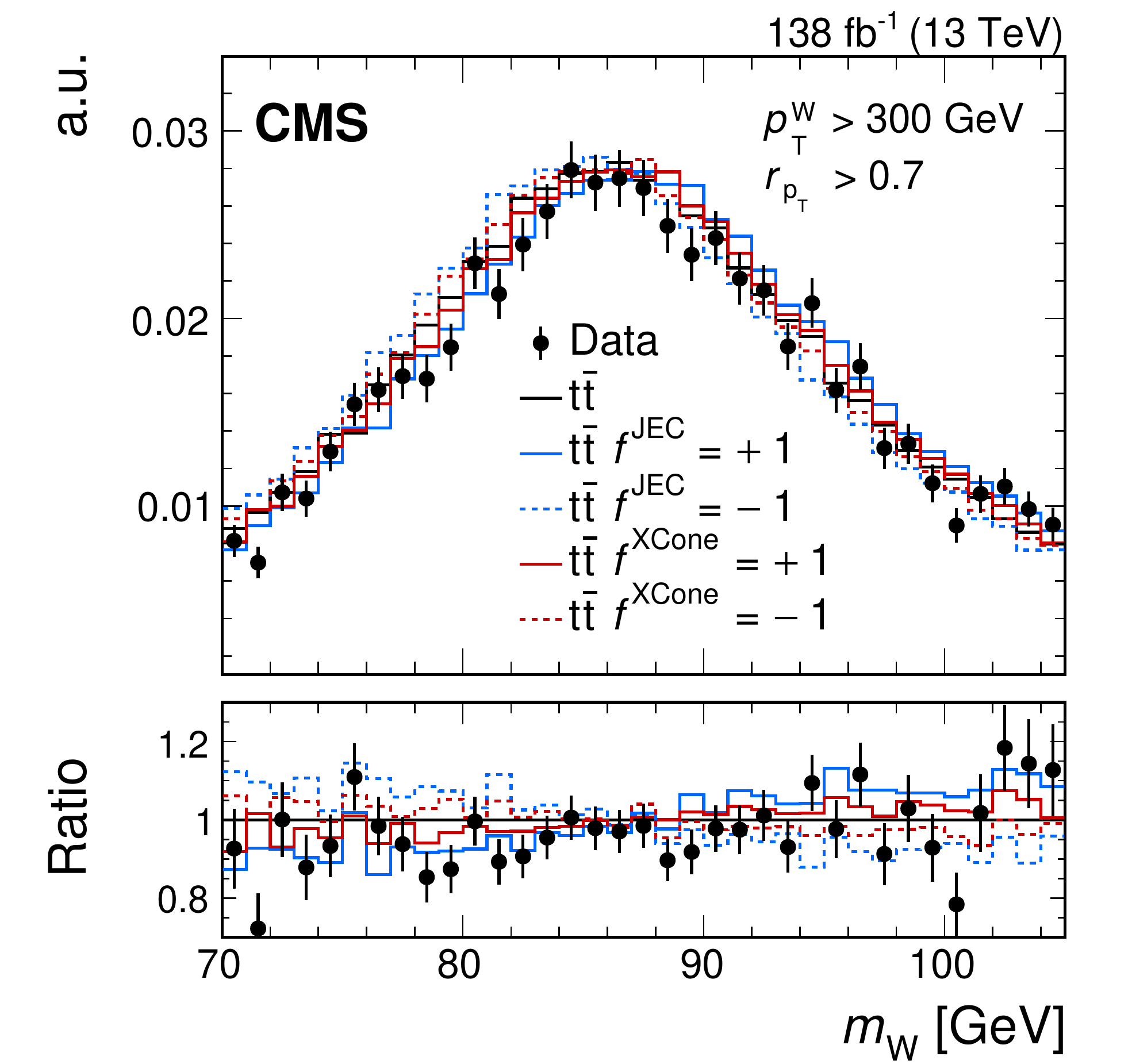}
  \caption{Peak region of the reconstructed \PW boson mass in the four regions 
    $\ptW<300\GeV$ and $\ptratio<0.7$~(upper left),
    $\ptW<300\GeV$ and $\ptratio>0.7$~(upper right), 
    $\ptW>300\GeV$ and $\ptratio<0.7$~(lower left), and
    $\ptW>300\GeV$ and $\ptratio>0.7$~(lower right).
    The background-subtracted data and the \ttbar simulation are normalised to unit area. 
    For illustration, the \ttbar simulation is also shown with the JEC and XCone correction 
    factors varied by one standard deviation. 
    The lower panels show the ratios to the nominal \ttbar simulation.
  \label{fig:mWregions}}
\end{figure*}

For the JMS calibration, the same selection as for the measurement is applied. 
The \PW boson decay is reconstructed using two of the three XCone subjets 
from the XCone jet initiated by the hadronic top quark decay. 
We identify the XCone subjet originating from the fragmentation of the \PQb quark
using the \textsc{DeepJet} algorithm on AK4 jets.
First, the AK4 jet with the largest value of the \textsc{DeepJet} \PQb discriminant 
among those with angular distance ${\Delta}R<1.2$ to the XCone jet is selected. 
In a second step, the XCone subjet with the smallest ${\Delta}R$ to the 
selected \PQb-tagged AK4 jet is assigned to originate from the \PQb quark. 
This XCone subjet is rejected, and the measurement of the JMS is 
performed using the invariant mass of the other two XCone subjets. 
Data from the two lepton flavours and three different years are combined 
for the JMS calibration. 

{\tolerance=800
The JMS in simulation is adjusted by introducing two factors, \fjec and \fxcone, 
that vary the jet energy scale in the AK4 JECs and the additional XCone-jet corrections, respectively. 
The factors are constructed such that values of 0, ${+}1$ and ${-}1$ represent the nominal correction, 
and the up and down shifts by one standard deviation, respectively. With these two factors, 
the squared XCone jet mass becomes 
\begin{linenomath}
\ifthenelse{\boolean{cms@external}}
{
\begin{multline}
\mjet^2 = \bigg( 
\sum_{i=1}^3
p_i \left( \cjec(\pti, \eta_i) + \fjec \sigmajec(\pti, \eta_i)  \right) \\
\left( \cxcone(\pti, \eta_i) + \fxcone \sigmaxcone(\pti, \eta_i) \right) \bigg)^2, 
\end{multline}
}
{
\begin{equation}
\mjet^2 = \left( 
\sum_{i=1}^3
p_i \left( \cjec(\pti, \eta_i) + \fjec \sigmajec(\pti, \eta_i)  \right)
\left( \cxcone(\pti, \eta_i) + \fxcone \sigmaxcone(\pti, \eta_i) \right)
\right)^2, 
\end{equation}
}
\end{linenomath}
where $p_i$ are the three subjet four-momenta before the application of the JEC and XCone corrections, 
$\cjec(\pti, \eta_i)$ and $\cxcone(\pti, \eta_i)$ denote the JEC and XCone corrections, respectively, 
and $\sigmajec(\pti, \eta_i)$ and $\sigmaxcone(\pti, \eta_i)$ are the uncertainties in these corrections. 
The JES and XCone corrections and the corresponding uncertainties depend on the \pt and $\eta$ of the uncorrected 
subjet four-momentum. 
The additional corrections proportional to \fjec and \fxcone allow \mjet to float, 
while retaining the shape and functional form of the JEC and XCone uncertainties in \pt and $\eta$. 
This JMS correction is constructed to change only the XCone jet mass but not the 
three-momentum that is calibrated with established methods.  
The decoupling of the JMS correction from the three-momentum calibration allows the JMS correction to target effects 
which change only the jet mass and not the three-momentum, like splitting and merging of calorimeter clusters. 
\par}

The measurement is performed in four regions that are defined in the two-dimensional plane 
of the \pt of the reconstructed \PW boson, \ptW, and the ratio $\ptratio = \pt^{\text{s}_1}/\ptW$, 
defined as the ratio of the \pt carried by the highest \pt XCone subjet $\text{s}_1$ to \ptW.
These regions are constructed to reduce correlations between \fjec and \fxcone, 
because these factors can cancel each other in an inclusive measurement of the JMS. 
We find an improvement by a factor of 1.6 in the obtained precision of \fjec and \fxcone 
when using these four regions, compared to a calibration using the inclusive \mw distribution. 
Because of the different size of $\sigmajec(\pt, \eta)$ and 
$\sigmaxcone(\pt, \eta)$ in subjet \pt and $\eta$, 
the effects of \fjec and \fxcone are different in the four regions defined by \ptW and \ptratio, 
such that these two factors can be determined simultaneously. 
Figure~\ref{fig:mWregions} shows the four distributions in the reconstructed \PW boson mass \mw 
in the vicinity of their peaks in the four regions, 
defined by \ptW larger or smaller than 300\GeV and \ptratio larger or smaller than 0.7. 
We consider only regions around the peak position with bins populated by more than 100 events 
in the background-subtracted data for a bin width of 1\GeV. 
This requirement leads to the following \mw ranges in the four regions: 
70--105\GeV for $\ptW>300\GeV$ and $\ptratio>0.7$;
75--104\GeV for $\ptW>300\GeV$ and $\ptratio<0.7$;
62--98\GeV for $\ptW<300\GeV$ and $\ptratio>0.7$; and 
63--101\GeV for $\ptW<300\GeV$ and $\ptratio<0.7$. 
These ranges are used to exclude tails in the \mw distributions, 
which originate from a wrong assignment of subjets to the reconstructed \PW boson. 
In total, 138 bins are used in the JMS calibration.
The distributions of background-subtracted data and \ttbar signal have been 
normalised to unit area and are given in arbitrary units (a.u.), 
such that only shapes are considered and the total yield does not affect the measurement. 
The \ttbar simulation is shown for the different variations in the jet 
corrections, parametrised by \fjec and \fxcone. 
The peak in \mw is shifted in the four regions by 0.42--0.61\GeV and 
by 0.17--0.25\GeV for the \fjec and \fxcone variations, respectively.

In each bin $i$ of the \mw distribution, a linear prediction $g_i$ as a function of 
\fjec and \fxcone is defined, 
\begin{linenomath}
  \begin{equation}
    g_i(\fjec,\fxcone) = a_i + b_i \fjec + c_i \fxcone,
    \label{eq:jmsfit}
  \end{equation}
\end{linenomath}
with the free parameters $a_i$, $b_i$, and $c_i$. The free parameters are obtained from 
a fit to simulation in the \fjec-\fxcone plane in each bin $i$.
We have verified that a linear fit in both factors describes the 
dependence of \mw on \fjec and \fxcone sufficiently well, with a fit quality 
matching the expectation of statistical fluctuations only. 

To verify that the statistical uncertainties in the simulation do not bias the result, 
we have performed a test where we increased the bin size in the four \mw distributions by a factor of three, to 3\GeV. 
This results in 47 bins and reduces the fluctuations in the four \mw distributions. 
We find that the linear parametrisations of Eq.~\eqref{eq:jmsfit} provide a good 
description of the variations in \fjec and \fxcone. 
Performing the JMS calibration with these larger bins and reduced statistical 
uncertainties in $g_i$ gives a similar result with respect to the nominal fit with 138 bins. 
The reduced information in the fit with 47 bins results in an increased correlation between \fxcone and \fjec compared to the nominal fit. 

\begin{figure}[tb]
  \centering
    \includegraphics[width=0.49\textwidth]{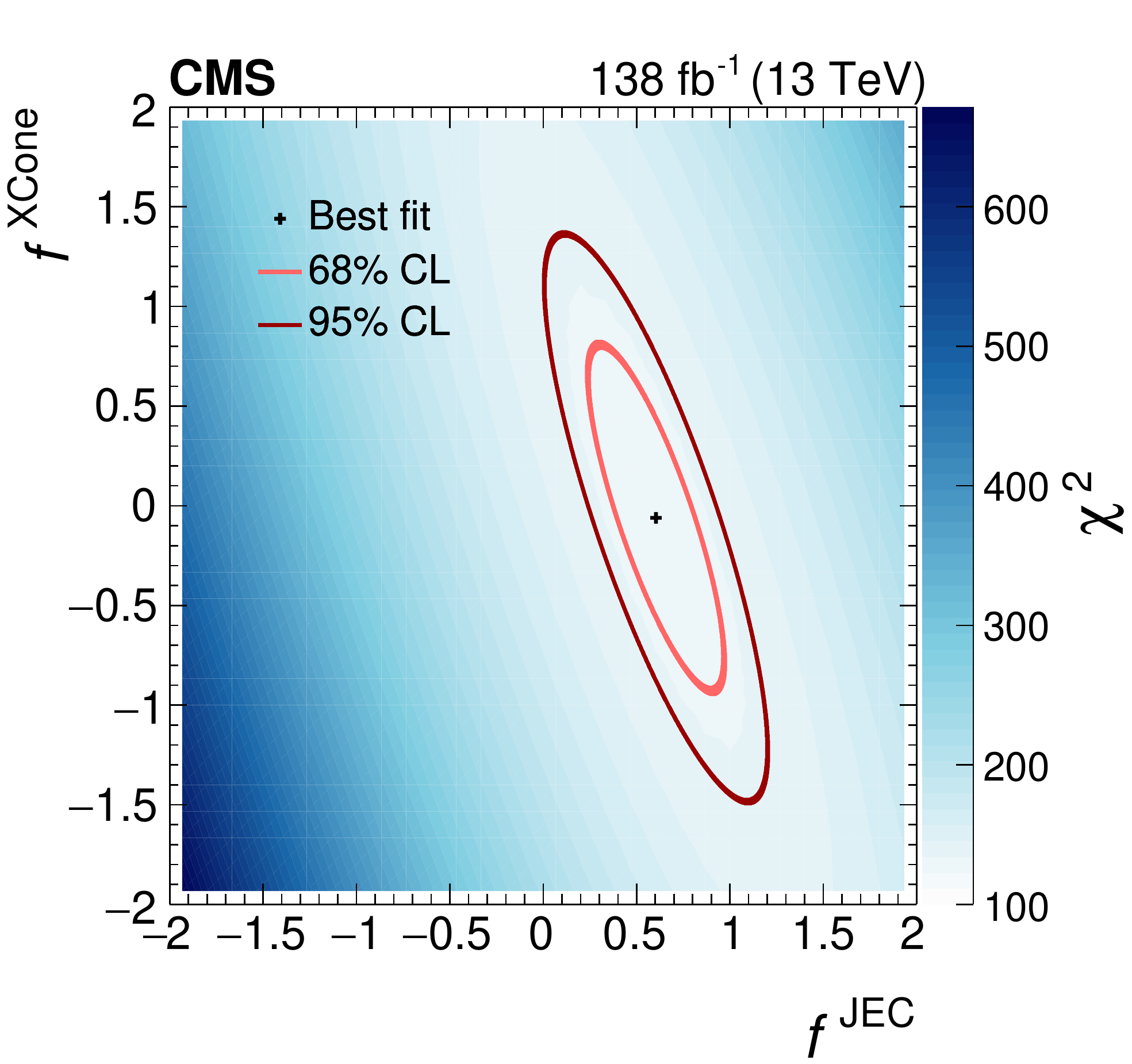}
    \caption{The two-dimensional $\chi^2$ as a function of \fjec and \fxcone, 
    obtained from a comparison of background-subtracted data with the predictions from \ttbar production 
    in the reconstructed \mw distributions. 
    The minimum is indicated by a black cross, and the borders of the 68 and 95\% \CL
    intervals are shown by the light and dark red ellipses, respectively.
    \label{fig:jms_chi2}}
\end{figure}

The factors \fjec and \fxcone are obtained from a fit to the data, where 
a two-dimensional $\chi^2$ function is constructed, 
\begin{linenomath}
  \begin{equation}
    \chi^2 =  d^T V^{-1} d.
    \label{eq:chisq}
  \end{equation}
\end{linenomath} 
The vector $d$ is built from the differences between the predictions $g_i(\fjec,\fxcone)$ 
and the back\-ground-sub\-tracted data in each bin $i$ of all four regions in \ptW and \ptratio. 
The covariance matrix $V$ includes the statistical uncertainty in data, also 
considering correlations from the normalisation to unit area, and 
the uncertainties in the functions $g_i$ from the fit to simulation. 
The latter are estimated from the statistical uncertainty of the simulated \ttbar sample. 
We also include the leading systematic uncertainties, namely the JER uncertainty, 
modelling uncertainties from the \ttbar simulation, and uncertainties 
from the background subtraction. 
These uncertainties are treated as fully correlated across all bins as well as the four regions. 
We find that the statistical uncertainties are the dominant uncertainties in the 
calibration of the JMS, followed by the JER uncertainties. All other uncertainties are 
small in comparison. 

{\tolerance=800
Figure~\ref{fig:jms_chi2} shows the evaluated two-dimensional $\chi^2$, 
as a function of \fjec and \fxcone. 
The minimum of the $\chi^2$ function lies within the one-standard deviation intervals 
of the correction factors. The global minimum has a value of $\chi^2 = 130$ for 132 degrees of freedom.
We find the best-fit values $\fjec = 0.60 \pm 0.24$ and $\fxcone = -0.06 \pm 0.57$ with 
a linear correlation coefficient of ${-}0.66$. 
The JMS uncertainty obtained from the two-dimensional 68\% confidence 
level (\CL) interval is reduced compared to the variations of \fjec and \fxcone in the intervals 
between ${-}1$ and ${+}1$. 
In order to construct variations of one standard deviation in one dimension for the evaluation 
of systematic uncertainties, the endpoints of the minor axis are chosen. 
These result in the largest shift in the \mjet distribution, because 
along the minor axis both factors \fjec and \fxcone shift the value of \mjet in the same direction. 
Changes of \fjec and \fxcone along the major axis result in shifts in opposite 
directions, which cancel to a large part.  
The extracted value pairs in $(\fjec, \fxcone)$, with the nominal value pair of $(0.60, -0.06)$, 
are $(0.78, 0.01)$ and $(0.42, -0.13)$, which are used in the determination of systematic uncertainties. 
These pairs of values are referred to as JMS correction in the following, with the corresponding uncertainties. 
We have verified that variations of \mtop in the \ttbar simulation do not alter this result. 
Additionally, we have tested that the results obtained from the electron and muon channels are 
compatible. The final results of the \mjet measurement agree within the uncertainties 
if the JMS calibration is carried out in the electron channel and applied to the muon channel, and vice versa.
\par} 

\begin{figure}[tb] 
  \centering
    \includegraphics[width=0.49\textwidth]{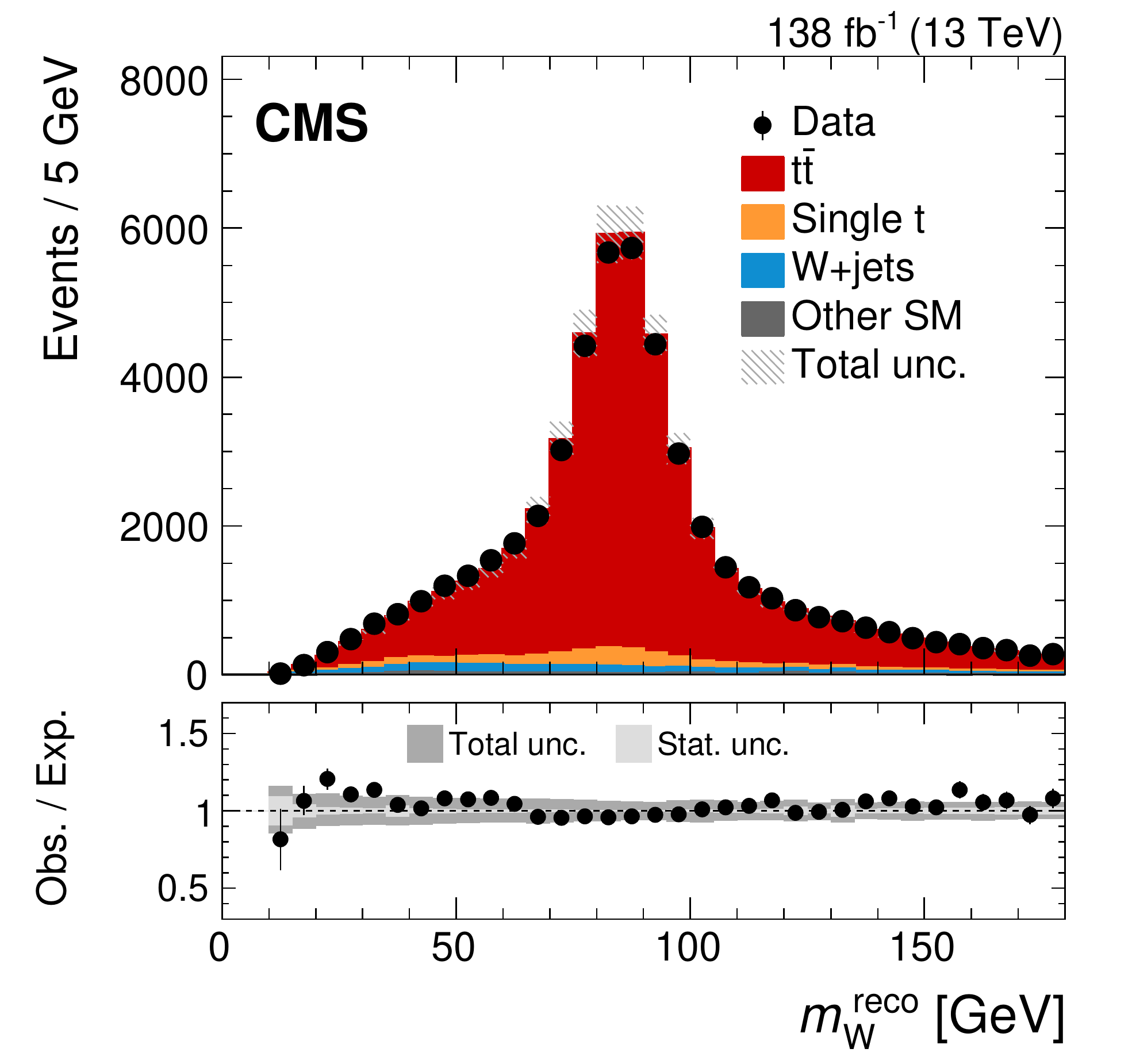}
    \caption{Jet mass distribution of hadronic decays of the \PW boson, reconstructed from 
      two XCone subjets.
      The vertical bars on the markers show the statistical uncertainty.
      The hatched regions show the total uncertainty in the
      simulation, including the statistical and experimental systematic uncertainties. The lower panel 
      shows the ratio of the data to the simulation. 
      The uncertainty bands include the experimental systematic uncertainties 
      and statistical uncertainties in the simulation.
      The statistical (light grey) and total (dark grey)
      uncertainties are shown separately in the ratio.
      \label{fig:wmass}}
\end{figure}
\begin{figure}[tb]
  \centering  
  \includegraphics[width=0.49\textwidth]{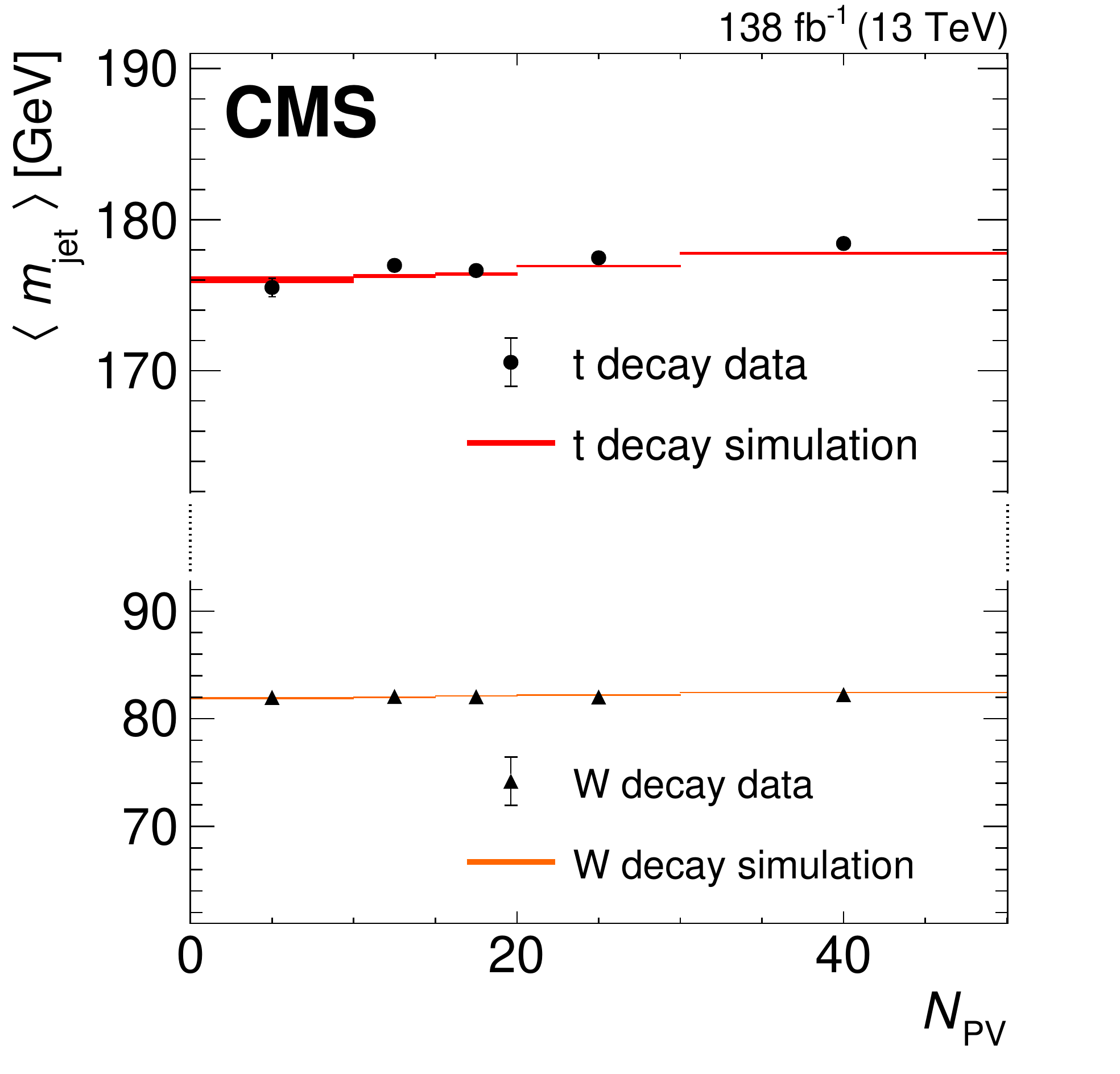} 
  \includegraphics[width=0.49\textwidth]{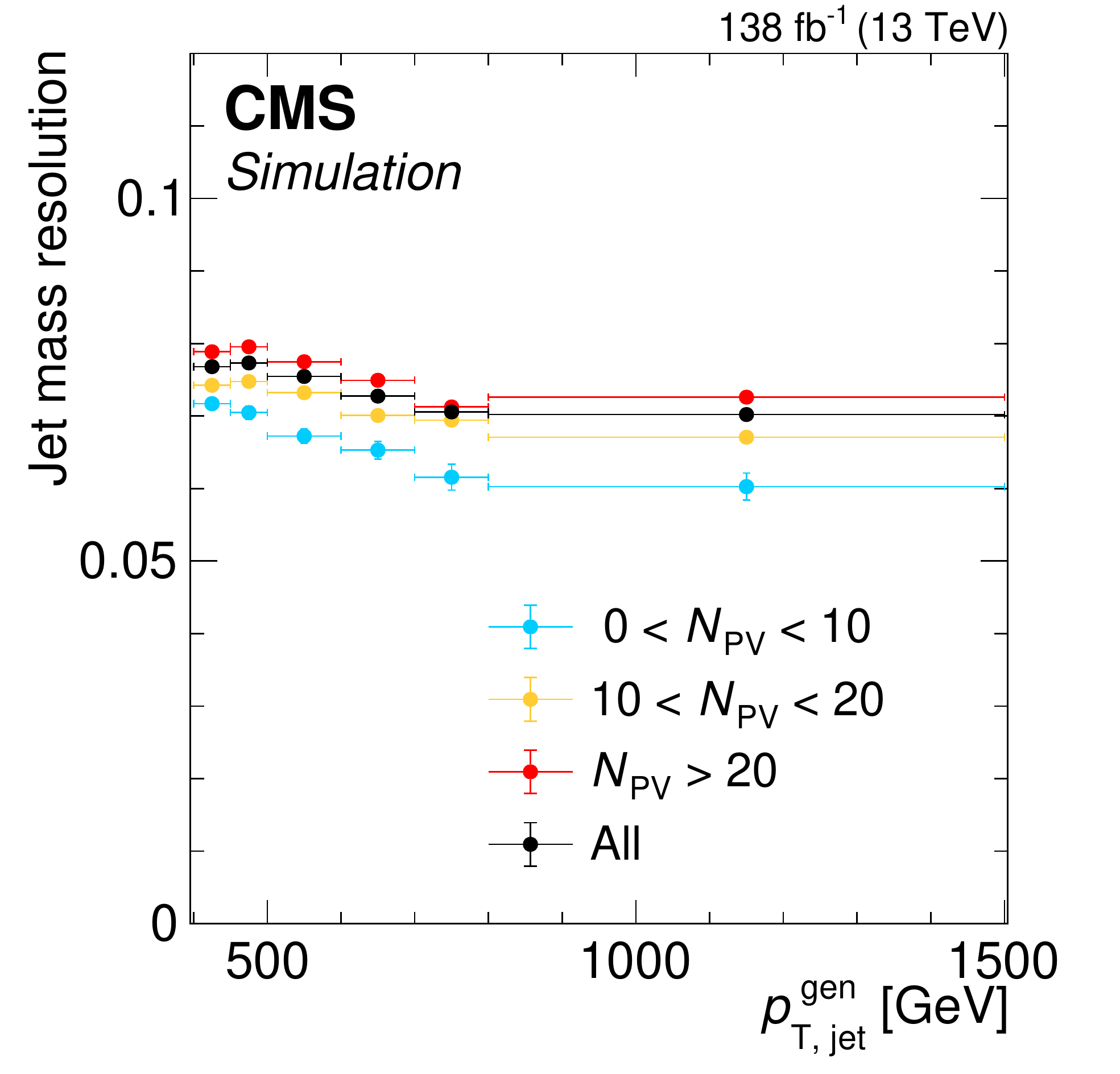}
  \caption{Mean values of the \mjet distribution for \PQt and \PW boson decays, 
  as a function of the number of primary vertices $N_{\text{PV}}$ (\cmsLeft). 
  Data~(markers) are compared with \ttbar simulation~(filled areas). 
  The vertical bars and size of the filled areas show the statistical uncertainties in the 
  calculation of the mean values. 
  Jet mass resolution in simulation as a function of particle-level XCone-jet \pt, 
  given for different intervals in the number of primary vertices (\cmsRight). 
  The vertical bars indicate the statistical uncertainties and the horizontal bars indicate
  the bin width. 
  \label{fig:mjetPUandReso}} 
\end{figure}  
Figure~\ref{fig:wmass} shows the reconstructed \mw distribution after applying the JMS correction.
The data are well described by the simulation over the full distribution in \mjet. 
The mean values of \mjet for the reconstructed top quark and \PW boson masses are shown as a 
function of the number of primary vertices in Fig.~\ref{fig:mjetPUandReso}~(\cmsLeft). 
The values for the top quark mass are obtained using all three XCone subjets, while the \PW boson 
mass is calculated from the two subjets not matched to the \PQb-tagged AK4 jet. 
The mean values of \mjet are larger than the parameters \mtop and \mw 
used in the simulation by about 4 and 2\GeV, 
respectively, because of contributions from the UE and pileup interactions. 
The slope of the mean value of \mjet as a function of the number of pileup interactions is small, indicating that 
the XCone reconstruction and calibration remove most of the contributions from pileup. 
The mean values and the slopes are well described by the simulation.
The achieved resolution in \mjet is displayed in Fig.~\ref{fig:mjetPUandReso}~(\cmsRight).
We calculate the resolution as the width parameter of a Gaussian function, fitted to 
distributions in $\mjet^\text{rec} / \mjet^\text{gen}$, where $\mjet^\text{rec}$ denotes the reconstructed 
value of \mjet at the detector level and $\mjet^\text{gen}$ is the jet mass at the 
particle level. 
The achieved resolution is below 8\% over the full range in \pt.
For an inclusive selection in the number of primary vertices, 
the mass resolution improves from 7.7\% at $\pt=400\GeV$ to 7\% for $\pt>800\GeV$.  
For a selection with less than 10 primary vertices, the resolution is 
about one percentage point better than for a selection with more than 20 primary vertices.  

\section{Studies of the final state radiation}
\label{sec:FSR}

The uncertainty in the modelling of FSR was the dominant model uncertainty 
in the previous \mjet measurement at 13\TeV~\cite{Sirunyan:2019rfa}. 
There, the energy scale parameter $\mu$, which enters into the definition of the strong 
coupling \asFSRmu, was changed by factors of 0.5 and 2 in the FSR simulation to estimate this uncertainty. 
This is equivalent to changing the value of the effective strong coupling 
at the mass of the \PZ boson from $\asFSRZ=0.1365$, as used in the parton shower and UE event tune 
CUETP8M2T4 for the simulation of 2016 data, to values of 0.1556 and 0.1217, respectively. 
While the data are well described using the central value, 
we find that the large uncertainty variations do not describe the data in the 
fiducial region of this measurement. 
For the simulation of 2017 and 2018 data, the CP5 tune is used with $\asFSRZ=0.118$, 
which is not the optimal choice for the modelling of jet substructure observables in \ttbar production, 
where a larger value is preferred~\cite{Sirunyan:2019dfx}.
To remedy this situation, we perform a study of the FSR modelling and find the value of $\asFSRZ$ 
that fits the data best. The study is performed separately for the two samples with 
different tunes, namely for the year 2016, and for the combination of the years 2017 and 2018. 
The uncertainties in $\asFSRZ$ from this study are propagated to 
the FSR uncertainty in the \mjet measurement. 

As a starting point, we modify the energy scale in the FSR simulation 
by a factor \fFSR. With this definition, the FSR modelling uncertainty 
as used in the previous measurement is obtained by setting $\fFSR=0.5$ and 2.  
The prediction becomes a function of \fFSR, which we use to determine the best fit value of \fFSR 
through a comparison of distributions between data and simulation 
in the $N$-subjettiness ratio $\tauratio=\tau_3 / \tau_2$~\cite{Thaler:2010tr,Thaler:2011gf}. 
The distributions in \tauratio are sensitive to the angular distribution of the energy density 
inside jets and are thus well suited for determining \fFSR.

We use the same event selection as used for the \mjet measurement, described in Section~\ref{sec:selection}.
Instead of XCone jets, we use AK8 jets to study the \tauratio distributions. 
These have a higher sensitivity to effects from FSR, because AK8 jets are obtained without 
jet grooming, unlike the XCone jets clustered with the two-step procedure. 
The AK8 jet that is within ${\Delta}R < 0.8$ of the XCone jet is selected, 
provided it has $\mjet>140\GeV$. This requirement on \mjet ensures that only 
jets including all particles from the hadronic \PQt decay are accepted.

\begin{figure}[tb]
  \centering
    \includegraphics[width=0.49\textwidth]{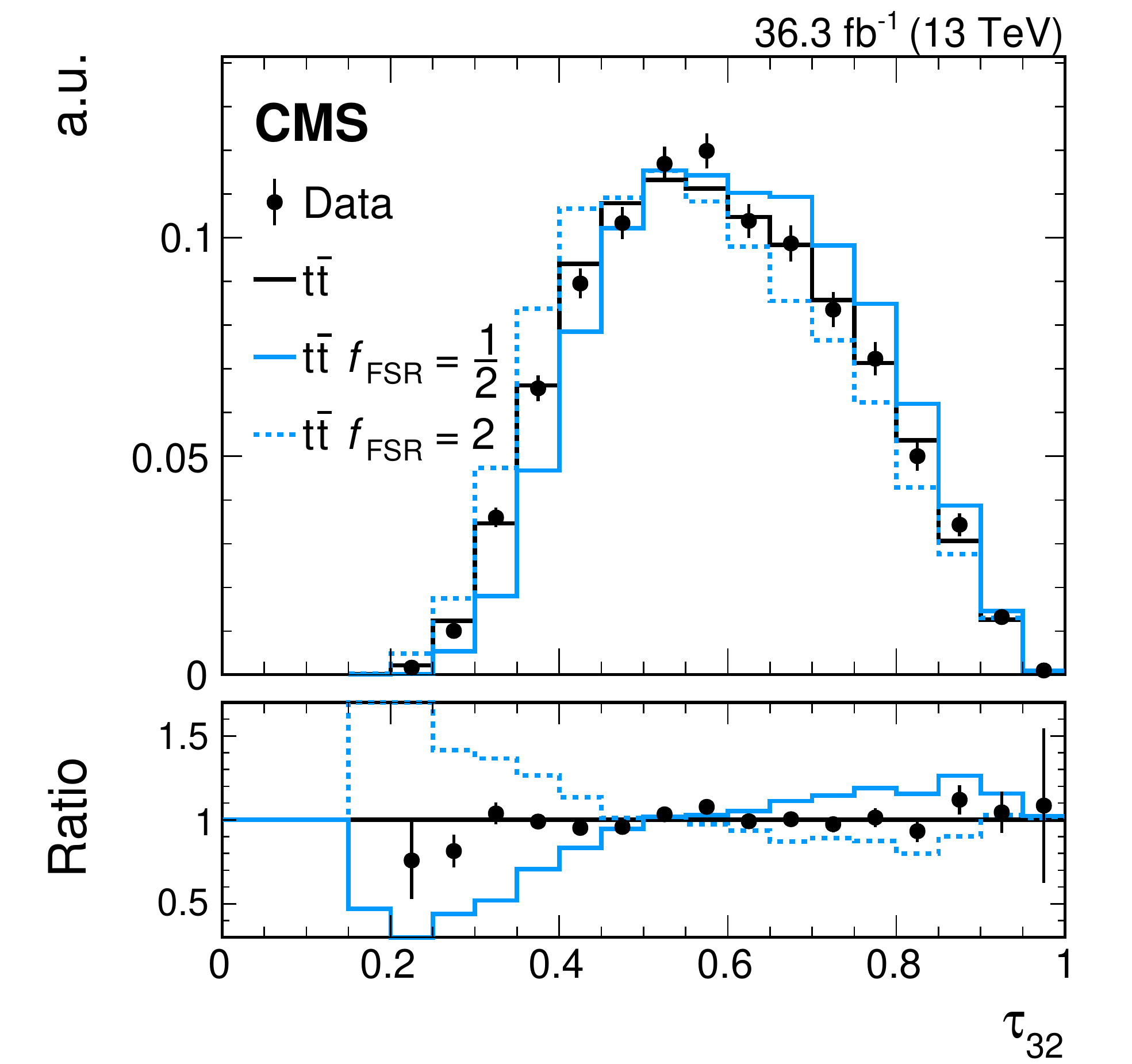}
    \includegraphics[width=0.49\textwidth]{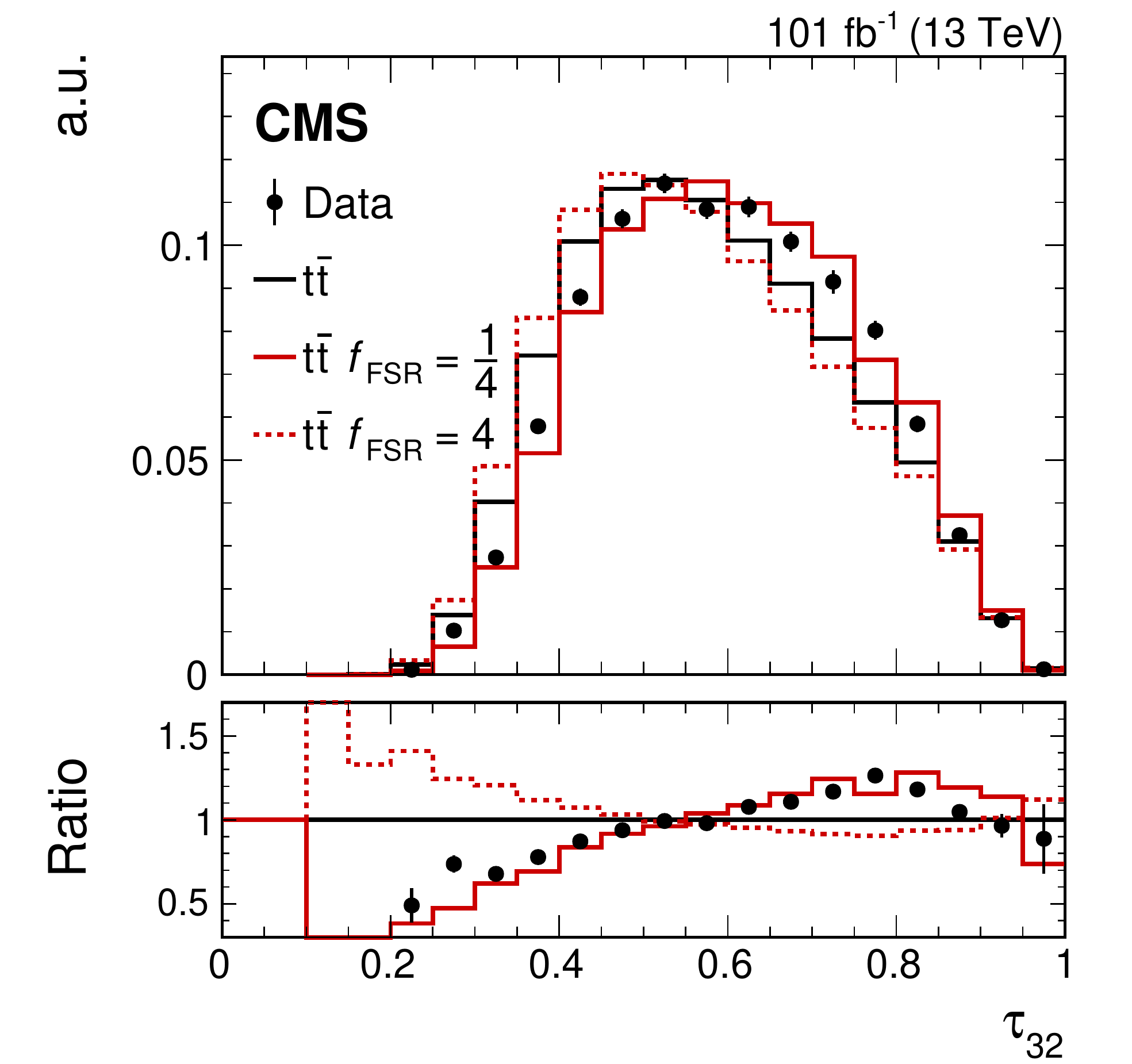}
    \caption{The normalised distributions in \tauratio for AK8 jets with $\mjet>140\GeV$ 
    from the hadronic decay of boosted top quarks. Shown are distributions for  
    2016~(\cmsLeft) and the combination of 2017 and 2018~(\cmsRight). 
    The background-subtracted data are compared to \ttbar simulations with 
    the UE tunes CUETP8M2T4 for 2016 and CP5 for the combination of 2017 and 2018, 
    and different values of \fFSR are shown as well. 
    The lower panels show the ratio to the \ttbar simulation with $\fFSR=1$.
    \label{fig:tau32}}
\end{figure}
Figure~\ref{fig:tau32} shows the normalised distributions in \tauratio 
for 2016~(\cmsLeft), and the combination of 2017 and 2018~(\cmsRight). 
In both cases, larger values of \fFSR shift the distributions to lower values in \tauratio, 
and smaller values of \fFSR lead to a larger average value of \tauratio. 
This is compatible with the expectation of less radiation for larger values of \fFSR, 
corresponding to smaller values of \asFSR. Without additional radiation, \tauratio 
becomes small and compatible with a three-prong decay. If radiation is added to the 
jet, the value of $\tau_3$ increases, and shifts the average \tauratio to larger values. 

{\tolerance=800
The sensitivity of the \tauratio distribution to FSR can be used to determine 
the value of \fFSR that is most compatible with the data. 
We construct predictions $g_i(\fFSR)$ in each bin $i$ of the normalised \tauratio distributions, 
\begin{linenomath}
  \begin{equation}
    g_i(\fFSR) = a_i + b_i \log{\fFSR^{-2}} + c_i  \fFSR^{-2},
  \end{equation}
\end{linenomath}
with the free parameters $a_i$, $b_i$, and $c_i$. The functional form of $g_i$ is inspired 
by the logarithmic dependence of \asFSR on the square of the modified energy scale $(\fFSR \mu)^2$. 
The values of the free parameters 
are determined in a fit to simulation, sampled at the points $\fFSR \in \{\frac{1}{2}, 1, 2\}$ in 2016 
and $\fFSR \in \{ \frac{1}{4}, \frac{1}{2}, \frac{1}{\sqrt{2}}, 1, \sqrt{2}, 2, 4 \}$ in 2017 and 2018. 
\par}

{\tolerance=2000
The compatibility with the data is tested with a $\chi^2$ function, equivalent to the definition 
in Eq.~\eqref{eq:chisq}. The vector of differences is built from the normalised background-subtracted data, 
and the predictions $g_i(\fFSR)$.
The uncertainties taken into account by the covariance matrix include statistical uncertainties 
from data with correlations from the normalisation, 
and systematic uncertainties in the JECs and in the predictions $g_i(\fFSR)$.
The latter are conservatively estimated by using the largest statistical uncertainty in a given bin $i$ 
from any of the points obtained from the simulated samples with different values of \fFSR.
This choice was made because the point with $\fFSR=4$ has the smallest statistical 
precision due to the presence of a large spread of weights in the simulation.
The statistical uncertainty in data is the dominant uncertainty in this measurement. 
\par}

The best fit value of \fFSR is obtained by minimising the $\chi^2$ function. 
Uncertainties corresponding to one standard deviation are evaluated at $\chi^2_\text{min}+1$.
We obtain the best fit values $\fFSR = 0.97 \pm 0.07$ for 2016, 
and $\fFSR = 0.33 \pm 0.02$ for the combined data of 2017 and 2018.
The uncertainties in $\fFSR$ take into account statistical and leading systematic sources, 
where the latter are dominated by changes of the modelling in simulation, 
as described in Section~\ref{sec:uncertainties}. 
The modelling uncertainties included are uncertainties in the initial state radiation (ISR), 
the colour reconnection model, 
the underlying event tune, and the matching between matrix element and the parton shower. 
Experimental uncertainties considered are uncertainties in the JECs, the additional 
XCone-jet corrections, and JMS. 
We have found that the \tauratio distributions obtained with different values of \mtop are 
compatible within the statistical precision of the simulated \ttbar samples, 
and therefore we do not consider changes of \mtop in this study. 
We find that the statistical uncertainties from data and the limited size of the simulated 
samples constitute the largest source of uncertainty in the determination of $\fFSR$. 

{\tolerance=800
The best fit values of \fFSR can be translated to values of \asFSRZ. This gives 
$\asFSRZ = 0.1373_{-0.0018}^{+0.0017}$ for 2016 and 
$\asFSRZ = 0.1416_{-0.0018}^{+0.0019}$ for the combination of 2017 and 2018, 
evaluated using five active flavours in the four-loop evolution of \alpS~\cite{Herren:2017osy}. 
We note that these values do not represent a generally valid measurement of \asFSR, which would need a 
different treatment of theory uncertainties from missing higher orders, but the results can be used to calibrate the 
two different tunes used for the \ttbar simulation with \POWHEG{}+\PYTHIA. 
In fact, the two values are compatible and much closer to each other than the values used in the 
CUETP8M2T4 and CP5 tunes. 
The uncertainty for 2016 is comparable to the one from the combination of 2017 and 2018, 
which constitutes a larger data set, because the latter is dominated by 
statistical uncertainties in the simulation originating from a large spread of weights 
used to obtain the samples with changes in \fFSR. 
The data are well described by the nominal simulation in 2016, but prefer a larger value of 
\asFSR in the 2017 and 2018 simulations. 
We have checked that the 2017 and 2018 data are equally well or better described by the 
adjusted simulations with $\fFSR = 0.33$ in all distributions relevant for this analysis. 
The change in the 2016 simulation is insubstantial, 
with changes in distributions that are consistent with the statistical 
uncertainties of the simulated \ttbar sample.
We have verified that extracting \fFSR from different intervals in \mjet and \pt 
leads to compatible results, validating the calibration of the FSR modelling in the full 
fiducial region of this measurement. 
\par}

\section{Unfolding}
\label{sec:unfolding}

The data are unfolded to the particle level using regularised unfolding as 
implemented in the \textsc{TUnfold}~\cite{Schmitt:2012kp} framework.
We have chosen the curvature regularisation condition, such that the second 
derivative of the unfolded result is regularised.
This option introduces the smallest model dependencies in this measurement.
The optimal regularisation strength is found by minimising the average global correlation 
coefficient in the output bins~\cite{Schmitt:2016orm}. 
In addition to the measurement phase space defined in Section~\ref{sec:selection}, five sideband regions 
are constructed by loosening the most important selection steps. 
These regions include events where the XCone jet has $350<\pt<400\GeV$, the lepton has $55<\pt<60\GeV$, 
at least one of the XCone subjets has $10<\pt<30\GeV$, \mjet is less than the invariant 
mass of the sum of the second XCone jet and lepton, and the AK4 jet passes a 
\PQb-tagging requirement with a misidentification rate of 1\%, but not the tight requirement with 0.1\%. 
Additionally, the measurement region and the region with XCone jet $350<\pt<400\GeV$ are divided into bins of \pt. 
The two bins in the peak region of \mjet with bin boundaries at 152, 172 and 192\GeV are split into four 
bins in the unfolding, but merged afterwards to avoid large bin-to-bin correlations.
The splitting into regions of \ptjet and the subdivision of \mjet bins result in a reduced dependence 
on the modelling parameters in the \ttbar simulation and help to reduce the corresponding uncertainties.
In addition, this procedure ensures that the most important migrations between the detector and particle levels 
into and out of the fiducial region of the measurement are included in the unfolding and not purely estimated from simulation. 
In total, the response matrix includes 200 bins at the detector level and 72 bins at the particle level.

We unfold the three years individually in order to check for a potential bias originating from the different tunes 
in the \ttbar simulation that is used to construct the response matrix. 
With the dedicated calibration of the FSR parameter in the simulation, all three years are compatible 
and agree within one standard deviation.
We have also ensured that unfolding the electron and muon channels separately 
leads to a consistent result.
For the final measurement, all data and simulated samples are combined before the unfolding.

\section{Uncertainties}
\label{sec:uncertainties}

Several sources of statistical and systematic uncertainties are considered in the measurement of \mjet.
These are split into four categories: statistical, experimental, model, and theory uncertainties.

Statistical uncertainties are defined as the uncertainties due to the finite
statistical precision of the data. 
With respect to the previous measurement~\cite{Sirunyan:2019rfa}, the statistical precision 
is increased by including data from 2017 and 2018, which increases 
the size of the data set by a factor of almost four. 
The statistical uncertainties are propagated through the unfolding process 
using Gaussian error propagation.

Experimental uncertainties encompass uncertainties in correction factors that are 
connected to the calibration of physics objects.
These include the JECs~\cite{Khachatryan:2016kdb}, JER, 
additional XCone-jet corrections, JMS, as well as the factors correcting for the 
efficiencies in the trigger selection~\cite{Khachatryan:2016bia}, 
lepton identification~\cite{Sirunyan:2018fpa, CMS:2020uim}, 
and \PQb tagging~\cite{Sirunyan:2017ezt}. 
The JMS correction has been obtained by calibrating \mjet in the 
reconstructed \mw, which is dominated by XCone subjets originating from light-flavour quarks. 
To account for a possible difference in the detector response to XCone 
subjets originating from the fragmentation of \PQb quarks, 
an additional flavour uncertainty~\cite{Khachatryan:2016kdb} is applied to 
XCone subjets matched to AK4 \PQb-tagged jets (JMS \PQb flavour uncertainty), 
where the matching is identical to the procedure outlined in Section~\ref{sec:jms}. 
This JMS \PQb flavour uncertainty is obtained from the response difference 
of \PQb jets in \PYTHIA and \HERWIG~\cite{Bahr:2008pv, Gieseke:2012ft}. 
In addition, it is studied in a \PZ{}+\PQb{}-jet 
sample where the \PQb jet response can be studied in data~\cite{Khachatryan:2016kdb}. 
The uncertainties in the reweighting of the pileup profile are considered. 
The experimental uncertainties are calculated by changing the corrections up and down by one standard deviation, 
and the difference with respect to the nominal response matrix is then propagated to the unfolded distribution.
The uncertainty in the measurement of the integrated luminosity is estimated to be 1.6\%~\cite{lumi16, lumi17, lumi18}
and is assigned to the unfolded distribution directly. 
Statistical uncertainties from the limited size of the simulated samples, 
denoted by ``MC stat'', are included in the experimental uncertainties. 
The simulated samples for 2017 and 2018 increase the statistical precision 
of the unfolding compared to the previous measurement with 2016 data only, 
because of the higher statistical precision in the response matrix, 
which is obtained using simulated \ttbar events.  
Simulated background processes are used to 
estimate the amount of background events and are subtracted from data. 
The corresponding statistical uncertainties in the background samples 
are much smaller than the uncertainties in the cross sections of these processes, 
which are 19\% for \Wjets production,
23\% for single top quark production and 100\% for other SM 
backgrounds~\cite{Khachatryan:2016kzg, Chatrchyan:2014mua, Sirunyan:2016cdg,         
Kidonakis:2012rm, Gehrmann:2014fva, Khachatryan:2016tgp}. 
The statistical uncertainties from the limited size of the MC samples are found 
to be a factor of more than three smaller compared to the ones from data. 

{\tolerance=800
Model uncertainties arise from the choice of parameters in the event simulation.
These parameters include the factorisation and renormalisation scales \muf and \mur, 
the top quark mass, the colour reconnection, the UE tune, and the choice of PDFs.
Uncertainties in the parton shower are estimated by changing the energy scales for the ISR and the FSR, 
and varying the parameter that controls the matching between matrix element and parton shower (\hdamp)~\cite{Sirunyan:2019dfx}. 
These variations cover all observed differences between data and simulation in distributions relevant for this measurement. 
The uncertainty in the fragmentation of the \PQb quark has been estimated by changing its \pt distribution 
in the \POWHEG{}+\PYTHIA{} \ttbar simulation. It was found to have a negligible effect.
\par}

We do not consider an additional uncertainty from a comparison to an alternative parton shower simulation, 
as for example implemented in the \HERWIG event generator. 
Simulated \ttbar events using \POWHEG{}+\HERWIG~\cite{Bahr:2008pv} version 7.1 with tune CH3~\cite{CMS:2020dqt} 
do not describe the data as well as events produced with \POWHEG{}+\PYTHIA. 
Furthermore, an uncertainty derived from the difference between these simulations would result in an 
overestimation of the parton shower uncertainty and in a double counting of uncertainty sources. 
Instead, accounting for the different sources of parton shower uncertainties (ISR, FSR, \hdamp) 
provides a means to trace the relevant modelling uncertainties for this measurement. 
All model parameters are varied within their uncertainties and the corresponding 
uncertainties in the \mjet measurement are estimated as described in the following. 

The values of \muf, \mur, and the ISR scales are varied by factors from 0.5 to 2. The parameter 
\hdamp and the UE tune are varied within their uncertainties~\cite{Sirunyan:2019dfx}.
For \muf and \mur, there are eight possible combinations to vary the scales. 
We find that the simultaneous up and down variations of both scales have the largest effects.
In order to estimate the uncertainty in the \muf and \mur scales, 
we thus only consider simultaneous shifts of \muf and \mur.
In order to estimate the uncertainty in the colour reconnection model,
three different models~\cite{Sjostrand:1987su, Argyropoulos:2014zoa, Christiansen:2015yqa} 
are considered as variations. 
The uncertainty due to the choice of PDFs has been found to be negligible in the last 
measurements of \mjet~\cite{Sirunyan:2017yar, Sirunyan:2019rfa} because \mjet in 
fully merged top quark decays is sensitive to the decay of the top quark, 
but not to the dynamics of its production. 
Therefore, we do not follow the recommendation for estimating PDF uncertainties 
using different PDF sets~\cite{Butterworth:2015oua}, 
but we estimate the PDF uncertainty by using 100 variations of 
the NNPDF sets versions 3.0~\cite{Ball:2014uwa} and 3.1~\cite{NNPDF:2017mvq}.

For all model variations, the simulated \mjet distribution at the detector level is unfolded to the particle level 
using the same setup as for data. 
Differences between the true distribution at the particle level and the unfolded simulation with model variations 
indicate a potential bias in the unfolding setup and are treated as uncertainties. 
For uncertainties in the ISR scale, the \muf and \mur scales, the \hdamp parameter, and the UE tune 
the average bias of the up and down variations is calculated in each bin and taken as an uncertainty.
In the case of the colour reconnection model, the impact of a change in the model is calculated by 
taking the difference in the mean of \mjet between the true distribution at the particle level and the unfolded distribution. 
The model with the largest difference is chosen, and we take the resulting bias as the 
uncertainty from the colour reconnection model.

\begin{figure}[tb]
  \centering
  \includegraphics[width=0.47\textwidth]{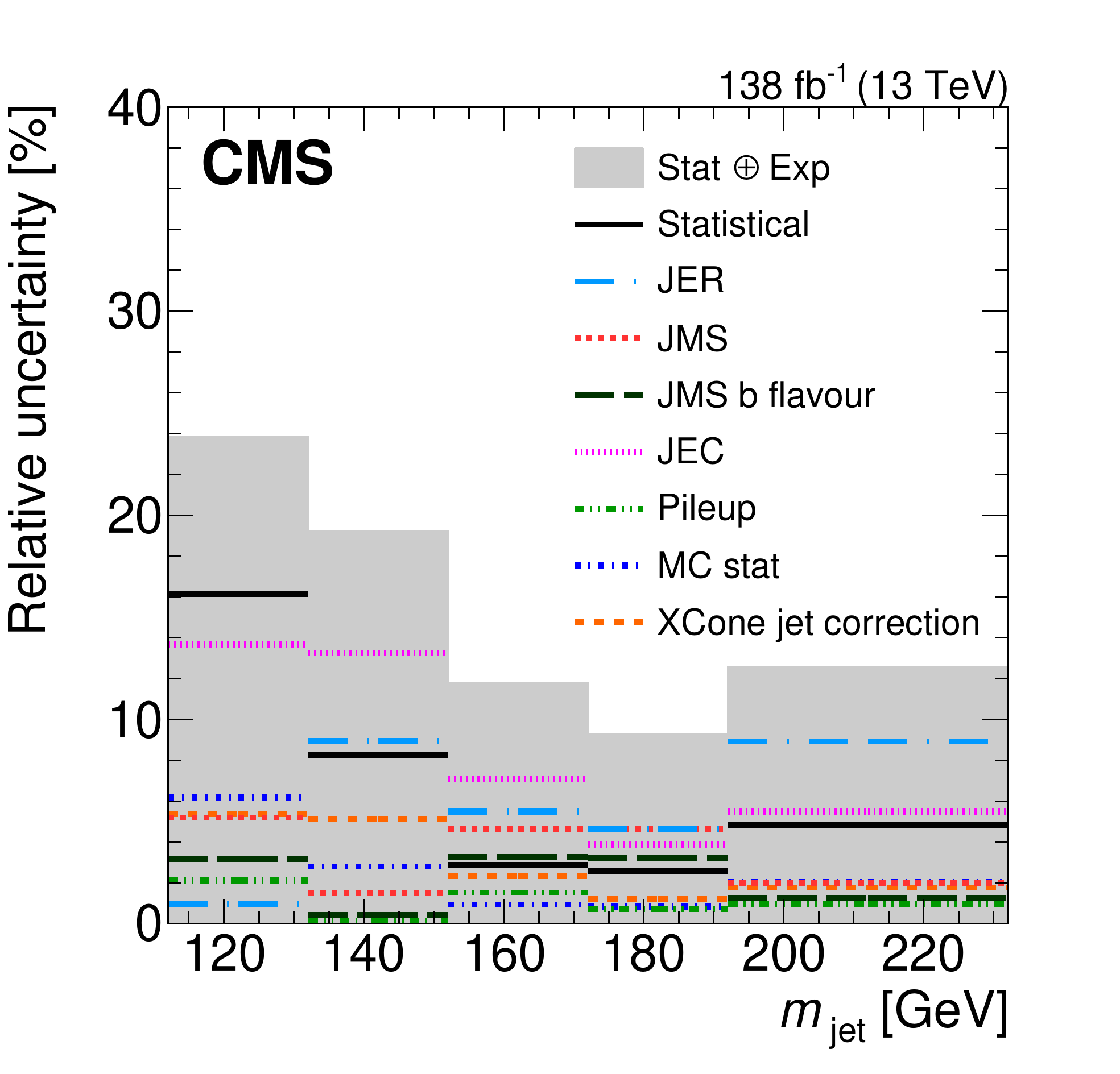}
  \includegraphics[width=0.47\textwidth]{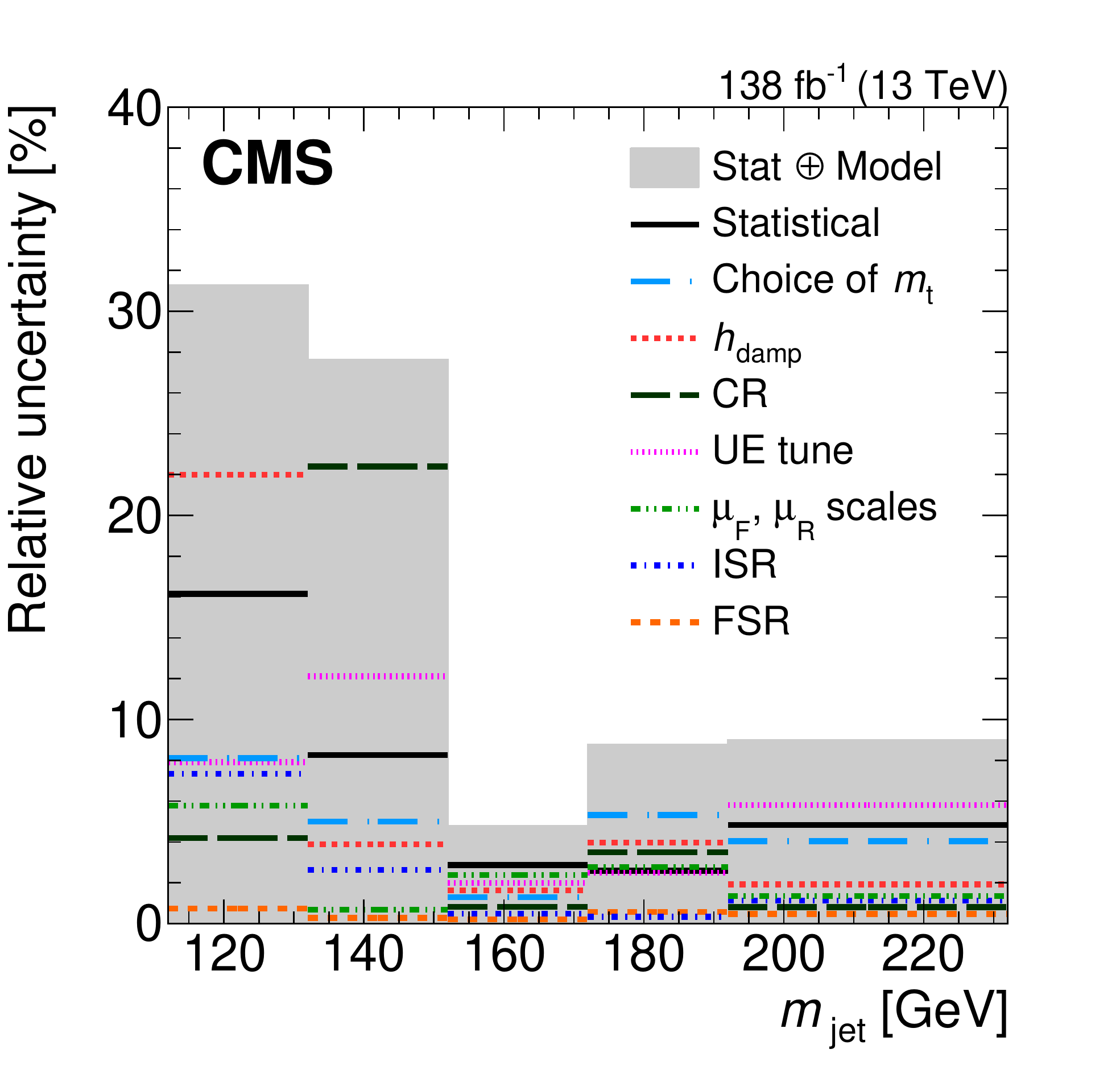}
  \caption{Relative experimental~(\cmsLeft) and model~(\cmsRight) uncertainties in the measurement of \mjet.
  Various sources are displayed as coloured lines and compared to the total experimental or model 
  uncertainty, respectively.
  The uncertainty sources are calculated as the square root of the diagonal entries from the respective covariance matrix, and do not include bin-to-bin correlations.
  \label{fig:uncerts}}
\end{figure}
The uncertainty due to the choice of \mtop in the \ttbar simulation used to unfold the data 
is calculated using samples with different values of \mtop.
The difference between the unfolded distribution and the true particle-level distribution is parametrised 
in each bin of the unfolded distribution. We use a linear function with \mtop as its argument 
to describe the difference. The parameters of this function are obtained 
using the \ttbar samples with $\mtop = 169.5$, 171.5, 173.5, and 175.5\GeV.
The uncertainty is then evaluated from the linear function at $\mtop = 172.5 \pm 1\GeV$. 
This procedure has the advantage of being less susceptible to statistical fluctuations in 
the individual samples, therefore resulting in a more reliable estimate of this uncertainty. 
The interval of ${\pm}1\GeV$ has been found to be sufficient, because larger variations do not 
agree with the data at the detector level. 

We use the same method to calculate the uncertainty in the modelling of FSR. The simulated samples 
with different choices of \fFSR are unfolded, and the differences between the true 
distribution at the particle level and the unfolded 
distributions are parametrised as a function of \fFSR in each bin. The uncertainty is obtained 
by evaluating the parametrisation at the values obtained in the studies described in Section~\ref{sec:FSR}.

Figure~\ref{fig:uncerts} summarises the experimental and model uncertainties in the measurement of \mjet.
The largest experimental uncertainties arise from the JES and JER corrections.
In the \mjet peak region (the third and fourth bins) the largest sources of model uncertainties are 
from the UE tune, the \hdamp parameter, and the choice of \mtop. 
In the first two bins, the limited statistical precision of the samples with model variations, 
in combination with a smaller number of observed events than in the peak region, 
leads to statistical fluctuations in the estimation 
of model uncertainties. This results in large uncertainties from \hdamp and the colour reconnection 
models in the first and second bins of the measurement, respectively.  
Because the sensitivity to \mtop of the \mjet measurement comes from the peak region, 
these uncertainties have a minor effect on the determination of \mtop. 

\begin{figure}[tb]
  \centering
  \includegraphics[width=0.47\textwidth]{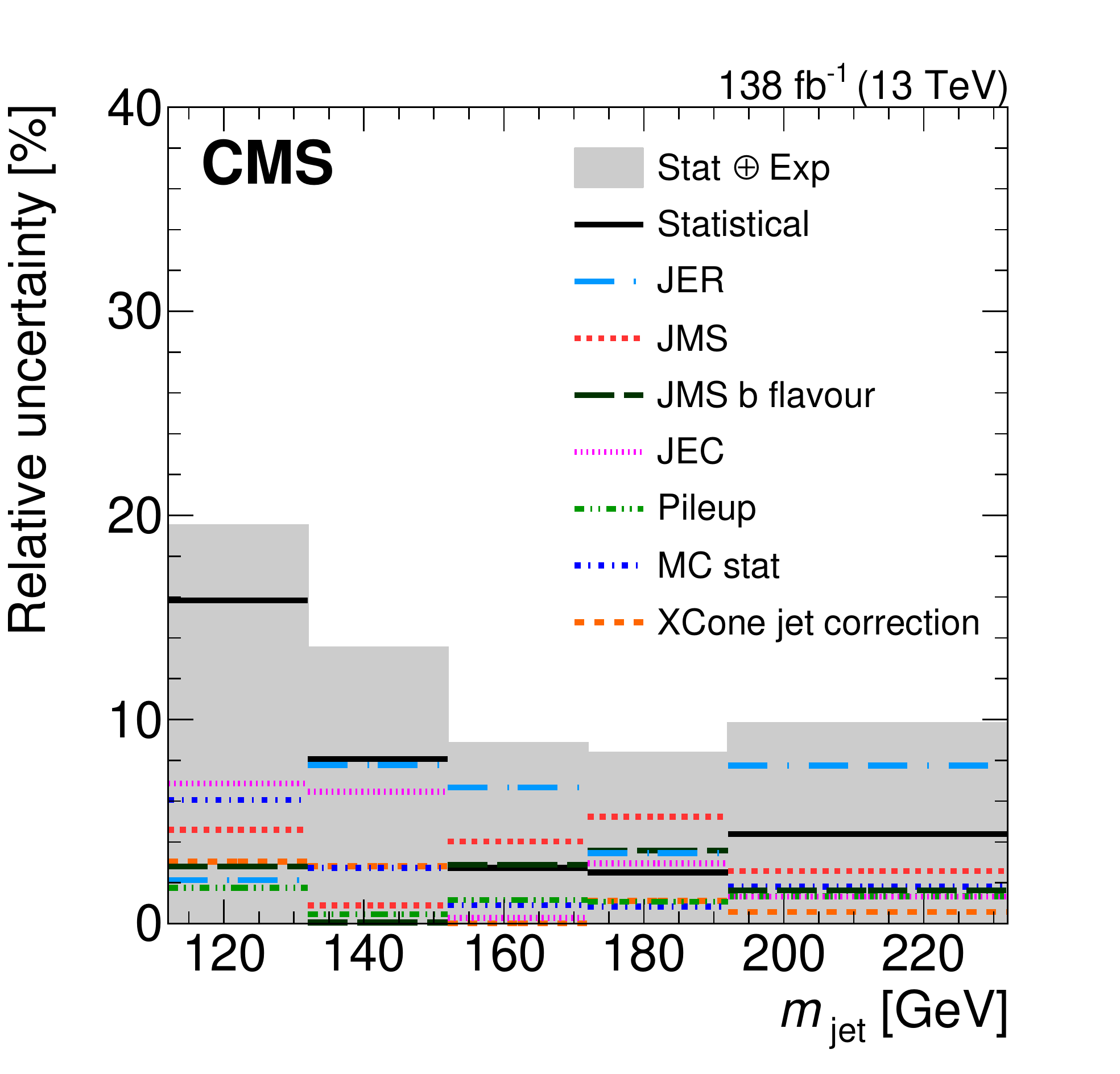}
  \includegraphics[width=0.47\textwidth]{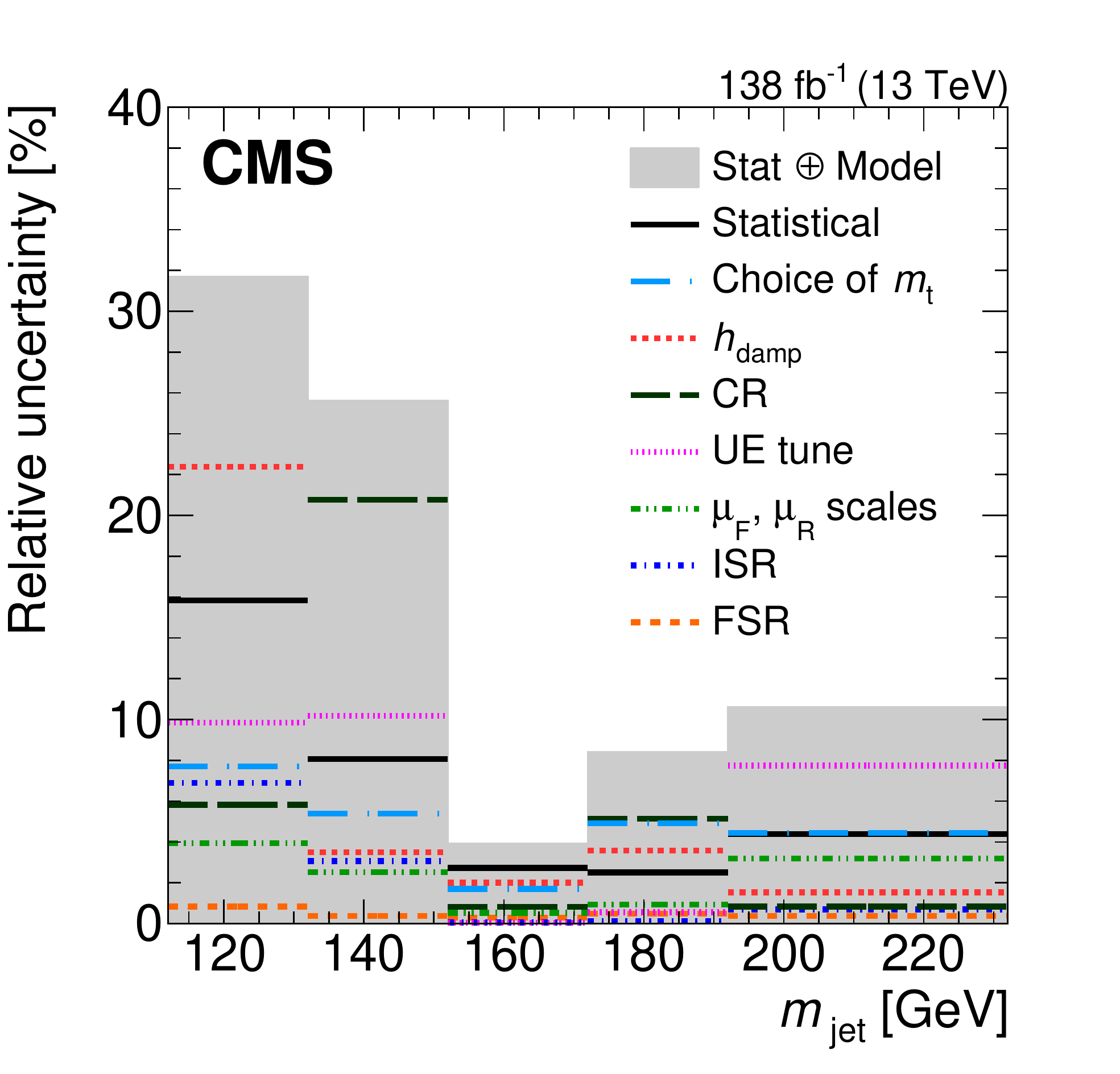} 
  \caption{Relative experimental~(\cmsLeft) and model~(\cmsRight) uncertainties after normalising the 
  measurement to the total cross section.
  Various sources are displayed as coloured lines and compared to the total 
  experimental or model uncertainty, respectively.
  The uncertainty sources are calculated as the square root of the diagonal entries from the respective covariance matrix, and do not include bin-to-bin correlations.
  \label{fig:uncerts_norm}} 
\end{figure}
When normalising the unfolded distribution, systematic uncertainties cancel fully or partially. 
For example, the uncertainty in the integrated luminosity cancels completely as it affects all bins by 
an equal amount. 
The uncertainty component in the JEC that changes only the three-vector 
predominantly changes the XCone jet \pt, and thus affects 
the selection efficiency of the measurement. This uncertainty cancels to a large part when 
normalising the measurement and becomes negligible. 
The uncertainties in the normalised measurement are summarized in Fig.~\ref{fig:uncerts_norm}. 
In the peak region, the dominant experimental uncertainties originate from JER and JMS corrections. 
The dominant model uncertainties are the same as for the absolute cross section measurement. 

Theory uncertainties are those uncertainties that apply to predictions at the particle level. 
The scales for FSR, ISR, as well as \muf and \mur are varied by factors of 0.5 and 2.
The UE tune and the \hdamp parameter are varied within their uncertainties.
All three models of colour reconnection are used to calculate the corresponding uncertainty. 
For each source, the uncertainty in each bin is estimated by the largest difference to the 
nominal prediction at the particle level.

\section{Results and determination of the top quark mass}
\label{sec:mt}

\begin{figure}[tb]
  \centering
    \includegraphics[width=0.49\textwidth]{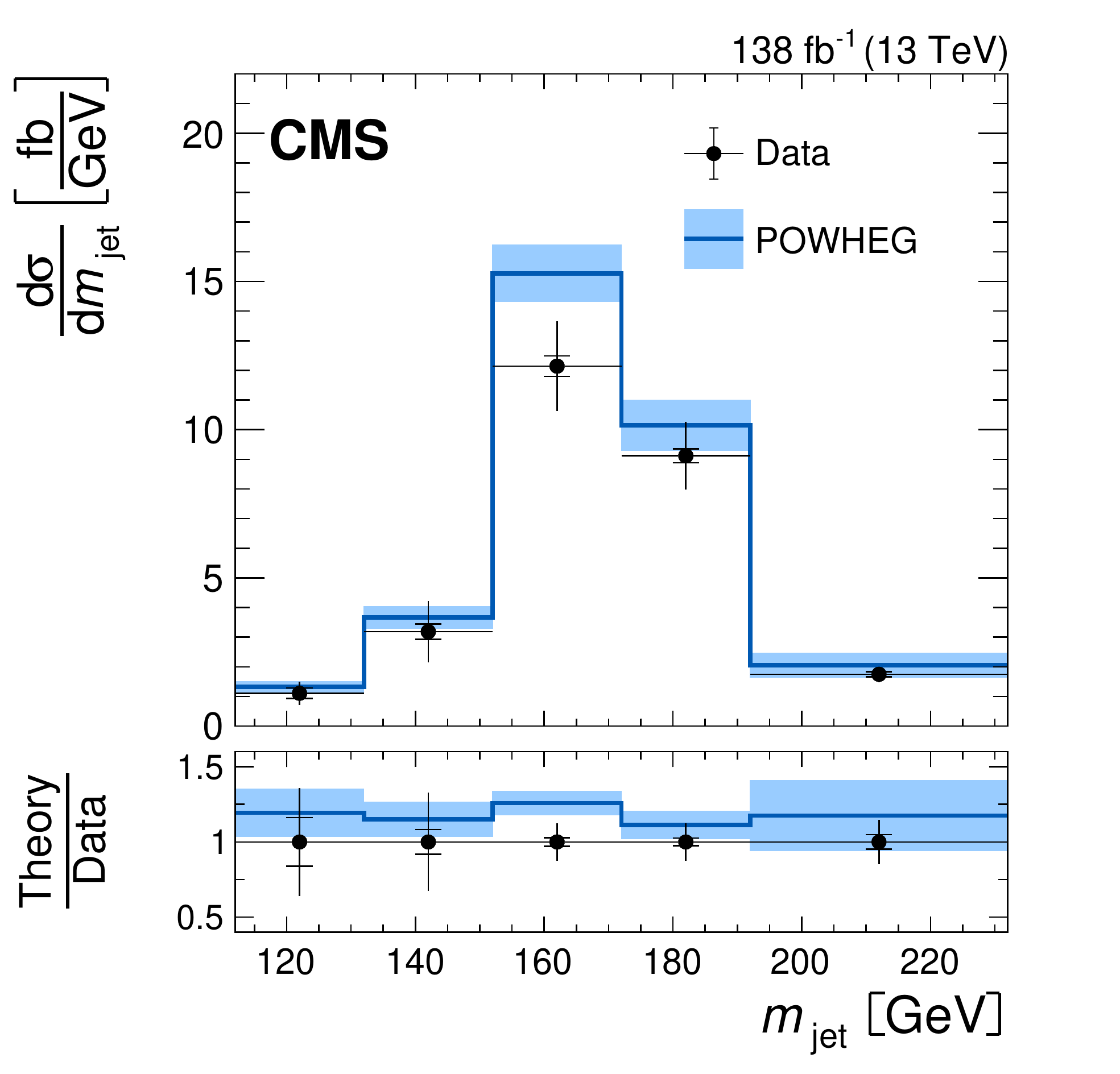}
    \includegraphics[width=0.49\textwidth]{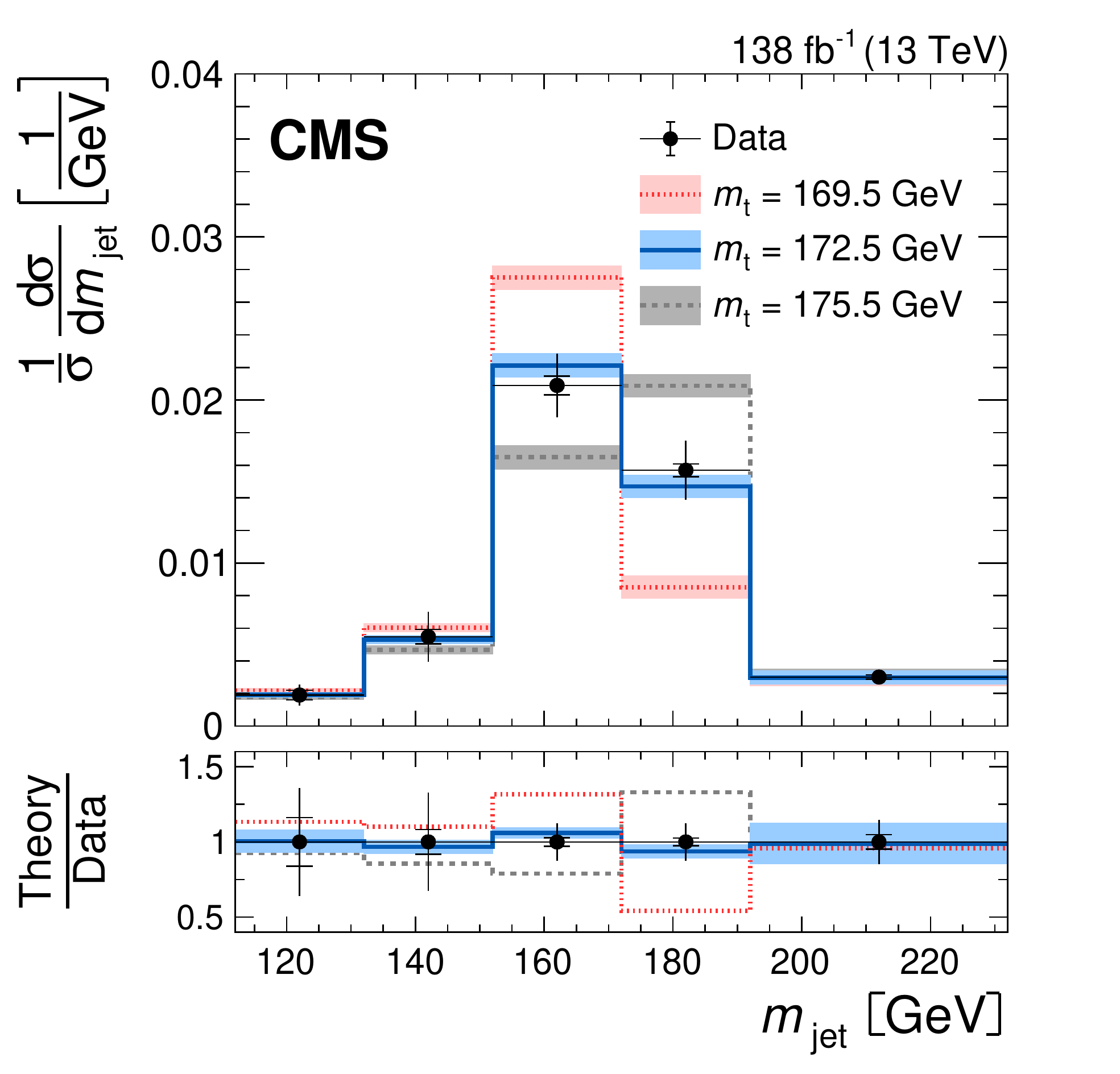}
    \caption{Differential \ttbar production cross section as a function of \mjet 
      compared to predictions obtained with \POWHEG: absolute (\cmsLeft) and normalised (\cmsRight).
      For the normalised measurement, the data are compared to predictions with different \mtop.
      The vertical bars represent the total uncertainties, and the statistical uncertainties 
      are shown by short horizontal bars.  
      The long horizontal bars reflect the bin widths.
      Theoretical uncertainties in the prediction are indicated by the bands.
      The lower panels show the ratio of the theoretical prediction to data.
    \label{fig:unfold}}
\end{figure}
The three different years, as well as the electron and muon channels, are combined before the unfolding, 
but are also processed individually to validate their consistency. 
Figure~\ref{fig:unfold} (\cmsLeft) shows the differential \ttbar cross section in the fiducial region as a function of \mjet, 
measured in data and compared to simulation.
The \ttbar production cross section in the fiducial region is measured to be 
$581 \pm 8 \stat \pm 46 \exper \pm 19 \model\fb$. This can be compared to the prediction 
from the \POWHEG simulation, $690 \pm 59\fb$. 
The smaller value of the measured cross section compared to the prediction from \POWHEG at NLO 
has been observed in other analyses for top quark $\pt>400\GeV$~\cite{CMS:2021vhb, ATLAS:2022xfj, ATLAS:2022mlu}, 
where NNLO calculations describe the shape of the top quark \pt distribution better.

We determine the value of \mtop from the normalised differential \ttbar production cross section as a function of \mjet. 
This enables a measurement using the shape of the \mjet distribution 
without sensitivity to uncertainties in the normalisation. 
Figure~\ref{fig:unfold} (\cmsRight) shows the normalised measurement compared to predictions from \POWHEG with different values of \mtop.
In order to extract \mtop, a fit is performed based on $\chi_m^2 = \dm^T \Vm^{-1} \dm$, 
where \dm is the vector of differences between the measured normalised differential cross section and the 
\POWHEG simulation with different values of \mtop. 
Four of the five bins in \mjet are used in the calculation of \dm, 
because of the normalisation of the measurement. 
The covariance matrix \Vm contains all statistical, experimental, model, and theory uncertainties. 
We use the Linear Template Fit~\cite{Britzger:2021ocj} package to parametrise the cross section 
as a function of \mtop and obtain the best fit value with the corresponding uncertainties analytically. 

\begin{figure}[tb]
  \centering
  \includegraphics[width=0.49\textwidth]{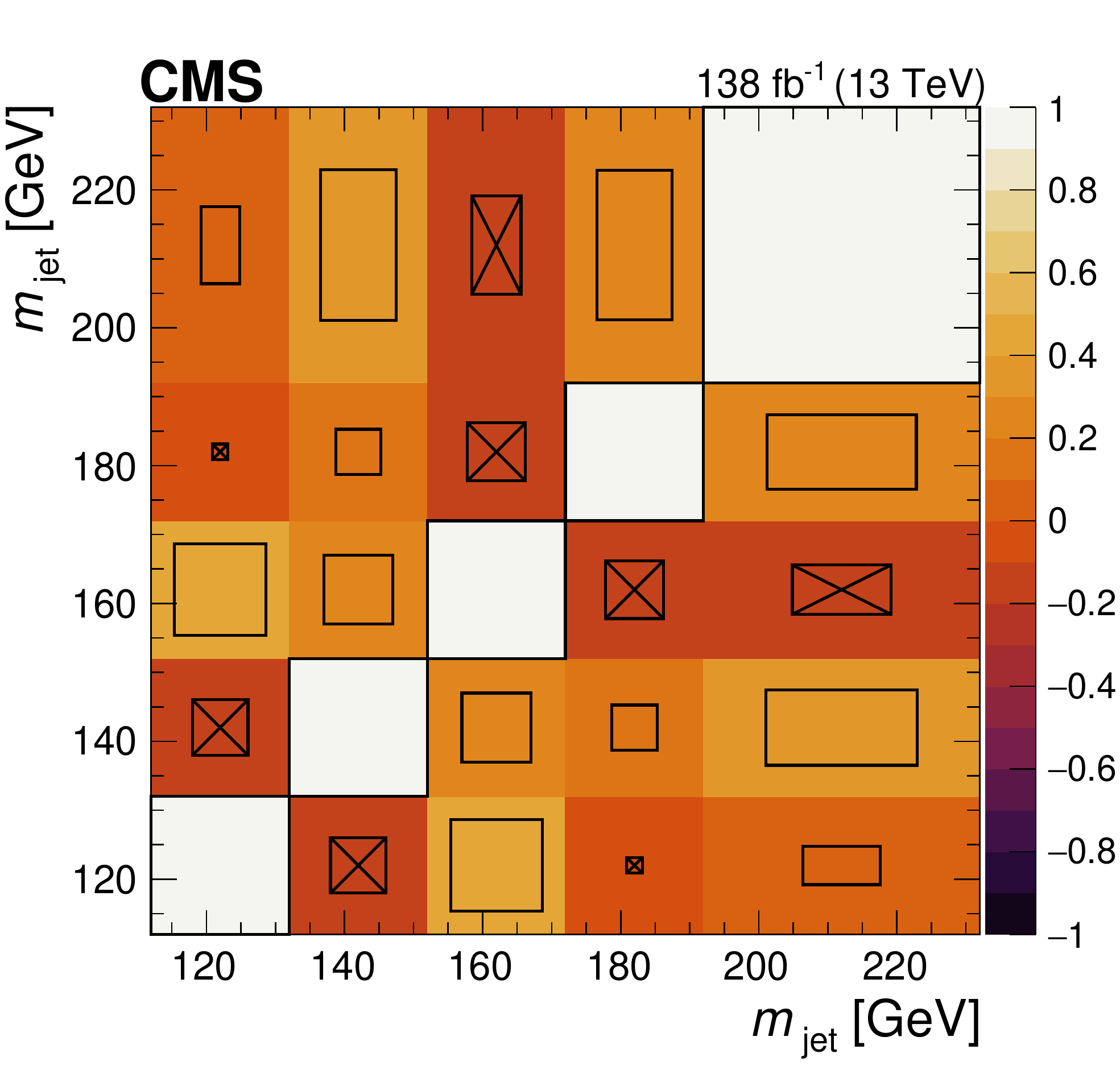} 
  \includegraphics[width=0.49\textwidth]{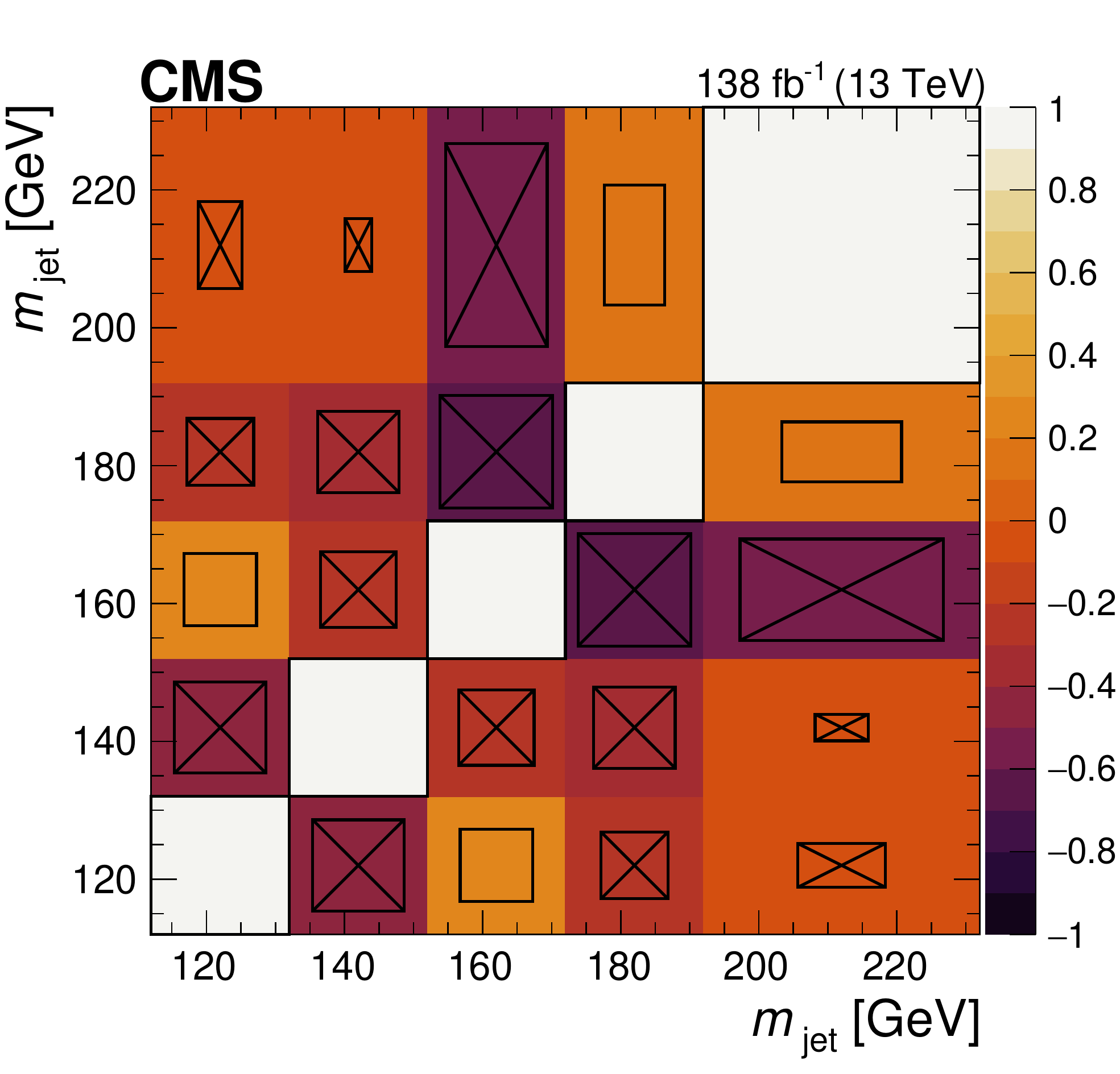} 
  \caption{Correlations between the bins in the unfolding before~(\cmsLeft) and after~(\cmsRight) normalising the distribution to the total cross section.
  Boxes with crosses indicate negative values of the correlation coefficient. 
  \label{fig:correlations}} 
\end{figure}
The bin-to-bin correlations in the measurement calculated from \Vm, including statistical, experimental, and 
model contributions, are displayed in Fig.~\ref{fig:correlations}.
Negative correlations between neighbouring bins originate from migrations at the detector level, 
which have been corrected for by the unfolding and result in anticorrelated statistical uncertainties. 
The systematic variations that shift the peak of the \mjet distribution, for example the JMS, 
also contribute to the negative correlations.

\begin{figure}[tb]
  \centering
  \includegraphics[width=0.49\textwidth]{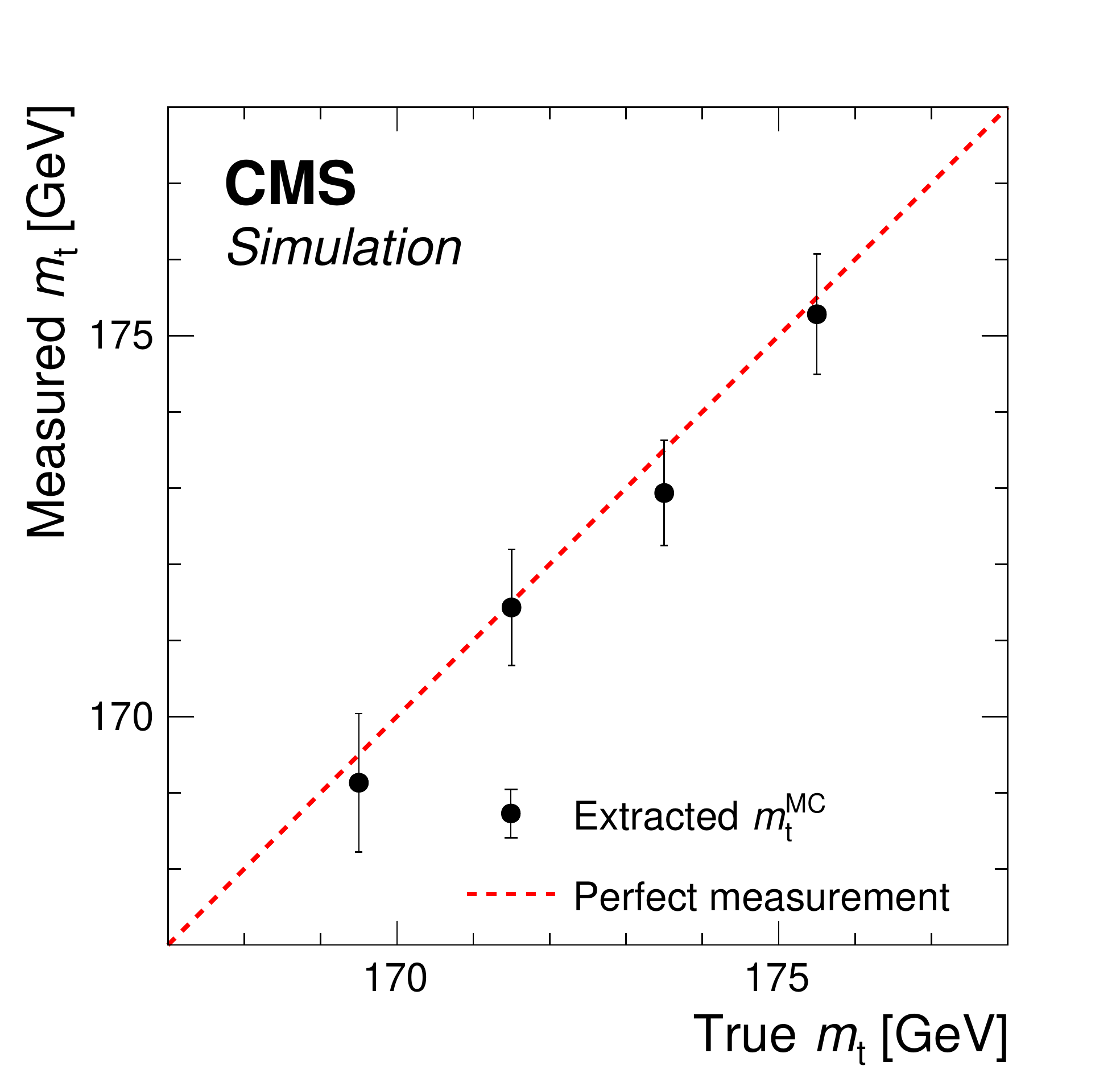}
  \caption{Extracted top quark mass from simulation compared to the true value.
    The vertical error bars show the total uncertainty in the extraction of \mtop.
  \label{fig:masscalib}}
\end{figure}

In order to validate that the determination of \mtop is unbiased, we perform the \mtop 
measurement using simulated samples with various values of \mtop. 
The obtained value of \mtop is compared to the true value in Fig.~\ref{fig:masscalib}.
In this comparison, all extracted values agree with the respective true values of \mtop, 
demonstrating the validity of the mass extraction.

Performing the extraction on collision data and considering all sources of uncertainties, 
we extract \mtop using the \POWHEG{}+\PYTHIA simulation, 
\begin{linenomath}
\ifthenelse{\boolean{cms@external}}
{
\begin{equation*}
\begin{aligned}
\mtop &= 173.06 \pm 0.24\stat \pm 0.61 \exper \\
      &\pm 0.47 \model \pm 0.23 \thy \GeV \\
      &= 173.06 \pm 0.84 \GeV.
\end{aligned} 
\end{equation*}
}
{
\begin{equation*}
\begin{aligned}
\mtop &= 173.06 \pm 0.24\stat \pm 0.61 \exper \pm 0.47 \model \pm 0.23 \thy \GeV \\
      &= 173.06 \pm 0.84 \GeV.
\end{aligned} 
\end{equation*}
}
\end{linenomath}
With respect to the previous CMS measurement at 13\TeV~\cite{Sirunyan:2019rfa}, this corresponds to 
an improvement by more than a factor of three in terms of precision.
This measurement from boosted top quark production has an uncertainty comparable with the most 
precise \mtop extractions from fully resolved final states~\cite{Aaboud:2016igd, Aaboud:2017mae, Aaboud:2018zbu,
Khachatryan:2015hba, Sirunyan:2017idq, Sirunyan:2018gqx, Sirunyan:2018mlv}.

When unfolding the 2016, 2017 and 2018 data separately and extracting \mtop from these three independent measurements, 
we find agreement between the extracted values of \mtop to better than one standard deviation. All three values 
are compatible with the combined value to better than one half standard deviation. We find the same when unfolding the 
electron and muon channels separately. 

\begin{table}[tb]
\centering
\topcaption{Total and individual uncertainties in the extraction of \mtop from the normalised differential cross section.
The uncertainties are grouped into experimental, model, theory, and statistical uncertainties. 
Uncertainties from the choice of the PDF, \PQb tagging, the luminosity measurement, and the lepton triggers, identification 
and reconstruction are smaller than 0.01\GeV and are not listed. 
\label{tab:masscontribs}}
\begin{tabular}{l r}
  Source & Uncertainty [{\GeVns}] \\
  \hline
  Jet energy resolution & 0.38 \\
  Jet mass scale & 0.37 \\
  Jet mass scale \PQb flavour & 0.26 \\
  MC stat & 0.09\\
  Pileup & 0.08\\
  Jet energy scale & 0.07\\
  Additional XCone corrections & 0.01\\
  Backgrounds & 0.01 \\[\cmsTabSkip]
 {Experimental total} &  {0.61} \\
  \hline\\[\cmsTabSkip]
  Choice of \mtop & 0.41 \\
  Colour reconnection & 0.17\\
  \hdamp & 0.09 \\
  Underlying event tune & 0.09\\
  \muf, \mur scales & 0.08\\
  ISR & 0.02\\
  FSR & 0.02\\[\cmsTabSkip]
 {Model total} &  {0.47} \\
  \hline\\[\cmsTabSkip]
  Underlying event tune & 0.13 \\
  FSR & 0.11 \\
  \muf, \mur scales & 0.10\\
  Colour reconnection & 0.09\\
  \hdamp & 0.04\\
  ISR & 0.04 \\[\cmsTabSkip]
 {Theory total} &  {0.23} \\
  \hline\\[\cmsTabSkip]
 {Statistical} &  {0.24} \\
 {Total} &  {0.84} \\
 \hline
\end{tabular}
\end{table}
The individual sources of uncertainty and their impact on the mass extraction are detailed in Table~\ref{tab:masscontribs}.
The dominant experimental uncertainties are connected to the calibration of the JER, 
the JMS calibration, and the JMS \PQb flavour uncertainty, 
also visible in Fig.~\ref{fig:uncerts_norm}. 
The dominant modelling uncertainties arise from the choice of the \mtop and \hdamp parameters in the \ttbar simulation.
Compared to the previous measurement, the dedicated measurement of the JMS 
leads to an uncertainty reduced by a factor of 5 in the jet calibration.
By constraining the simulation of FSR with data, this previously dominant model uncertainty becomes small.
The use of about four times the data, corresponding to an integrated luminosity of 138\fbinv, 
leads to a reduction in the statistical uncertainty by a factor of 2. 

The improvements described in this article result in a considerable gain in precision, allowing for a determination 
of \mtop from \ttbar production at high \pt with an uncertainty comparable to the one achieved 
in measurements close to the \ttbar production threshold with fully resolved final state objects. 
The measurement also provides important information on the modelling of the jet mass in decays of boosted top quarks, 
which is the most important substructure variable for the identification of large-radius jets~\cite{CMS:2020poo}.   

\section{Conclusions}
\label{sec:summary}

A measurement of the differential top quark pair (\ttbar) production cross section as a function of the jet 
mass \mjet in hadronic decays of boosted top quarks has been presented. 
The normalised distribution in \mjet is sensitive to the top quark mass \mtop, which is 
measured to be $173.06 \pm 0.84\GeV$. This value is compatible with earlier precision measurements in 
fully resolved final states~\cite{Aaboud:2018zbu, Sirunyan:2018gqx, Sirunyan:2018mlv}. 
With respect to an earlier CMS analysis~\cite{Sirunyan:2019rfa}, the precision is improved by a 
factor of more than three. This has been achieved by a dedicated calibration of the jet mass scale, 
a study of the effects of final state radiation inside large-radius jets, and about 4 times more data. 
With these improvements, the uncertainty in the extraction of \mtop at high top quark boosts becomes 
comparable to direct measurements close to the \ttbar production threshold. 
The sources of the leading systematic uncertainties are very different, highlighting the 
complementarity of this measurement. 
In addition, the study of boosted top quarks offers the possibility to directly compare the 
distribution in \mjet to analytic calculations~\cite{Hoang:2017kmk}.
When these calculations become available, the unfolded \mjet distribution 
can be used to measure the top quark pole mass directly. 
The precisely measured differential cross section as a function of \mjet 
represents an important step towards understanding and resolving the ambiguities 
between the top quark mass extracted from a direct reconstruction of \mtop, 
and the top quark pole mass.

\begin{acknowledgments}
We congratulate our colleagues in the CERN accelerator departments for the excellent performance of the LHC and thank the technical and administrative staffs at CERN and at other CMS institutes for their contributions to the success of the CMS effort. In addition, we gratefully acknowledge the computing centres and personnel of the Worldwide LHC Computing Grid and other centres for delivering so effectively the computing infrastructure essential to our analyses. Finally, we acknowledge the enduring support for the construction and operation of the LHC, the CMS detector, and the supporting computing infrastructure provided by the following funding agencies: BMBWF and FWF (Austria); FNRS and FWO (Belgium); CNPq, CAPES, FAPERJ, FAPERGS, and FAPESP (Brazil); MES and BNSF (Bulgaria); CERN; CAS, MoST, and NSFC (China); MINCIENCIAS (Colombia); MSES and CSF (Croatia); RIF (Cyprus); SENESCYT (Ecuador); MoER, ERC PUT and ERDF (Estonia); Academy of Finland, MEC, and HIP (Finland); CEA and CNRS/IN2P3 (France); BMBF, DFG, and HGF (Germany); GSRI (Greece); NKFIH (Hungary); DAE and DST (India); IPM (Iran); SFI (Ireland); INFN (Italy); MSIP and NRF (Republic of Korea); MES (Latvia); LAS (Lithuania); MOE and UM (Malaysia); BUAP, CINVESTAV, CONACYT, LNS, SEP, and UASLP-FAI (Mexico); MOS (Montenegro); MBIE (New Zealand); PAEC (Pakistan); MES and NSC (Poland); FCT (Portugal); MESTD (Serbia); MCIN/AEI and PCTI (Spain); MOSTR (Sri Lanka); Swiss Funding Agencies (Switzerland); MST (Taipei); MHESI and NSTDA (Thailand); TUBITAK and TENMAK (Turkey); NASU (Ukraine); STFC (United Kingdom); DOE and NSF (USA).

\hyphenation{Rachada-pisek} Individuals have received support from the Marie-Curie programme and the European Research Council and Horizon 2020 Grant, contract Nos.\ 675440, 724704, 752730, 758316, 765710, 824093, 884104, and COST Action CA16108 (European Union); the Leventis Foundation; the Alfred P.\ Sloan Foundation; the Alexander von Humboldt Foundation; the Belgian Federal Science Policy Office; the Fonds pour la Formation \`a la Recherche dans l'Industrie et dans l'Agriculture (FRIA-Belgium); the Agentschap voor Innovatie door Wetenschap en Technologie (IWT-Belgium); the F.R.S.-FNRS and FWO (Belgium) under the ``Excellence of Science -- EOS" -- be.h project n.\ 30820817; the Beijing Municipal Science \& Technology Commission, No. Z191100007219010; the Ministry of Education, Youth and Sports (MEYS) of the Czech Republic; the Hellenic Foundation for Research and Innovation (HFRI), Project Number 2288 (Greece); the Deutsche Forschungsgemeinschaft (DFG), under Germany's Excellence Strategy -- EXC 2121 ``Quantum Universe" -- 390833306, and under project number 400140256 - GRK2497; the Hungarian Academy of Sciences, the New National Excellence Program - \'UNKP, the NKFIH research grants K 124845, K 124850, K 128713, K 128786, K 129058, K 131991, K 133046, K 138136, K 143460, K 143477, 2020-2.2.1-ED-2021-00181, and TKP2021-NKTA-64 (Hungary); the Council of Science and Industrial Research, India; the Latvian Council of Science; the Ministry of Education and Science, project no. 2022/WK/14, and the National Science Center, contracts Opus 2021/41/B/ST2/01369 and 2021/43/B/ST2/01552 (Poland); the Funda\c{c}\~ao para a Ci\^encia e a Tecnologia, grant CEECIND/01334/2018 (Portugal); the National Priorities Research Program by Qatar National Research Fund; MCIN/AEI/10.13039/501100011033, ERDF ``a way of making Europe", and the Programa Estatal de Fomento de la Investigaci{\'o}n Cient{\'i}fica y T{\'e}cnica de Excelencia Mar\'{\i}a de Maeztu, grant MDM-2017-0765 and Programa Severo Ochoa del Principado de Asturias (Spain); the Chulalongkorn Academic into Its 2nd Century Project Advancement Project, and the National Science, Research and Innovation Fund via the Program Management Unit for Human Resources \& Institutional Development, Research and Innovation, grant B05F650021 (Thailand); the Kavli Foundation; the Nvidia Corporation; the SuperMicro Corporation; the Welch Foundation, contract C-1845; and the Weston Havens Foundation (USA).
\end{acknowledgments}

\bibliography{auto_generated}

\providecommand{\href}[2]{#2}\begingroup\raggedright\begin{thebibliography}{100}%
\makeatletter
\providecommand{\hrefCMSnoop }[0]{\@secondoftwo}%
\makeatother
\providecommand{\doi}{\texttt{doi:}\begingroup \urlstyle{tt}\Url}

\bibitem{Abe:1995hr}
\hrefCMSnoop {}{{CDF} Collaboration, ``{Observation of top quark production in
  $\bar{p}p$ collisions}'',} \textit{ Phys. Rev. Lett.} \textbf{ 74} (1995)
  2626,
  \href{http://dx.doi.org/10.1103/PhysRevLett.74.2626}{\doi{10.1103/PhysRevLett.74.2626}},
\href{http://www.arXiv.org/abs/hep-ex/9503002}{\texttt{arXiv:hep-ex/9503002}}.

\bibitem{D0:1995jca}
\hrefCMSnoop {}{{\DZERO} Collaboration, ``{Observation of the top quark}'',}
  \textit{ Phys. Rev. Lett.} \textbf{ 74} (1995) 2632,
  \href{http://dx.doi.org/10.1103/PhysRevLett.74.2632}{\doi{10.1103/PhysRevLett.74.2632}},
\href{http://www.arXiv.org/abs/hep-ex/9503003}{\texttt{arXiv:hep-ex/9503003}}.

\bibitem{ALEPH:2010aa}
\hrefCMSnoop {}{{ALEPH, CDF, D0, DELPHI, L3, OPAL, and SLD Collaborations, the
  LEP Electroweak Working Group, the Tevatron Electroweak Working Group, and
  the SLD Electroweak and Heavy Flavour Groups}, ``Precision electroweak
  measurements and constraints on the standard model'',} 2010.
  \href{http://www.arXiv.org/abs/1012.2367}{\texttt{arXiv:1012.2367}}.

\bibitem{Haller:2018nnx}
J.~Haller\hrefCMSnoop {}{ {et~al.}, ``Update of the global electroweak fit and
  constraints on two-{H}iggs-doublet models'',} \textit{ Eur. Phys. J. C}
  \textbf{ 78} (2018) 675,
  \href{http://dx.doi.org/10.1140/epjc/s10052-018-6131-3}{\doi{10.1140/epjc/s10052-018-6131-3}},
\href{http://www.arXiv.org/abs/1803.01853}{\texttt{arXiv:1803.01853}}.

\bibitem{PDG:2020}
\hrefCMSnoop {}{{Particle Data Group}, P.~A. Zyla {et~al.}, ``Review of
  particle physics'',} \textit{ Prog. Theor. Exp. Phys.} \textbf{ 2020} (2020)
  083C01,
  \href{http://dx.doi.org/10.1093/ptep/ptaa104}{\doi{10.1093/ptep/ptaa104}}.

\bibitem{Degrassi:2012ry}
G.~Degrassi\hrefCMSnoop {}{ {et~al.}, ``Higgs mass and vacuum stability in the
  standard model at {NNLO}'',} \textit{ JHEP} \textbf{ 08} (2012) 098,
  \href{http://dx.doi.org/10.1007/JHEP08(2012)098}{\doi{10.1007/JHEP08(2012)098}},
  \href{http://www.arXiv.org/abs/1205.6497}{\texttt{arXiv:1205.6497}}.

\bibitem{Bezrukov:2012sa}
\hrefCMSnoop {}{F.~Bezrukov, M.~Y. Kalmykov, B.~A. Kniehl, and M.~Shaposhnikov,
  ``{H}iggs boson mass and new physics'',} \textit{ JHEP} \textbf{ 10} (2012)
  140,
  \href{http://dx.doi.org/10.1007/JHEP10(2012)140}{\doi{10.1007/JHEP10(2012)140}},
\href{http://www.arXiv.org/abs/1205.2893}{\texttt{arXiv:1205.2893}}.

\bibitem{Bednyakov:2015sca}
\hrefCMSnoop {}{A.~V. Bednyakov, B.~A. Kniehl, A.~F. Pikelner, and O.~L.
  Veretin, ``Stability of the electroweak vacuum: {G}auge independence and
  advanced precision'',} \textit{ Phys. Rev. Lett.} \textbf{ 115} (2015)
  201802,
  \href{http://dx.doi.org/10.1103/PhysRevLett.115.201802}{\doi{10.1103/PhysRevLett.115.201802}},
  \href{http://www.arXiv.org/abs/1507.08833}{\texttt{arXiv:1507.08833}}.

\bibitem{Aaboud:2016igd}
\hrefCMSnoop {}{{ATLAS Collaboration}, ``Measurement of the top quark mass in
  the $\ttbar\rightarrow$ dilepton channel from $\sqrt{s}=8$ {TeV} {ATLAS}
  data'',} \textit{ Phys. Lett. B} \textbf{ 761} (2016) 350,
  \href{http://dx.doi.org/10.1016/j.physletb.2016.08.042}{\doi{10.1016/j.physletb.2016.08.042}},
\href{http://www.arXiv.org/abs/1606.02179}{\texttt{arXiv:1606.02179}}.

\bibitem{Aaboud:2017mae}
\hrefCMSnoop {}{{ATLAS Collaboration}, ``Top-quark mass measurement in the
  all-hadronic $\ttbar$ decay channel at $\sqrt{s}=8$ {TeV} with the {ATLAS}
  detector'',} \textit{ JHEP} \textbf{ 09} (2017) 118,
  \href{http://dx.doi.org/10.1007/JHEP09(2017)118}{\doi{10.1007/JHEP09(2017)118}},
\href{http://www.arXiv.org/abs/1702.07546}{\texttt{arXiv:1702.07546}}.

\bibitem{Aaboud:2018zbu}
\hrefCMSnoop {}{{ATLAS Collaboration}, ``Measurement of the top quark mass in
  the $\ttbar\rightarrow$ lepton+jets channel from $\sqrt{s}=8$ {TeV} {ATLAS}
  data and combination with previous results'',} \textit{ Eur. Phys. J. C}
  \textbf{ 79} (2019) 290,
  \href{http://dx.doi.org/10.1140/epjc/s10052-019-6757-9}{\doi{10.1140/epjc/s10052-019-6757-9}},
\href{http://www.arXiv.org/abs/1810.01772}{\texttt{arXiv:1810.01772}}.

\bibitem{Khachatryan:2015hba}
\hrefCMSnoop {}{{CMS Collaboration}, ``Measurement of the top quark mass using
  proton-proton data at ${\sqrt{s}} = 7$ and 8 {TeV}'',} \textit{ Phys. Rev. D}
  \textbf{ 93} (2016) 072004,
  \href{http://dx.doi.org/10.1103/PhysRevD.93.072004}{\doi{10.1103/PhysRevD.93.072004}},
\href{http://www.arXiv.org/abs/1509.04044}{\texttt{arXiv:1509.04044}}.

\bibitem{Sirunyan:2017idq}
\hrefCMSnoop {}{{CMS Collaboration}, ``Measurement of the top quark mass in the
  dileptonic \ttbar decay channel using the mass observables ${M}_{\cPqb\ell}$,
  ${M}_{ {\text T}2 }$, and ${M}_{\cPqb\ell\nu}$ in \pp collisions at
  $\sqrt{s}=8$ {TeV}'',} \textit{ Phys. Rev. D} \textbf{ 96} (2017) 032002,
  \href{http://dx.doi.org/10.1103/PhysRevD.96.032002}{\doi{10.1103/PhysRevD.96.032002}},
\href{http://www.arXiv.org/abs/1704.06142}{\texttt{arXiv:1704.06142}}.

\bibitem{Sirunyan:2018gqx}
\hrefCMSnoop {}{{CMS Collaboration}, ``Measurement of the top quark mass with
  lepton+jets final states using \pp collisions at $\sqrt{s}=13$ {TeV}'',}
  \textit{ Eur. Phys. J. C} \textbf{ 78} (2018) 891,
  \href{http://dx.doi.org/10.1140/epjc/s10052-018-6332-9}{\doi{10.1140/epjc/s10052-018-6332-9}},
\href{http://www.arXiv.org/abs/1805.01428}{\texttt{arXiv:1805.01428}}.

\bibitem{Sirunyan:2018mlv}
\hrefCMSnoop {}{{CMS Collaboration}, ``Measurement of the top quark mass in the
  all-jets final state at $\sqrt{s} = 13$ {TeV} and combination with the
  lepton+jets channel'',} \textit{ Eur. Phys. J. C} \textbf{ 79} (2019) 313,
  \href{http://dx.doi.org/10.1140/epjc/s10052-019-6788-2}{\doi{10.1140/epjc/s10052-019-6788-2}},
\href{http://www.arXiv.org/abs/1812.10534}{\texttt{arXiv:1812.10534}}.

\bibitem{Hoang:2017suc}
A.~H. Hoang\hrefCMSnoop {}{ {et~al.}, ``The {MSR} mass and the
  $\mathcal{O}\left({\Lambda}_{\mathrm{qcd}}\right)$ renormalon sum rule'',}
  \textit{ JHEP} \textbf{ 04} (2018) 003,
  \href{http://dx.doi.org/10.1007/JHEP04(2018)003}{\doi{10.1007/JHEP04(2018)003}},
  \href{http://www.arXiv.org/abs/1704.01580}{\texttt{arXiv:1704.01580}}.

\bibitem{Hoang:2020iah}
\hrefCMSnoop {}{A.~H. Hoang, ``What is the top quark mass?'',} \textit{ Ann.
  Rev. Nucl. Part. Sci.} \textbf{ 70} (2020) 225,
  \href{http://dx.doi.org/10.1146/annurev-nucl-101918-023530}{\doi{10.1146/annurev-nucl-101918-023530}},
  \href{http://www.arXiv.org/abs/2004.12915}{\texttt{arXiv:2004.12915}}.

\bibitem{Abazov:2011pta}
\hrefCMSnoop {}{{\DZERO} Collaboration, ``Determination of the pole and
  $\overline{MS}$ masses of the top quark from the \ttbar cross section'',}
  \textit{ Phys. Lett. B} \textbf{ 703} (2011) 422,
  \href{http://dx.doi.org/10.1016/j.physletb.2011.08.015}{\doi{10.1016/j.physletb.2011.08.015}},
\href{http://www.arXiv.org/abs/1104.2887}{\texttt{arXiv:1104.2887}}.

\bibitem{Abazov:2016ekt}
\hrefCMSnoop {}{{\DZERO} Collaboration, ``Measurement of the inclusive \ttbar
  production cross section in $\rm{p\bar{p}}$ collisions at $\sqrt{s}=1.96$
  {TeV} and determination of the top quark pole mass'',} \textit{ Phys. Rev. D}
  \textbf{ 94} (2016) 092004,
  \href{http://dx.doi.org/10.1103/PhysRevD.94.092004}{\doi{10.1103/PhysRevD.94.092004}},
\href{http://www.arXiv.org/abs/1605.06168}{\texttt{arXiv:1605.06168}}.

\bibitem{Aad:2014kva}
\hrefCMSnoop {}{{ATLAS Collaboration}, ``Measurement of the \ttbar production
  cross-section using $e\mu$ events with b-tagged jets in \pp collisions at
  $\sqrt{s} = 7$ and 8 {TeV} with the {ATLAS} detector'',} \textit{ Eur. Phys.
  J. C} \textbf{ 74} (2014) 3109,
  \href{http://dx.doi.org/10.1140/epjc/s10052-014-3109-7}{\doi{10.1140/epjc/s10052-014-3109-7}},
  \href{http://www.arXiv.org/abs/1406.5375}{\texttt{arXiv:1406.5375}}.
[Addendum: \DOI{10.1140/epjc/s10052-016-4501-2}].

\bibitem{Aaboud:2017ujq}
\hrefCMSnoop {}{{ATLAS Collaboration}, ``Measurement of lepton differential
  distributions and the top quark mass in \ttbar production in \pp collisions
  at $\sqrt{s}=8$ {TeV} with the {ATLAS} detector'',} \textit{ Eur. Phys. J. C}
  \textbf{ 77} (2017) 804,
  \href{http://dx.doi.org/10.1140/epjc/s10052-017-5349-9}{\doi{10.1140/epjc/s10052-017-5349-9}},
\href{http://www.arXiv.org/abs/1709.09407}{\texttt{arXiv:1709.09407}}.

\bibitem{ATLAS:2019hau}
\hrefCMSnoop {}{{ATLAS Collaboration}, ``Measurement of the \ttbar production
  cross-section and lepton differential distributions in $e\mu$ dilepton events
  from \pp collisions at $\sqrt{s}=13$ {TeV} with the {ATLAS} detector'',}
  \textit{ Eur. Phys. J. C} \textbf{ 80} (2020) 528,
  \href{http://dx.doi.org/10.1140/epjc/s10052-020-7907-9}{\doi{10.1140/epjc/s10052-020-7907-9}},
  \href{http://www.arXiv.org/abs/1910.08819}{\texttt{arXiv:1910.08819}}.

\bibitem{Chatrchyan:2013haa}
\hrefCMSnoop {}{{CMS Collaboration}, ``Determination of the top-quark pole mass
  and strong coupling constant from the \ttbar production cross section in \pp
  collisions at $\sqrt{s}$ = 7 {TeV}'',} \textit{ Phys. Lett. B} \textbf{ 728}
  (2014) 496,
  \href{http://dx.doi.org/10.1016/j.physletb.2013.12.009}{\doi{10.1016/j.physletb.2013.12.009}},
  \href{http://www.arXiv.org/abs/1307.1907}{\texttt{arXiv:1307.1907}}.
[Erratum: \DOI{10.1016/j.physletb.2014.08.040}].

\bibitem{Khachatryan:2016mqs}
\hrefCMSnoop {}{{CMS Collaboration}, ``Measurement of the \ttbar production
  cross section in the e$\mu$ channel in proton-proton collisions at $\sqrt{s}
  = 7$ and 8 {TeV}'',} \textit{ JHEP} \textbf{ 08} (2016) 029,
  \href{http://dx.doi.org/10.1007/JHEP08(2016)029}{\doi{10.1007/JHEP08(2016)029}},
\href{http://www.arXiv.org/abs/1603.02303}{\texttt{arXiv:1603.02303}}.

\bibitem{Sirunyan:2018goh}
\hrefCMSnoop {}{{CMS Collaboration}, ``Measurement of the \ttbar production
  cross section, the top quark mass, and the strong coupling constant using
  dilepton events in \pp collisions at $\sqrt{s} = 13$ {TeV}'',} \textit{ Eur.
  Phys. J. C} \textbf{ 79} (2019) 368,
  \href{http://dx.doi.org/10.1140/epjc/s10052-019-6863-8}{\doi{10.1140/epjc/s10052-019-6863-8}},
\href{http://www.arXiv.org/abs/1812.10505}{\texttt{arXiv:1812.10505}}.

\bibitem{ATLAS:2015pfy}
\hrefCMSnoop {}{{ATLAS Collaboration}, ``Determination of the top-quark pole
  mass using $\ttbar$+1-jet events collected with the {ATLAS} experiment in 7
  {TeV} ${\Pp\Pp}$ collisions'',} \textit{ JHEP} \textbf{ 10} (2015) 121,
  \href{http://dx.doi.org/10.1007/JHEP10(2015)121}{\doi{10.1007/JHEP10(2015)121}},
  \href{http://www.arXiv.org/abs/1507.01769}{\texttt{arXiv:1507.01769}}.

\bibitem{ATLAS:2017dhr}
\hrefCMSnoop {}{{ATLAS Collaboration}, ``Measurement of lepton differential
  distributions and the top quark mass in $\ttbar$ production in ${\Pp\Pp}$
  collisions at $\sqrt{s}=8$ {TeV} with the {ATLAS} detector'',} \textit{ Eur.
  Phys. J. C} \textbf{ 77} (2017) 804,
  \href{http://dx.doi.org/10.1140/epjc/s10052-017-5349-9}{\doi{10.1140/epjc/s10052-017-5349-9}},
  \href{http://www.arXiv.org/abs/1709.09407}{\texttt{arXiv:1709.09407}}.

\bibitem{ATLAS:2019guf}
\hrefCMSnoop {}{{ATLAS Collaboration}, ``Measurement of the top-quark mass in
  $\ttbar$+1-jet events collected with the {ATLAS} detector in ${\Pp\Pp}$
  collisions at $\sqrt{s}=8$ {TeV}'',} \textit{ JHEP} \textbf{ 11} (2019) 150,
  \href{http://dx.doi.org/10.1007/JHEP11(2019)150}{\doi{10.1007/JHEP11(2019)150}},
  \href{http://www.arXiv.org/abs/1905.02302}{\texttt{arXiv:1905.02302}}.

\bibitem{CMS:2022emx}
\hrefCMSnoop {}{{CMS Collaboration}, ``Measurement of the top quark pole mass
  using {\ttbar}+jet events in the dilepton final state in proton-proton
  collisions at $\sqrt{s}$ = 13~{TeV}'',} 2022.
  \href{http://www.arXiv.org/abs/2207.02270}{\texttt{arXiv:2207.02270}}.
  Submitted to \textit{JHEP}.

\bibitem{CMS:2019esx}
\hrefCMSnoop {}{{CMS Collaboration}, ``Measurement of \ttbar normalised
  multi-differential cross sections in \pp collisions at $\sqrt{s}=13$ {TeV},
  and simultaneous determination of the strong coupling strength, top quark
  pole mass, and parton distribution functions'',} \textit{ Eur. Phys. J. C}
  \textbf{ 80} (2020) 658,
  \href{http://dx.doi.org/10.1140/epjc/s10052-020-7917-7}{\doi{10.1140/epjc/s10052-020-7917-7}},
  \href{http://www.arXiv.org/abs/1904.05237}{\texttt{arXiv:1904.05237}}.

\bibitem{Larkoski:2017jix}
\hrefCMSnoop {}{A.~J. Larkoski, I.~Moult, and B.~Nachman, ``Jet substructure at
  the {L}arge {H}adron {C}ollider: {A} review of recent advances in theory and
  machine learning'',} \textit{ Phys. Rep.} \textbf{ 841} (2020) 1,
  \href{http://dx.doi.org/10.1016/j.physrep.2019.11.001}{\doi{10.1016/j.physrep.2019.11.001}},
\href{http://www.arXiv.org/abs/1709.04464}{\texttt{arXiv:1709.04464}}.

\bibitem{Asquith:2018igt}
R.~Kogler\hrefCMSnoop {}{ {et~al.}, ``Jet substructure at the large hadron
  collider'',} \textit{ Rev. Mod. Phys.} \textbf{ 91} (2019) 045003,
  \href{http://dx.doi.org/10.1103/RevModPhys.91.045003}{\doi{10.1103/RevModPhys.91.045003}},
\href{http://www.arXiv.org/abs/1803.06991}{\texttt{arXiv:1803.06991}}.

\bibitem{Kogler:2021kkw}
R.~Kogler, ``Advances in jet substructure at the {LHC}: {A}lgorithms,
  measurements and searches for new physical phenomena'', volume 284 of
  \textit{ Springer Tracts Mod. Phys.}
\newblock Springer, 2021.
\newblock
  \href{http://dx.doi.org/10.1007/978-3-030-72858-8}{\doi{10.1007/978-3-030-72858-8}},
  ISBN~978-3-030-72857-1, 978-3-030-72858-8.

\bibitem{Hoang:2017kmk}
\hrefCMSnoop {}{A.~H. Hoang, S.~Mantry, A.~Pathak, and I.~W. Stewart,
  ``Extracting a short distance top mass with light grooming'',} \textit{ Phys.
  Rev. D} \textbf{ 100} (2019) 074021,
  \href{http://dx.doi.org/10.1103/PhysRevD.100.074021}{\doi{10.1103/PhysRevD.100.074021}},
\href{http://www.arXiv.org/abs/1708.02586}{\texttt{arXiv:1708.02586}}.

\bibitem{Sirunyan:2017yar}
\hrefCMSnoop {}{{CMS Collaboration}, ``Measurement of the jet mass in highly
  boosted \ttbar events from {\pp} collisions at $\sqrt{s}=8$ {TeV}'',}
  \textit{ Eur. Phys. J. C} \textbf{ 77} (2017) 467,
  \href{http://dx.doi.org/10.1140/epjc/s10052-017-5030-3}{\doi{10.1140/epjc/s10052-017-5030-3}},
\href{http://www.arXiv.org/abs/1703.06330}{\texttt{arXiv:1703.06330}}.

\bibitem{Sirunyan:2019rfa}
\hrefCMSnoop {}{{CMS Collaboration}, ``Measurement of the jet mass distribution
  and top quark mass in hadronic decays of boosted top quarks in \pp collisions
  at $\sqrt{s} =13$ {TeV}'',} \textit{ Phys. Rev. Lett.} \textbf{ 124} (2020)
  202001,
  \href{http://dx.doi.org/10.1103/PhysRevLett.124.202001}{\doi{10.1103/PhysRevLett.124.202001}},
  \href{http://www.arXiv.org/abs/1911.03800}{\texttt{arXiv:1911.03800}}.

\bibitem{CMS:2011shu}
\hrefCMSnoop {}{{CMS Collaboration}, ``Determination of jet energy calibration
  and transverse momentum resolution in {CMS}'',} \textit{ JINST} \textbf{ 6}
  (2011) P11002,
  \href{http://dx.doi.org/10.1088/1748-0221/6/11/P11002}{\doi{10.1088/1748-0221/6/11/P11002}},
  \href{http://www.arXiv.org/abs/1107.4277}{\texttt{arXiv:1107.4277}}.

\bibitem{Khachatryan:2016kdb}
\hrefCMSnoop {}{{CMS Collaboration}, ``Jet energy scale and resolution in the
  {CMS} experiment in {\pp} collisions at 8 {TeV}'',} \textit{ JINST} \textbf{
  12} (2017) P02014,
  \href{http://dx.doi.org/10.1088/1748-0221/12/02/P02014}{\doi{10.1088/1748-0221/12/02/P02014}},
\href{http://www.arXiv.org/abs/1607.03663}{\texttt{arXiv:1607.03663}}.

\bibitem{Thaler:2010tr}
\hrefCMSnoop {}{J.~Thaler and K.~Van~Tilburg, ``Identifying boosted objects
  with ${N}$-subjettiness'',} \textit{ JHEP} \textbf{ 03} (2011) 015,
  \href{http://dx.doi.org/10.1007/JHEP03(2011)015}{\doi{10.1007/JHEP03(2011)015}},
\href{http://www.arXiv.org/abs/1011.2268}{\texttt{arXiv:1011.2268}}.

\bibitem{Thaler:2011gf}
\hrefCMSnoop {}{J.~Thaler and K.~Van~Tilburg, ``Maximizing boosted top
  identification by minimizing ${N}$-subjettiness'',} \textit{ JHEP} \textbf{
  02} (2012) 093,
  \href{http://dx.doi.org/10.1007/JHEP02(2012)093}{\doi{10.1007/JHEP02(2012)093}},
  \href{http://www.arXiv.org/abs/1108.2701}{\texttt{arXiv:1108.2701}}.

\bibitem{Skands:2014pea}
\hrefCMSnoop {}{P.~Skands, S.~Carrazza, and J.~Rojo, ``Tuning {PYTHIA 8.1}:
  {T}he {Monash 2013 Tune}'',} \textit{ Eur. Phys. J. C} \textbf{ 74} (2014)
  3024,
  \href{http://dx.doi.org/10.1140/epjc/s10052-014-3024-y}{\doi{10.1140/epjc/s10052-014-3024-y}},
\href{http://www.arXiv.org/abs/1404.5630}{\texttt{arXiv:1404.5630}}.

\bibitem{CMS:2018ypj}
\hrefCMSnoop {}{{CMS Collaboration}, ``Measurement of jet substructure
  observables in \ttbar events from proton-proton collisions at
  $\sqrt{s}=13$~{TeV}'',} \textit{ Phys. Rev. D} \textbf{ 98} (2018) 092014,
  \href{http://dx.doi.org/10.1103/PhysRevD.98.092014}{\doi{10.1103/PhysRevD.98.092014}},
  \href{http://www.arXiv.org/abs/1808.07340}{\texttt{arXiv:1808.07340}}.

\bibitem{hepdata}
\hrefCMSnoop {}{}{HEPD}ata record for this analysis, 2022.
\newblock
  \href{http://dx.doi.org/10.17182/hepdata.130712}{\doi{10.17182/hepdata.130712}}.

\bibitem{Chatrchyan:2008zzk}
\hrefCMSnoop {}{{CMS Collaboration}, ``The {CMS} experiment at the {CERN}
  {LHC}'',} \textit{ JINST} \textbf{ 3} (2008) S08004,
  \href{http://dx.doi.org/10.1088/1748-0221/3/08/S08004}{\doi{10.1088/1748-0221/3/08/S08004}}.

\bibitem{CMSTrackerGroup:2020edz}
\hrefCMSnoop {}{{Tracker Group of the CMS} Collaboration, ``The {CMS} {P}hase-1
  pixel detector upgrade'',} \textit{ JINST} \textbf{ 16} (2021) P02027,
  \href{http://dx.doi.org/10.1088/1748-0221/16/02/P02027}{\doi{10.1088/1748-0221/16/02/P02027}},
  \href{http://www.arXiv.org/abs/2012.14304}{\texttt{arXiv:2012.14304}}.

\bibitem{CMS:2020cmk}
\hrefCMSnoop {}{{CMS Collaboration}, ``Performance of the {CMS} {L}evel-1
  trigger in proton-proton collisions at $\sqrt{s} =13$~{TeV}'',} \textit{
  JINST} \textbf{ 15} (2020) P10017,
  \href{http://dx.doi.org/10.1088/1748-0221/15/10/P10017}{\doi{10.1088/1748-0221/15/10/P10017}},
  \href{http://www.arXiv.org/abs/2006.10165}{\texttt{arXiv:2006.10165}}.

\bibitem{Khachatryan:2016bia}
\hrefCMSnoop {}{{CMS Collaboration}, ``The {CMS} trigger system'',} \textit{
  JINST} \textbf{ 12} (2017) P01020,
  \href{http://dx.doi.org/10.1088/1748-0221/12/01/P01020}{\doi{10.1088/1748-0221/12/01/P01020}},
\href{http://www.arXiv.org/abs/1609.02366}{\texttt{arXiv:1609.02366}}.

\bibitem{lumi16}
\hrefCMSnoop {}{{CMS Collaboration}, ``Precision luminosity measurement in
  proton-proton collisions at $\sqrt{s} =$ 13 {TeV} in 2015 and 2016 at
  {CMS}'',} \textit{ Eur. Phys. J. C} \textbf{ 81} (2021) 800,
  \href{http://dx.doi.org/10.1140/epjc/s10052-021-09538-2}{\doi{10.1140/epjc/s10052-021-09538-2}},
  \href{http://www.arXiv.org/abs/2104.01927}{\texttt{arXiv:2104.01927}}.

\bibitem{lumi17}
\href {https://cds.cern.ch/record/2621960}{{CMS Collaboration}, ``{CMS}
  luminosity measurement for the 2017 data-taking period at $\sqrt{s} =
  13~\mathrm{TeV}$'',} CMS Physics Analysis Summary CMS-PAS-LUM-17-004, 2018.

\bibitem{lumi18}
\href {https://cds.cern.ch/record/2676164}{{CMS Collaboration}, ``{CMS}
  luminosity measurement for the 2018 data-taking period at $\sqrt{s} =
  13~\mathrm{TeV}$'',} CMS Physics Analysis Summary CMS-PAS-LUM-18-002, 2019.

\bibitem{Nason:2004rx}
\hrefCMSnoop {}{P.~Nason, ``A new method for combining {NLO QCD} with shower
  {Monte Carlo} algorithms'',} \textit{ JHEP} \textbf{ 11} (2004) 040,
  \href{http://dx.doi.org/10.1088/1126-6708/2004/11/040}{\doi{10.1088/1126-6708/2004/11/040}},
\href{http://www.arXiv.org/abs/hep-ph/0409146}{\texttt{arXiv:hep-ph/0409146}}.

\bibitem{Frixione:2007vw}
\hrefCMSnoop {}{S.~Frixione, P.~Nason, and C.~Oleari, ``Matching {NLO QCD}
  computations with parton shower simulations: the {POWHEG} method'',} \textit{
  JHEP} \textbf{ 11} (2007) 070,
  \href{http://dx.doi.org/10.1088/1126-6708/2007/11/070}{\doi{10.1088/1126-6708/2007/11/070}},
\href{http://www.arXiv.org/abs/0709.2092}{\texttt{arXiv:0709.2092}}.

\bibitem{Alioli:2010xd}
\hrefCMSnoop {}{S.~Alioli, P.~Nason, C.~Oleari, and E.~Re, ``A general
  framework for implementing {NLO} calculations in shower {Monte Carlo}
  programs: the {POWHEG BOX}'',} \textit{ JHEP} \textbf{ 06} (2010) 043,
  \href{http://dx.doi.org/10.1007/JHEP06(2010)043}{\doi{10.1007/JHEP06(2010)043}},
\href{http://www.arXiv.org/abs/1002.2581}{\texttt{arXiv:1002.2581}}.

\bibitem{Frixione:2007nw}
\hrefCMSnoop {}{S.~Frixione, P.~Nason, and G.~Ridolfi, ``A positive-weight
  next-to-leading-order {Monte Carlo} for heavy flavour hadroproduction'',}
  \textit{ JHEP} \textbf{ 09} (2007) 126,
  \href{http://dx.doi.org/10.1088/1126-6708/2007/09/126}{\doi{10.1088/1126-6708/2007/09/126}},
\href{http://www.arXiv.org/abs/0707.3088}{\texttt{arXiv:0707.3088}}.

\bibitem{Alioli:2009je}
\hrefCMSnoop {}{S.~Alioli, P.~Nason, C.~Oleari, and E.~Re, ``{NLO} single-top
  production matched with shower in {POWHEG}: $s$- and $t$-channel
  contributions'',} \textit{ JHEP} \textbf{ 09} (2009) 111,
  \href{http://dx.doi.org/10.1088/1126-6708/2009/09/111}{\doi{10.1088/1126-6708/2009/09/111}},
  \href{http://www.arXiv.org/abs/0907.4076}{\texttt{arXiv:0907.4076}}.
[Erratum: \DOI{10.1007/JHEP02(2010)011}].

\bibitem{Re:2010bp}
\hrefCMSnoop {}{E.~Re, ``Single-top {Wt}-channel production matched with parton
  showers using the {POWHEG} method'',} \textit{ Eur. Phys. J. C} \textbf{ 71}
  (2011) 1547,
  \href{http://dx.doi.org/10.1140/epjc/s10052-011-1547-z}{\doi{10.1140/epjc/s10052-011-1547-z}},
\href{http://www.arXiv.org/abs/1009.2450}{\texttt{arXiv:1009.2450}}.

\bibitem{Czakon:2011xx}
\hrefCMSnoop {}{M.~Czakon and A.~Mitov, ``Top++: {A} program for the
  calculation of the top-pair cross-section at hadron colliders'',} \textit{
  Comput. Phys. Commun.} \textbf{ 185} (2014) 2930,
  \href{http://dx.doi.org/10.1016/j.cpc.2014.06.021}{\doi{10.1016/j.cpc.2014.06.021}},
  \href{http://www.arXiv.org/abs/1112.5675}{\texttt{arXiv:1112.5675}}.

\bibitem{Alwall:2014hca}
J.~Alwall\hrefCMSnoop {}{ {et~al.}, ``The automated computation of tree-level
  and next-to-leading order differential cross sections, and their matching to
  parton shower simulations'',} \textit{ JHEP} \textbf{ 07} (2014) 079,
  \href{http://dx.doi.org/10.1007/JHEP07(2014)079}{\doi{10.1007/JHEP07(2014)079}},
\href{http://www.arXiv.org/abs/1405.0301}{\texttt{arXiv:1405.0301}}.

\bibitem{Frixione:2002ik}
\hrefCMSnoop {}{S.~Frixione and B.~R. Webber, ``Matching {NLO QCD} computations
  and parton shower simulations'',} \textit{ JHEP} \textbf{ 06} (2002) 029,
  \href{http://dx.doi.org/10.1088/1126-6708/2002/06/029}{\doi{10.1088/1126-6708/2002/06/029}},
\href{http://www.arXiv.org/abs/hep-ph/0204244}{\texttt{arXiv:hep-ph/0204244}}.

\bibitem{Kidonakis:2010ux}
\hrefCMSnoop {}{N.~Kidonakis, ``Two-loop soft anomalous dimensions for single
  top quark associated production with a ${\PWm}$ or ${\PH}^-$'',} \textit{
  Phys. Rev. D} \textbf{ 82} (2010) 054018,
  \href{http://dx.doi.org/10.1103/PhysRevD.82.054018}{\doi{10.1103/PhysRevD.82.054018}},
  \href{http://www.arXiv.org/abs/1005.4451}{\texttt{arXiv:1005.4451}}.

\bibitem{Kidonakis:2013zqa}
\hrefCMSnoop {}{N.~Kidonakis, ``Top quark production'',} in \textit{ Helmholtz
  International Summer School on Physics of Heavy Quarks and Hadrons}.
\newblock 2013.
\newblock \href{http://www.arXiv.org/abs/1311.0283}{\texttt{arXiv:1311.0283}}.
\newblock
  \href{http://dx.doi.org/10.3204/DESY-PROC-2013-03/Kidonakis}{\doi{10.3204/DESY-PROC-2013-03/Kidonakis}}.

\bibitem{Aliev:2010zk}
M.~Aliev\hrefCMSnoop {}{ {et~al.}, ``{HATHOR} --- {HA}dronic {T}op and {H}eavy
  quarks cr{O}ss section calculato{R}'',} \textit{ Comput. Phys. Commun.}
  \textbf{ 182} (2011) 1034,
  \href{http://dx.doi.org/10.1016/j.cpc.2010.12.040}{\doi{10.1016/j.cpc.2010.12.040}},
  \href{http://www.arXiv.org/abs/1007.1327}{\texttt{arXiv:1007.1327}}.

\bibitem{Li:2012wna}
\hrefCMSnoop {}{Y.~Li and F.~Petriello, ``Combining {QCD} and electroweak
  corrections to dilepton production in {FEWZ}'',} \textit{ Phys. Rev. D}
  \textbf{ 86} (2012) 094034,
  \href{http://dx.doi.org/10.1103/PhysRevD.86.094034}{\doi{10.1103/PhysRevD.86.094034}},
\href{http://www.arXiv.org/abs/1208.5967}{\texttt{arXiv:1208.5967}}.

\bibitem{Sjostrand:2014zea}
T.~Sj{\"o}strand\hrefCMSnoop {}{ {et~al.}, ``An introduction to {PYTHIA
  8.2}'',} \textit{ Comput. Phys. Commun.} \textbf{ 191} (2015) 159,
  \href{http://dx.doi.org/10.1016/j.cpc.2015.01.024}{\doi{10.1016/j.cpc.2015.01.024}},
\href{http://www.arXiv.org/abs/1410.3012}{\texttt{arXiv:1410.3012}}.

\bibitem{Ball:2014uwa}
\hrefCMSnoop {}{{NNPDF} Collaboration, ``Parton distributions for the {LHC}
  {Run II}'',} \textit{ JHEP} \textbf{ 04} (2015) 040,
  \href{http://dx.doi.org/10.1007/JHEP04(2015)040}{\doi{10.1007/JHEP04(2015)040}},
\href{http://www.arXiv.org/abs/1410.8849}{\texttt{arXiv:1410.8849}}.

\bibitem{NNPDF:2017mvq}
\hrefCMSnoop {}{{NNPDF} Collaboration, ``Parton distributions from
  high-precision collider data'',} \textit{ Eur. Phys. J. C} \textbf{ 77}
  (2017) 663,
  \href{http://dx.doi.org/10.1140/epjc/s10052-017-5199-5}{\doi{10.1140/epjc/s10052-017-5199-5}},
  \href{http://www.arXiv.org/abs/1706.00428}{\texttt{arXiv:1706.00428}}.

\bibitem{Frederix:2012ps}
\hrefCMSnoop {}{R.~Frederix and S.~Frixione, ``Merging meets matching in
  {MC@NLO}'',} \textit{ JHEP} \textbf{ 12} (2012) 061,
  \href{http://dx.doi.org/10.1007/JHEP12(2012)061}{\doi{10.1007/JHEP12(2012)061}},
\href{http://www.arXiv.org/abs/1209.6215}{\texttt{arXiv:1209.6215}}.

\bibitem{Alwall:2007fs}
J.~Alwall\hrefCMSnoop {}{ {et~al.}, ``Comparative study of various algorithms
  for the merging of parton showers and matrix elements in hadronic
  collisions'',} \textit{ Eur. Phys. J. C} \textbf{ 53} (2008) 473,
  \href{http://dx.doi.org/10.1140/epjc/s10052-007-0490-5}{\doi{10.1140/epjc/s10052-007-0490-5}},
\href{http://www.arXiv.org/abs/0706.2569}{\texttt{arXiv:0706.2569}}.

\bibitem{Sirunyan:2019dfx}
\hrefCMSnoop {}{{CMS Collaboration}, ``Extraction and validation of a new set
  of {CMS} {PYTHIA8} tunes from underlying-event measurements'',} \textit{ Eur.
  Phys. J. C} \textbf{ 80} (2020) 4,
  \href{http://dx.doi.org/10.1140/epjc/s10052-019-7499-4}{\doi{10.1140/epjc/s10052-019-7499-4}},
\href{http://www.arXiv.org/abs/1903.12179}{\texttt{arXiv:1903.12179}}.

\bibitem{Khachatryan:2015pea}
\hrefCMSnoop {}{{CMS Collaboration}, ``Event generator tunes obtained from
  underlying event and multiparton scattering measurements'',} \textit{ Eur.
  Phys. J. C} \textbf{ 76} (2016) 155,
  \href{http://dx.doi.org/10.1140/epjc/s10052-016-3988-x}{\doi{10.1140/epjc/s10052-016-3988-x}},
\href{http://www.arXiv.org/abs/1512.00815}{\texttt{arXiv:1512.00815}}.

\bibitem{Agostinelli:2002hh}
\hrefCMSnoop {}{{GEANT4} Collaboration, ``{\GEANTfour} --- {A} simulation
  toolkit'',} \textit{ Nucl. Instrum. Meth. A} \textbf{ 506} (2003) 250,
\href{http://dx.doi.org/10.1016/S0168-9002(03)01368-8}{\doi{10.1016/S0168-9002(03)01368-8}}.

\bibitem{Allison:2006ve}
\hrefCMSnoop {}{J.~Allison {et~al.}, ``{\GEANTfour} developments and
  applications'',} \textit{ IEEE Trans. Nucl. Sci.} \textbf{ 53} (2006) 270,
\href{http://dx.doi.org/10.1109/TNS.2006.869826}{\doi{10.1109/TNS.2006.869826}}.

\bibitem{Sirunyan:2018nqx}
\hrefCMSnoop {}{{CMS Collaboration}, ``Measurement of the inelastic
  proton-proton cross section at $ \sqrt{s}=13 $ {TeV}'',} \textit{ JHEP}
  \textbf{ 07} (2018) 161,
  \href{http://dx.doi.org/10.1007/JHEP07(2018)161}{\doi{10.1007/JHEP07(2018)161}},
\href{http://www.arXiv.org/abs/1802.02613}{\texttt{arXiv:1802.02613}}.

\bibitem{Sirunyan:2017ulk}
\hrefCMSnoop {}{{CMS Collaboration}, ``Particle-flow reconstruction and global
  event description with the {CMS} detector'',} \textit{ JINST} \textbf{ 12}
  (2017) P10003,
  \href{http://dx.doi.org/10.1088/1748-0221/12/10/P10003}{\doi{10.1088/1748-0221/12/10/P10003}},
\href{http://www.arXiv.org/abs/1706.04965}{\texttt{arXiv:1706.04965}}.

\bibitem{Cacciari:2008gp}
\hrefCMSnoop {}{M.~Cacciari, G.~P. Salam, and G.~Soyez, ``The anti-\kt jet
  clustering algorithm'',} \textit{ JHEP} \textbf{ 04} (2008) 063,
  \href{http://dx.doi.org/10.1088/1126-6708/2008/04/063}{\doi{10.1088/1126-6708/2008/04/063}},
\href{http://www.arXiv.org/abs/0802.1189}{\texttt{arXiv:0802.1189}}.

\bibitem{Cacciari:2011ma}
\hrefCMSnoop {}{M.~Cacciari, G.~P. Salam, and G.~Soyez, ``Fastjet user
  manual'',} \textit{ Eur. Phys. J. C} \textbf{ 72} (2012) 1896,
  \href{http://dx.doi.org/10.1140/epjc/s10052-012-1896-2}{\doi{10.1140/epjc/s10052-012-1896-2}},
\href{http://www.arXiv.org/abs/1111.6097}{\texttt{arXiv:1111.6097}}.

\bibitem{CMS-TDR-15-02}
\href {http://cds.cern.ch/record/2020886}{{CMS Collaboration}, ``Technical
  proposal for the {P}hase-{II} upgrade of the {C}ompact {M}uon {S}olenoid'',}
  CMS Technical Proposal CERN-LHCC-2015-010, CMS-TDR-15-02, 2015.

\bibitem{Sirunyan:2018fpa}
\hrefCMSnoop {}{{CMS Collaboration}, ``Performance of the {CMS} muon detector
  and muon reconstruction with proton-proton collisions at $\sqrt{s}=13$
  {TeV}'',} \textit{ JINST} \textbf{ 13} (2018) P06015,
  \href{http://dx.doi.org/10.1088/1748-0221/13/06/P06015}{\doi{10.1088/1748-0221/13/06/P06015}},
\href{http://www.arXiv.org/abs/1804.04528}{\texttt{arXiv:1804.04528}}.

\bibitem{CMS:2020uim}
\hrefCMSnoop {}{{CMS Collaboration}, ``Electron and photon reconstruction and
  identification with the {CMS} experiment at the {CERN LHC}'',} \textit{
  JINST} \textbf{ 16} (2021) P05014,
  \href{http://dx.doi.org/10.1088/1748-0221/16/05/P05014}{\doi{10.1088/1748-0221/16/05/P05014}},
  \href{http://www.arXiv.org/abs/2012.06888}{\texttt{arXiv:2012.06888}}.

\bibitem{Stewart:2015waa}
I.~W. Stewart\hrefCMSnoop {}{ {et~al.}, ``{XCone}: ${N}$-jettiness as an
  exclusive cone jet algorithm'',} \textit{ JHEP} \textbf{ 11} (2015) 072,
  \href{http://dx.doi.org/10.1007/JHEP11(2015)072}{\doi{10.1007/JHEP11(2015)072}},
\href{http://www.arXiv.org/abs/1508.01516}{\texttt{arXiv:1508.01516}}.

\bibitem{Thaler:2015xaa}
\hrefCMSnoop {}{J.~Thaler and T.~F. Wilkason, ``Resolving boosted jets with
  {XCone}'',} \textit{ JHEP} \textbf{ 12} (2015) 051,
  \href{http://dx.doi.org/10.1007/JHEP12(2015)051}{\doi{10.1007/JHEP12(2015)051}},
\href{http://www.arXiv.org/abs/1508.01518}{\texttt{arXiv:1508.01518}}.

\bibitem{Krohn:2009th}
\hrefCMSnoop {}{D.~Krohn, J.~Thaler, and L.-T. Wang, ``Jet trimming'',}
  \textit{ JHEP} \textbf{ 02} (2010) 084,
  \href{http://dx.doi.org/10.1007/JHEP02(2010)084}{\doi{10.1007/JHEP02(2010)084}},
\href{http://www.arXiv.org/abs/0912.1342}{\texttt{arXiv:0912.1342}}.

\bibitem{CACluster1}
\hrefCMSnoop {}{Y.~L. Dokshitzer, G.~D. Leder, S.~Moretti, and B.~R. Webber,
  ``{Better jet clustering algorithms}'',} \textit{ JHEP} \textbf{ 08} (1997)
  001,
  \href{http://dx.doi.org/10.1088/1126-6708/1997/08/001}{\doi{10.1088/1126-6708/1997/08/001}},
\href{http://www.arXiv.org/abs/hep-ph/9707323}{\texttt{arXiv:hep-ph/9707323}}.

\bibitem{CACluster2}
\href
  {https://inspirehep.net/record/484872/files/arXiv:hep-ph_9907280.pdf}{M.~Wobisch
  and T.~Wengler, ``{Hadronization corrections to jet cross sections in
  deep-inelastic scattering}'',} in \textit{ {Workshop on Monte Carlo
  generators for HERA physics}}.
\newblock DESY, Hamburg, Germany, 1998.
\newblock
\href{http://www.arXiv.org/abs/hep-ph/9907280}{\texttt{arXiv:hep-ph/9907280}}.
\newblock

\bibitem{Sirunyan:2018ryr}
\hrefCMSnoop {}{{CMS Collaboration}, ``Search for resonant \ttbar production in
  proton-proton collisions at $\sqrt{s}=13$ {TeV}'',} \textit{ JHEP} \textbf{
  04} (2019) 031,
  \href{http://dx.doi.org/10.1007/JHEP04(2019)031}{\doi{10.1007/JHEP04(2019)031}},
\href{http://www.arXiv.org/abs/1810.05905}{\texttt{arXiv:1810.05905}}.

\bibitem{Sirunyan:2018rfo}
\hrefCMSnoop {}{{CMS Collaboration}, ``Search for a heavy resonance decaying to
  a top quark and a vector-like top quark in the lepton+jets final state in \pp
  collisions at $\sqrt{s} = 13$ {TeV}'',} \textit{ Eur. Phys. J. C} \textbf{
  79} (2019) 208,
  \href{http://dx.doi.org/10.1140/epjc/s10052-019-6688-5}{\doi{10.1140/epjc/s10052-019-6688-5}},
\href{http://www.arXiv.org/abs/1812.06489}{\texttt{arXiv:1812.06489}}.

\bibitem{CMS-DP-2018-033}
\href {https://cds.cern.ch/record/2627468}{{CMS Collaboration}, ``Performance
  of b tagging algorithms in proton-proton collisions at 13 {TeV} with {P}hase
  1 {CMS} detector'',} CMS Detector Performance Note CMS-DP-2018-033, 2018.

\bibitem{Bols:2020bkb}
E.~Bols\hrefCMSnoop {}{ {et~al.}, ``Jet flavour classification using
  {DeepJet}'',} \textit{ JINST} \textbf{ 15} (2020) P12012,
  \href{http://dx.doi.org/10.1088/1748-0221/15/12/P12012}{\doi{10.1088/1748-0221/15/12/P12012}},
  \href{http://www.arXiv.org/abs/2008.10519}{\texttt{arXiv:2008.10519}}.

\bibitem{Sirunyan:2019kia}
\hrefCMSnoop {}{{CMS Collaboration}, ``Performance of missing transverse
  momentum reconstruction in proton-proton collisions at $\sqrt{s} = 13$ {TeV}
  using the {CMS} detector'',} \textit{ JINST} \textbf{ 14} (2019) P07004,
  \href{http://dx.doi.org/10.1088/1748-0221/14/07/P07004}{\doi{10.1088/1748-0221/14/07/P07004}},
\href{http://www.arXiv.org/abs/1903.06078}{\texttt{arXiv:1903.06078}}.

\bibitem{CMS:2021vhb}
\hrefCMSnoop {}{{CMS Collaboration}, ``Measurement of differential $\ttbar$
  production cross sections in the full kinematic range using lepton+jets
  events from proton-proton collisions at $\sqrt {s}$ = 13\,{TeV}'',} \textit{
  Phys. Rev. D} \textbf{ 104} (2021) 092013,
  \href{http://dx.doi.org/10.1103/PhysRevD.104.092013}{\doi{10.1103/PhysRevD.104.092013}},
  \href{http://www.arXiv.org/abs/2108.02803}{\texttt{arXiv:2108.02803}}.

\bibitem{ATLAS:2022xfj}
\hrefCMSnoop {}{{ATLAS Collaboration}, ``Measurements of differential
  cross-sections in top-quark pair events with a high transverse momentum top
  quark and limits on beyond the standard model contributions to top-quark pair
  production with the {ATLAS} detector at $\sqrt{s} = 13$~{TeV}'',} \textit{
  JHEP} \textbf{ 06} (2022) 063,
  \href{http://dx.doi.org/10.1007/JHEP06(2022)063}{\doi{10.1007/JHEP06(2022)063}},
  \href{http://www.arXiv.org/abs/2202.12134}{\texttt{arXiv:2202.12134}}.

\bibitem{ATLAS:2022mlu}
\hrefCMSnoop {}{{ATLAS Collaboration}, ``Differential $\ttbar$ cross-section
  measurements using boosted top quarks in the all-hadronic final state with
  139 fb$^{-1}$ of {ATLAS} data'',} 2022.
  \href{http://www.arXiv.org/abs/2205.02817}{\texttt{arXiv:2205.02817}}.

\bibitem{Herren:2017osy}
\hrefCMSnoop {}{F.~Herren and M.~Steinhauser, ``Version 3 of {RunDec} and
  {CRunDec}'',} \textit{ Comput. Phys. Commun.} \textbf{ 224} (2018) 333,
  \href{http://dx.doi.org/10.1016/j.cpc.2017.11.014}{\doi{10.1016/j.cpc.2017.11.014}},
  \href{http://www.arXiv.org/abs/1703.03751}{\texttt{arXiv:1703.03751}}.

\bibitem{Schmitt:2012kp}
\hrefCMSnoop {}{S.~Schmitt, ``{TUnfold}: {An} algorithm for correcting
  migration effects in high energy physics'',} \textit{ JINST} \textbf{ 7}
  (2012) T10003,
  \href{http://dx.doi.org/10.1088/1748-0221/7/10/T10003}{\doi{10.1088/1748-0221/7/10/T10003}},
\href{http://www.arXiv.org/abs/1205.6201}{\texttt{arXiv:1205.6201}}.

\bibitem{Schmitt:2016orm}
\href
  {https://inspirehep.net/record/1495996/files/arXiv:1611.01927.pdf}{S.~Schmitt,
  ``Data unfolding methods in high energy physics'',} in \textit{ {12th
  Conference on Quark Confinement and the Hadron Spectrum}}.
\newblock Confinement XII, Thessaloniki, Greece, 2016.
\newblock
\href{http://www.arXiv.org/abs/1611.01927}{\texttt{arXiv:1611.01927}}.
\newblock

\bibitem{Sirunyan:2017ezt}
\hrefCMSnoop {}{{CMS Collaboration}, ``Identification of heavy-flavour jets
  with the {CMS} detector in {\pp} collisions at 13 {TeV}'',} \textit{ JINST}
  \textbf{ 13} (2018) P05011,
  \href{http://dx.doi.org/10.1088/1748-0221/13/05/P05011}{\doi{10.1088/1748-0221/13/05/P05011}},
\href{http://www.arXiv.org/abs/1712.07158}{\texttt{arXiv:1712.07158}}.

\bibitem{Bahr:2008pv}
\hrefCMSnoop {}{M.~Bahr {et~al.}, ``Herwig++ physics and manual'',} \textit{
  Eur. Phys. J. C} \textbf{ 58} (2008) 639,
  \href{http://dx.doi.org/10.1140/epjc/s10052-008-0798-9}{\doi{10.1140/epjc/s10052-008-0798-9}},
  \href{http://www.arXiv.org/abs/0803.0883}{\texttt{arXiv:0803.0883}}.

\bibitem{Gieseke:2012ft}
\hrefCMSnoop {}{S.~Gieseke, C.~Rohr, and A.~Siodmok, ``Colour reconnections in
  {H}erwig++'',} \textit{ Eur. Phys. J. C} \textbf{ 72} (2012) 2225,
  \href{http://dx.doi.org/10.1140/epjc/s10052-012-2225-5}{\doi{10.1140/epjc/s10052-012-2225-5}},
  \href{http://www.arXiv.org/abs/1206.0041}{\texttt{arXiv:1206.0041}}.

\bibitem{Khachatryan:2016kzg}
\hrefCMSnoop {}{{CMS Collaboration}, ``Measurement of the \ttbar production
  cross section using events in the {\Pe}$\mu$ final state in {\pp} collisions
  at $\sqrt{s} = 13$ {TeV}'',} \textit{ Eur. Phys. J. C} \textbf{ 77} (2017)
  172,
  \href{http://dx.doi.org/10.1140/epjc/s10052-017-4718-8}{\doi{10.1140/epjc/s10052-017-4718-8}},
\href{http://www.arXiv.org/abs/1611.04040}{\texttt{arXiv:1611.04040}}.

\bibitem{Chatrchyan:2014mua}
\hrefCMSnoop {}{{CMS Collaboration}, ``Measurement of inclusive {\PW} and {\PZ}
  boson production cross sections in {\pp} collisions at $\sqrt{s} = 8$
  {TeV}'',} \textit{ Phys. Rev. Lett.} \textbf{ 112} (2014) 191802,
  \href{http://dx.doi.org/10.1103/PhysRevLett.112.191802}{\doi{10.1103/PhysRevLett.112.191802}},
\href{http://www.arXiv.org/abs/1402.0923}{\texttt{arXiv:1402.0923}}.

\bibitem{Sirunyan:2016cdg}
\hrefCMSnoop {}{{CMS Collaboration}, ``Cross section measurement of $t$-channel
  single top quark production in {\pp} collisions at $\sqrt{s} = 13$ {TeV}'',}
  \textit{ Phys. Lett. B} \textbf{ 772} (2017) 752,
  \href{http://dx.doi.org/10.1016/j.physletb.2017.07.047}{\doi{10.1016/j.physletb.2017.07.047}},
\href{http://www.arXiv.org/abs/1610.00678}{\texttt{arXiv:1610.00678}}.

\bibitem{Kidonakis:2012rm}
\hrefCMSnoop {}{N.~Kidonakis, ``{NNLL} threshold resummation for top-pair and
  single-top production'',} \textit{ Phys. Part. Nucl.} \textbf{ 45} (2014)
  714,
  \href{http://dx.doi.org/10.1134/S1063779614040091}{\doi{10.1134/S1063779614040091}},
\href{http://www.arXiv.org/abs/1210.7813}{\texttt{arXiv:1210.7813}}.

\bibitem{Gehrmann:2014fva}
T.~Gehrmann\hrefCMSnoop {}{ {et~al.}, ``${\PW}^+{\PW}^-$ production at hadron
  colliders in next-to-next-to-leading order {QCD}'',} \textit{ Phys. Rev.
  Lett.} \textbf{ 113} (2014) 212001,
  \href{http://dx.doi.org/10.1103/PhysRevLett.113.212001}{\doi{10.1103/PhysRevLett.113.212001}},
\href{http://www.arXiv.org/abs/1408.5243}{\texttt{arXiv:1408.5243}}.

\bibitem{Khachatryan:2016tgp}
\hrefCMSnoop {}{{CMS Collaboration}, ``Measurement of the {WZ} production cross
  section in {\pp} collisions at $\sqrt{s} = 13$ {TeV}'',} \textit{ Phys. Lett.
  B} \textbf{ 766} (2017) 268,
  \href{http://dx.doi.org/10.1016/j.physletb.2017.01.011}{\doi{10.1016/j.physletb.2017.01.011}},
\href{http://www.arXiv.org/abs/1607.06943}{\texttt{arXiv:1607.06943}}.

\bibitem{CMS:2020dqt}
\hrefCMSnoop {}{{CMS Collaboration}, ``Development and validation of {HERWIG}~7
  tunes from {CMS} underlying-event measurements'',} \textit{ Eur. Phys. J. C}
  \textbf{ 81} (2021) 312,
  \href{http://dx.doi.org/10.1140/epjc/s10052-021-08949-5}{\doi{10.1140/epjc/s10052-021-08949-5}},
  \href{http://www.arXiv.org/abs/2011.03422}{\texttt{arXiv:2011.03422}}.

\bibitem{Sjostrand:1987su}
\hrefCMSnoop {}{T.~Sj{\"o}strand and M.~van Zijl, ``A multiple interaction
  model for the event structure in hadron collisions'',} \textit{ Phys. Rev. D}
  \textbf{ 36} (1987) 2019,
\href{http://dx.doi.org/10.1103/PhysRevD.36.2019}{\doi{10.1103/PhysRevD.36.2019}}.

\bibitem{Argyropoulos:2014zoa}
\hrefCMSnoop {}{S.~Argyropoulos and T.~Sj{\"o}strand, ``Effects of color
  reconnection on $\ttbar$ final states at the {LHC}'',} \textit{ JHEP}
  \textbf{ 11} (2014) 043,
  \href{http://dx.doi.org/10.1007/JHEP11(2014)043}{\doi{10.1007/JHEP11(2014)043}},
\href{http://www.arXiv.org/abs/1407.6653}{\texttt{arXiv:1407.6653}}.

\bibitem{Christiansen:2015yqa}
\hrefCMSnoop {}{J.~R. Christiansen and P.~Z. Skands, ``String formation beyond
  leading colour'',} \textit{ JHEP} \textbf{ 08} (2015) 003,
  \href{http://dx.doi.org/10.1007/JHEP08(2015)003}{\doi{10.1007/JHEP08(2015)003}},
\href{http://www.arXiv.org/abs/1505.01681}{\texttt{arXiv:1505.01681}}.

\bibitem{Butterworth:2015oua}
\hrefCMSnoop {}{J.~Butterworth {et~al.}, ``{PDF4LHC} recommendations for {LHC}
  {R}un {II}'',} \textit{ J. Phys. G} \textbf{ 43} (2016) 023001,
  \href{http://dx.doi.org/10.1088/0954-3899/43/2/023001}{\doi{10.1088/0954-3899/43/2/023001}},
  \href{http://www.arXiv.org/abs/1510.03865}{\texttt{arXiv:1510.03865}}.

\bibitem{Britzger:2021ocj}
\hrefCMSnoop {}{D.~Britzger, ``The linear template fit'',} \textit{ Eur. Phys.
  J. C} \textbf{ 82} (2022) 731,
  \href{http://dx.doi.org/10.1140/epjc/s10052-022-10581-w}{\doi{10.1140/epjc/s10052-022-10581-w}},
  \href{http://www.arXiv.org/abs/2112.01548}{\texttt{arXiv:2112.01548}}.

\bibitem{CMS:2020poo}
\hrefCMSnoop {}{{CMS Collaboration}, ``Identification of heavy, energetic,
  hadronically decaying particles using machine-learning techniques'',}
  \textit{ JINST} \textbf{ 15} (2020) P06005,
  \href{http://dx.doi.org/10.1088/1748-0221/15/06/P06005}{\doi{10.1088/1748-0221/15/06/P06005}},
  \href{http://www.arXiv.org/abs/2004.08262}{\texttt{arXiv:2004.08262}}.

\end{thebibliography}\endgroup
\cleardoublepage \appendix\section{The CMS Collaboration \label{app:collab}}\begin{sloppypar}\hyphenpenalty=5000\widowpenalty=500\clubpenalty=5000
\cmsinstitute{Yerevan Physics Institute, Yerevan, Armenia}
{\tolerance=6000
A.~Tumasyan\cmsAuthorMark{1}\cmsorcid{0009-0000-0684-6742}
\par}
\cmsinstitute{Institut f\"{u}r Hochenergiephysik, Vienna, Austria}
{\tolerance=6000
W.~Adam\cmsorcid{0000-0001-9099-4341}, J.W.~Andrejkovic, T.~Bergauer\cmsorcid{0000-0002-5786-0293}, S.~Chatterjee\cmsorcid{0000-0003-2660-0349}, K.~Damanakis\cmsorcid{0000-0001-5389-2872}, M.~Dragicevic\cmsorcid{0000-0003-1967-6783}, A.~Escalante~Del~Valle\cmsorcid{0000-0002-9702-6359}, P.S.~Hussain\cmsorcid{0000-0002-4825-5278}, M.~Jeitler\cmsAuthorMark{2}\cmsorcid{0000-0002-5141-9560}, N.~Krammer\cmsorcid{0000-0002-0548-0985}, L.~Lechner\cmsorcid{0000-0002-3065-1141}, D.~Liko\cmsorcid{0000-0002-3380-473X}, I.~Mikulec\cmsorcid{0000-0003-0385-2746}, P.~Paulitsch, F.M.~Pitters, J.~Schieck\cmsAuthorMark{2}\cmsorcid{0000-0002-1058-8093}, R.~Sch\"{o}fbeck\cmsorcid{0000-0002-2332-8784}, D.~Schwarz\cmsorcid{0000-0002-3821-7331}, S.~Templ\cmsorcid{0000-0003-3137-5692}, W.~Waltenberger\cmsorcid{0000-0002-6215-7228}, C.-E.~Wulz\cmsAuthorMark{2}\cmsorcid{0000-0001-9226-5812}
% ...
\par}
\vskip\cmsinstskip
\dag:~Deceased\\
$^{1}$Also at Yerevan State University, Yerevan, Armenia\\
$^{2}$Also at TU Wien, Vienna, Austria\\
% ...
\end{sloppypar}
\end{document}